\documentclass[%
 aip,
 jmp,%
 amsmath,amssymb,
 reprint,%
]{revtex4-1}

\usepackage{graphicx}
\usepackage{dcolumn}
\usepackage{bm}
\usepackage{color}
\usepackage{textcomp}
\usepackage{appendix}

\def\knat{Kugland2012a}
\def\krsi{Kugland2012}

\def\mclR{}
\def\mcln{}
\def\mcl{}

\def\rice{Department of Physics and Astronomy, Rice University, Houston, Texas 77005, USA}
\def\llnl{Lawrence Livermore National Laboratory, Livermore, California 94551, USA}
\def\ox{Clarendon Laboratory, University of Oxford, Parks Road, Oxford OX1 3PU, United Kingdom}
\def\EM{electromagnetic }
\def\dhe{$\mathrm{D}\ ^3\mathrm{He}$ }
\def\pr{\textit{PRIME }}

\def\ep{\mathcal{E}_{p}}
\def\tp{T_{p}}
\def\np{N_{p}}
\def\rp{r_{p}}
\def\vp{\vec{V}_{p}}
\def\lx{\ell_x}
\def\ly{\ell_y}
\def\lz{\ell_z}
\def\zs{z_s}
\def\zo{z_o}
\def\zi{z_i}
\def\pindex{\mathcal{P}}
\def\sindex{\mathcal{S}}
\def\dang{\alpha}

\begin{document}

\title{Development of \mclR{an} Interpretive Simulation Tool for the Proton Radiography Technique}

\author{M. C. Levy}
\email{levymc@stanford.edu}
\affiliation{\ox}
\affiliation{\llnl}
\author{D. D. Ryutov}
\affiliation{\llnl}
\author{S. C. Wilks}
\affiliation{\llnl}
\author{J. S. Ross}
\affiliation{\llnl}
\author{C. M. Huntington}
\affiliation{\llnl}
\author{F. Fiuza}
\affiliation{\llnl}
\author{D. A. Martinez}
\affiliation{\llnl}
\author{N. L. Kugland}
\affiliation{Lam Research Corporation, 4400 Cushing Parkway, Fremont, CA 94538 USA}
\author{M. G. Baring}
\affiliation{\rice}
\author{H-. S. Park}
\affiliation{\llnl}

\date{\today}

\begin{abstract}
Proton radiography is a useful diagnostic of high energy density (HED) plasmas under active theoretical and experimental development.
In this paper we describe a new simulation tool that interacts realistic laser-driven point-like proton sources with three dimensional \EM fields of arbitrary strength and structure and synthesizes the associated high resolution proton radiograph. The present tool's numerical approach captures all relevant physics effects, including effects related to the formation of caustics.
Electromagnetic fields can be imported from PIC or hydrodynamic codes in a streamlined fashion, and a library of \EM field `primitives' is also provided. 
This latter capability allows users to add a primitive, modify the field strength, rotate a primitive, and so on, while quickly generating a high resolution radiograph at each step.
In this way, our tool enables the user to deconstruct features in a radiograph and interpret them in connection to specific underlying \EM field elements.
We show an example application of the tool in connection to experimental observations of the Weibel instability in counterstreaming plasmas, using $\sim 10^8$ particles generated from a realistic laser-driven point-like proton source, imaging fields which cover volumes of $\sim10 $ mm$^3$. 
Insights derived from this application show that the tool can support understanding of  HED plasmas.

\end{abstract}

\maketitle

\section{Introduction}

Understanding the \EM field generation driven by intense laser-matter interactions is of fundamental importance to high energy density (HED) plasma physics\mcl{\cite{HEDP2003,Remington2006,Drake2006}}.
In this pursuit the proton radiography diagnostic technique\mcl{\cite{Hogan1999a,Borghesi2001,Borghesi2005,Pape2007,Borghesi2008a}} has enjoyed considerable success, providing insight into megagauss-scale \EM fields in inertial confinement fusion (ICF) implosions
\mcl{\cite{Mackinnon2006,Li2006a,Li2006,Kar2008,Rygg2008,Li2009a,Li2009,Gotchev2009,Borghesi2010,Sarri2010,Li2010,Manuel2012,Li2012,Zylstra2012}}
, large-scale self-organizing \EM field structures in high\mcln{-}velocity counter-streaming plasma flows\cite{\knat}, magnetic reconnection processes\mcl{\cite{Nilson2006,Li2007,Willingale2010}}, HED plasma instabilities\cite{Li2007a,Huntington2013,Fox2013,Gao2012,Gao2013} and more.

As implemented over the past decade, 
the proton radiography technique works by passing a low-density point-source-like proton beam through a HED plasma\mcl{\cite{Roth2002,Borghesi2003,Mackinnon2004,Borghesi2007,Romagnani2008,Cecchetti2009,Sokollik2009,Volpe2011,Quinn2012}}.  
The proton beam is typically generated using the \mcl{target normal sheath acceleration (TNSA)} process in which an ultraintense short pulse laser ($>10^{18}$ W cm$^{-2}$) irradiates a solid target, producing a polychromatic proton source with useful energies ranging from $\sim 5 - 60$ MeV\cite{Wilks2001}.  \mcln{L}aser-driven implosions of \dhe fusion capsules have also been employed to produce monoenergetic 3 and 14.7 MeV proton sources \cite{Li2006a,Li2006}.
The protons generated using either process propagate ballistically from the source to the interaction region containing the HED   plasma, deflect from the \EM fields according to the Lorentz force, then travel ballistically to a distant detector where the radiograph, a two dimensional fluence map, is recorded.
\mclR{
Collisional scattering is negligible across a broad range of plasma areal densities $10^{16}-10^{20}$ cm$^{-2}$.
As an example, the stopping power\cite{PSTAR} of a 10 MeV proton beam in Carbon is $\approx 41$ MeV cm$^{2}$ g$^{-1}$.  Thus collisional interactions over 1 mm of Carbon having number density $10^{20}$ cm$^{-3}$  (mass density 2 mg cm$^{-3}$) induce only a 0.1\% change in the proton beam energy.
Consequently, in these regimes the fluence map captures the \EM fields alone. 
}
Radiography generated in this way is a uniquely high performance diagnostic, imaging HED plasmas with extraordinary spatial resolution of \mcln{several micrometers} and temporal resolution of $1-10$ ps.

For the technique's virtues, the general question of how to interpret a radiograph in connection to its underlying \EM fields has remained open.
A key challenge stems from the fact that the radiographic image is not a one-to-one \EM field map, but rather forms a convolution of the three dimensional fields with the sampling proton properties.  
\mcl{Useful aspects of the field geometry have been deduced from qualitative inspection\cite{Maksimchuk2000,Borghesi2002,Borghesi2002a,borghesi2005fast,Pape2007,Borghesi2008a,Loupias2009,Willingale2010,Borghesi2010,Gregory2010,Chen2012}
, and by means of quantitative estimates based on scalings of the Lorentz force\mcl{\cite{jackson1999classical}} when features of the plasma are known.\cite{Nilson2006,Li2007,Li2008,Rygg2008,Petrasso2009,Sarri2009,Li2010,Li2010a,Willingale2010,Willingale2011,Seguin2012,Zylstra2012}.}
\mcl{Recently analytic theory describing the deconvolution has been developed\cite{\krsi}, but its application is constrained to simple field geometries and low field strengths, since the general mapping is nonlinear and degenerate.}

\begin{figure*}[t]
\begin{center}
\resizebox{15cm}{!}{\includegraphics{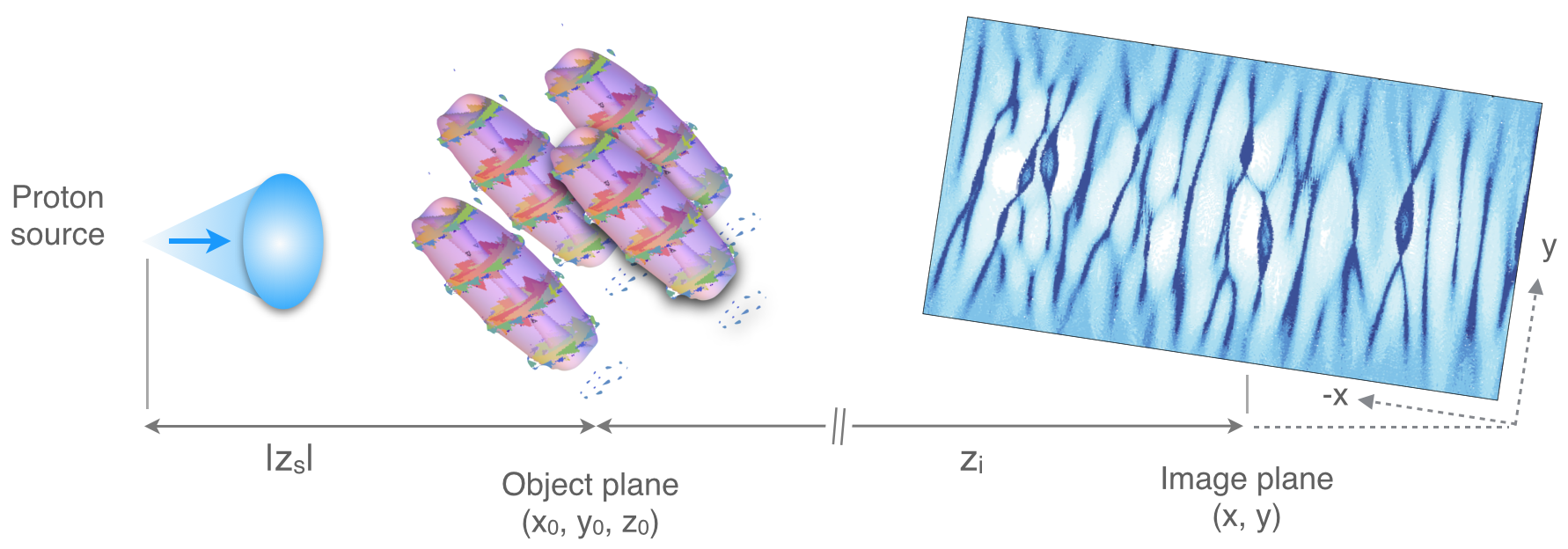}}
  \end{center}
  \caption{\mcln{Schematic of key aspects of the proton radiography simulation tool, following the propagation of protons along the $z$ axis from left to right.
  Parameters controlling the proton source are described in section \ref{sec:source}.  As an example,  specifying the source control vector $\vec{S}=(2,\ 14.7 \text{MeV},\ 10^9,\ 1  \text{cm})$ creates a 14.7 MeV monoenergetic proton source isotropically emitting 1 billion protons, imitating a realistic \dhe source, situated $|z_s|=1$ cm from the object plane containing the plasma \EM fields.  
  Depicted in the object plane at $z=0$ are four tilted ellipsoidal magnetic  filaments, each having form given by equation (\ref{eqn:bPhi}).
  These fields are created in \pr by specifying the single field control vector $\vec{L} = (4, 2, 2, 100 \mathrm{\mu m}, 50 \mathrm{\mu m}, 500 \mathrm{\mu m}, 0, 0, 1 \text{MG})$, as described in section \ref{sec:fieldInput}.
  A simulated proton radiograph created by the tool is shown in the image plane, situated at $z=z_i$.  Details on the field structure underpinning this radiograph 
  are covered in section \ref{sec:example}.
  }}
  \label{fig:schem} 
\end{figure*}

\mcl{
Numerical simulations can provide insight into a broader range of situations when  plasma-dynamical modeling and synthetic radiography modeling tools are used in concert.
In the former role, particle-in-cell (PIC) codes are typically employed when kinetic features must be resolved, and hydrodynamic codes when the plasma electron and ion collisional mean-free paths are small relative to the lengthscales of interest.
The latter role of simulating the proton radiograph, given the sampling proton properties and the configuration of plasma and \EM field, can \mcln{be} filled using either a `ray trace'
or Monte-Carlo code.\mclR{\cite{Aufderheide2000,MCNP5,Borghesi2003,Mackinnon2004,Li2006,Li2006a,Borghesi2007,Li2007,Borghesi2008a,Romagnani2008,Cecchetti2009,Sokollik2009,Sarri2010,Li2009,Manuel2012a,Manuel2012b,Quinn2012}}
}
\mcl{In the ray trace simulation model, a number of straight-line trajectories (rays) are created at the source some distance from the detector, connecting to the detector.  The \EM fields along a given ray are path-integrated and a corresponding net Lorentz deflection is applied to that ray's final position. Ray tracing codes have been widely used not only for protons, but also for neutrons, x-rays and so on, addressing other physical processes such as absorption and scattering.
The Monte-Carlo, or discretized, numerical approach by contrast represents protons as test particles having the appropriate mass and time-dependent phase space coordinates.  
As a consequence, all relevant physical processes can be included in the simulation. 
}

In this paper we describe a new simulation tool that interacts realistic laser-driven point-like proton sources with three dimensional \EM fields of arbitrary strength and structure, using the \mcl{discretized} method, and synthesizes the associated high resolution proton radiograph.  \mcl{The tool, called  \pr for Proton Radiograph IMage Exposition, has been developed to support regimes of operation matching those found in the emerging field of HED plasma science.} 
\mcln{A schematic of the tool's workings is shown in Fig. \ref{fig:schem}.} 
The present tool's implementation of the \mcl{discretized} numerical approach  captures all relevant physics effects, including \mcl{effects related to the formation of caustics \cite{\krsi}}.
Electromagnetic fields can be imported from PIC or hydrodynamic codes in a streamlined fashion. 
A library of \EM field `primitives' is also provided.  \mcl{These primitives can be considered `eigenvectors,' in effect spanning the basis of \EM fields, such that through linear combinations the user may construct realistic field topologies by hand.}
This capability allows users to add a primitive, modify the field strength, rotate a primitive, and so on, while quickly generating a high resolution radiograph at each step.
In this way, \pr enables the user to deconstruct features in a radiograph and interpret them in connection to specific underlying \EM field elements.
In this paper we show results from high resolution simulations performed in connection to experimental observations of the Weibel instability in counterstreaming plasmas\cite{Huntington2013}, using $\sim 10^8$ particles generated from a realistic laser-driven point-like proton source, imaging fields which cover volumes of $\sim10 $mm$^3$.  These results show that \pr can support understanding of \mcln{a broad range of} HED plasmas.

%


\section{Features of \pr}

\pr is a three dimensional simulation tool that we have been developing for modeling HED plasma situations.   
Both realistic TNSA and \dhe (14.7 and 3 MeV) laser-driven proton sources have been tested in experimentally-realistic configurations and are available to the user.   Additionally the user has the ability to specify a proton source having arbitrary spectral properties. 
We anticipate that this radiography tool will have two primary \mcl{uses}.  The first is in constructing \EM field structures using primitives, guided by the predictions of plasma physics theory and PIC and hydrodynamic simulation results.  This approach provides the advantage that fields are free of numerical noise, a key issue arising in kinetic simulations of millimeter and larger-scale plasmas.  Here the user also has the capability to add a primitive, modify the field strength, rotate a primitive, and so on, while quickly generating a high resolution radiograph at each step.
In this manner \pr should provide insights into the crucial question of how to interpret proton radiographs.
We also anticipate that synthetic radiographs produced by this tool should become particularly useful in cases where running PIC and hydrodynamic codes is computationally infeasible, and further to guide these expensive simulations towards larger scales. 
The second \mcl{use} of this tool will be in quickly and efficiently simulating a high resolution proton radiograph associated with \EM fields exported from PIC and hydrodynamic codes. For this purpose we have built in the capability to import fields directly from a variety of existing codes (e.g. OSIRIS\cite{Fonseca2008}).

Related to the first use,  the standard object description in \pr is a three dimensional \EM primitive describing the volumetric field structure.  \mclR{Descriptions, schematics and simulations of these primitives are given in Appendix A.}
The user has a number of high level options for inputting these fields, for example generating a lattice of primitives or programmatically including randomization effects, that are enumerated in section \ref{sec:fieldInput}.
\mclR{By combining primitives together the user can simulate fields representative of a large number of important HED processes including electrostatic shock waves, magnetized cylindrical shocks, \mcl{two-stream and other electrostatic instabilities,} intense laser-driven $\nabla n \times \nabla T$ `Biermann battery' magnetic fields (for plasma density $n$ and temperature $T$), magnetic fields creation by collisional current drive in interpenetrating plasma jets and filamentary magnetic field structures generated via the Weibel instability\cite{\knat,Ryutov2011,Huntington2013,Ryutov2013,Kugland2013}. }


With respect to numerical schemes, in \pr we have implemented a modular approach in order to accurately and efficiently simulate the proton radiography technique.  This is motivated by the disparate spatial scales characterizing the source -- plasma -- detector system.  The macroscopic volume is vast: the detector typically sweeps out an area $\sim 25$cm$^2$ and the axial distance between the source and detector, passing through the interaction region containing the HED plasma, can exceed $>10$cm.  At the same time the microscopic field structures associated with the plasma often have spatial scales of $\sim \mu$m.  Simulating the full volume of the cone connecting the source to the detector resolving the \EM fields would require $\sim 10^{14}$ grid cells. This situation clearly exceeds reasonable computational efforts.
Therefore to mitigate this issue in \pr we have divided the system into three regions. 
\mcl{The tool covers the source-to-plasma object region, region containing the plasma object itself, and plasma object-to-detector region, as well as the interfaces connecting them.}
In the plasma region we are currently using LSP\cite{Welch2004} for the particle push.  This provides the additional advantage that scattering models for dense plasmas as well as deflections due to \EM forces can be included.  
The modular approach in \pr allows a set of \EM fields to be specified, then different proton sources and different detectors to be `hooked up' to these fields in a streamlined manner.  
For example in section \ref{sec:example} we show several high resolution proton radiography results of filamentation-instability-driven fields, obtained by keeping the fields unchanged while swapping between realistic proton sources. By allowing users to quickly image the same field configuration using a TNSA proton source, and 3 MeV and 14.7 MeV \dhe proton sources, we show that \pr can help unravel the convolution between the properties of the source and those of the \EM fields.
The particle push and other parts of the code have been parallelized in order to take advantage of the Lawrence Livermore National Laboratory (LLNL) Livermore Computing (LC) Linux architecture, enabling efficient radiography simulations. 
\mclR{
As such, in order to access the tool at this stage, we ask that interested scientists  please correspond with the one of authors.
}

\subsection{Tools for constructing \EM fields\label{sec:fieldInput}}



A robust set of tools is available to the user for constructing \EM fields in \pr.
\mclR{The complete library of analytic \EM field primitives described in \mclR{ref. \cite{\krsi}} is available to the user, including electrostatic Gaussian ellipsoids, magnetic flux ropes and magnetostatic Gaussian ellipsoids.  Their associated functional forms and schematics are enumerated in Appendix A.
}
While the length scales of the primitives set the grid resolution, the particle push timestep is adjusted to the Courant condition\cite{birdsall2004plasma} evaluated using the velocity of the protons, enabling efficient and fast simulations.
Each primitive is controlled by a set of parameters governing the nominal peak electric (magnetic) field strength $E_0$ ($B_0$), the Cartesian position of the primitive's centroid $(x_0,y_0,z_0)$ with respect to the center of the region containing the HED plasma and two angles $\theta$ and $\psi$ indicating the primitive's polar and azimuthal angles relative to the proton propagation axis $\hat{z}$.  The sign of $E_0$ ($B_0$) determines whether protons interacting with the primitive will experience a focusing ($E_0, B_0 > 0$) or defocusing force.
Spatial extent is specified, taking the ellipsoids for example, using the parameters $a$ and $b$ representing the major and semi-major axes respectively.
By varying the ratio $a/b$ the user can produce field structures \mcl{representative of} Weibel instability-\mcln{driven} magnetic filaments, as well as advecting laser-driven Biermann battery-like magnetic `pancakes' \cite{\krsi,Kugland2013,Ryutov2013}.

 The user can construct a field topology featuring an arbitrary number of primitives, each having unique parameters.  
 A number of input methods describing configurations of several primitives are available to the user.
At the lowest level, the user specifies a list of $N$ field control vectors each having the form,
\begin{eqnarray}
\vec{G}_n = (\pindex, x_0, y_0, z_0, \theta, \psi, a, b, E_0 (B_0)  )
\label{eqn:basicVec}
\end{eqnarray}
which are then transformed by the tool into $N$ volumetric fields in the three dimensional simulation \mcl{($n\in[1,N]$)}.  The $\pindex$ element is an integer mapping to the desired primitive type \mclR{($\pindex=1$ corresponds to an electrostatic Gaussian ellipsoid, for example; see Appendix A for the complete enumeration)}.
In the simulation overlapping regions of field have $E$ and $B$ automatically summed.

Higher level input options are also available to the user.
To support modeling of periodic systems, a lattice of primitives can be generated programmatically by specifying a single field control vector of the form,
\begin{eqnarray}
\vec{L} = (\pindex, N_{rows}, N_{cols}, d_{rows}, d_{cols}, a, b, \theta, \psi, E_0 (B_0))
\label{eqn:latticeVec}
\end{eqnarray}
The tool transforms this vector into a body-centered rectangular prism lattice of $N_{rows} \times N_{cols}$ primitives of type $\pindex$.  The lattice is centered at the origin of the plasma region and the $n^{th}$ primitive has the centroid position $(x_{0,n},0,z_{0,n})$. Rows are oriented along $\hat{z}$ and columns are oriented along $\hat{x}$. $N_{rows}$ is thus the number of primitives in the lattice in $\hat{z}$ and $d_{rows}$ is the spacing between primitives in $\hat{z}$. Similarly the `cols' subscript corresponds to periodicity in $\hat{x}$. 

To support more realistic field configurations, high level input options that enable randomization effects are also available to the user.
By appending the elements $(\delta \theta, \delta \psi, \delta E_0 (\delta B_0) )$ to the lattice-generating vector $\vec{L}$, the user can programmatically make unique the $\theta, \psi$ and $E_0$ ($B_0$) values for each primitive.  Taking the altitude angle as an example, specifying $\delta \theta = 0$ (or omitting the $\delta$ elements) means that $\theta_n = \theta$ for the $n^{th}$ primitive. Randomization effects enter as specifying a nonzero $\delta \theta$ applies the mapping $\theta_n \rightarrow \theta + \delta \theta_n$, where $\delta \theta_n \in [-\delta \theta, \delta \theta]$ is sampled randomly within this interval for each primitive.  Individualized parameter effects can be as small or large as desired, and are generally quite important since they introduce a realistic asymmetry into the simulation. 
Indeed, section \ref{sec:example} below discusses the significant impact on the resulting proton radiographs of $\delta \theta$ and $\delta \psi$ effects in representing filamentation-instability-driven fields.




\subsection{Specifying source and detector properties\label{sec:source}}

Two methods of proton beam generation are supported, which together offer users the capability to specify sources with arbitrary spectral properties.  The first method produces point proton sources.  The user can choose a temperature $\tp$ representing TNSA-generated protons having a quasi-Maxwellian distribution.  Alternatively, with the first method users can specify an energy $\ep$ to generate a mononergetic point proton source.  Specifying $\ep= 3$ or 14.7  MeV reproduces the properties of protons generated through fusion reactions in intense laser-imploded \dhe capsules.   
In addition to setting the energy parameter, users also choose the number of protons to simulate $\np$ and the axial position of the source, $\zs < 0 $, relative to the object plane containing the HED plasma at $\zo = 0$.  The dimensions of the plasma region $\lx, \ly$ and $\lz$ (lengths in $x$, $y$ and $z$ respectively) are determined automatically such that they contain the plasma. This region is centered at $(0,0,0)$ and is situated between \mcl{$|x| \leq \lx/2, |y| \leq \ly/2$ and $|z| \leq \lz/2$.} The proton source is then instantiated in the simulation at the position $(0, 0, \zs)$
with a phase space distribution corresponding to a point source according to these parameters.
In short, a realistic point proton source is created in \pr by specifying a single source control vector of the form $\vec{S}=(\sindex, \tp (\ep), \np, \zs)$, where $\sindex = 1$ and the second element is $\tp$ for a TNSA source or $\sindex=2$ and the second element is $\ep$ for a monoenergetic source.

In the second proton generation method, the user specifies the proton source `spot' size $\rp$ in addition to $\np$ and $\zs$. The source is then instantiated in the simulation at $z=\zs$ with finite transverse size between $x^2 +y^2 \leq \rp^2$.
The proton beam divergence and energy distribution are specified through a combination of the beam thermal temperature $\tp$ and a vector drift velocity $\vp$. 
The user can specify spatial variations in both $\tp(x,y)$ and $\vp(x,y)$  across the source. This allows a high degree of customization of beam properties, \mclR{e.g., simulating temperature (Doppler) broadening and finite source size effects in an otherwise monoenergetic point proton source,} or reproducing a plane proton source when $\tp=0$ and $\vp(x,y) = (0, 0, const.)$.


To support a range of conditions, the user has the option to specify the detector properties in addition to the source properties.  
The user may choose the image plane axial position $\zi$ of the detector corresponding to the nominal magnification $M = - \zi / \zs$. The user also can specify the size of the detector and the binning resolution in each transverse direction.
If no detector attributes are chosen, the default detector will be instantiated in the simulation with infinite transverse dimensions at $\zi=10$cm, with $30\mu$m $\times 30\mu$m resolution in nominal object plane units.
\mclR{
By default the detector records the $(x,y)$ positions of the protons it collects at $z=\zi$, using weighting that is irrespective of energy.
The user may choose to expand the set of recorded quantities to include the proton velocity $\vec{v}=(v_x,v_y,v_z)$.  This capability enables the determination of proton energy deposition within a finite bandwidth, as is useful for many purposes, e.g., for a monoenergetic source, supporting the identification of magnetic  ($d/dt\ 1/2 m_p \vec{v}\cdot \vec{v} = 0$) versus electric  ($\neq 0$) deflections.
}

\section{Benchmarking against analytic theory\label{sec:analytics}}

Analytic theory describing the connection between \EM fields and the fluence images produced by sampling protons has been developed in Kugland \textit{et al.}\cite{\krsi}  In this section predictions of this formalism are compared to results produced by our numerical radiography tool.

\begin{figure*}[t]
\begin{center}
\resizebox{17cm}{!}{\includegraphics{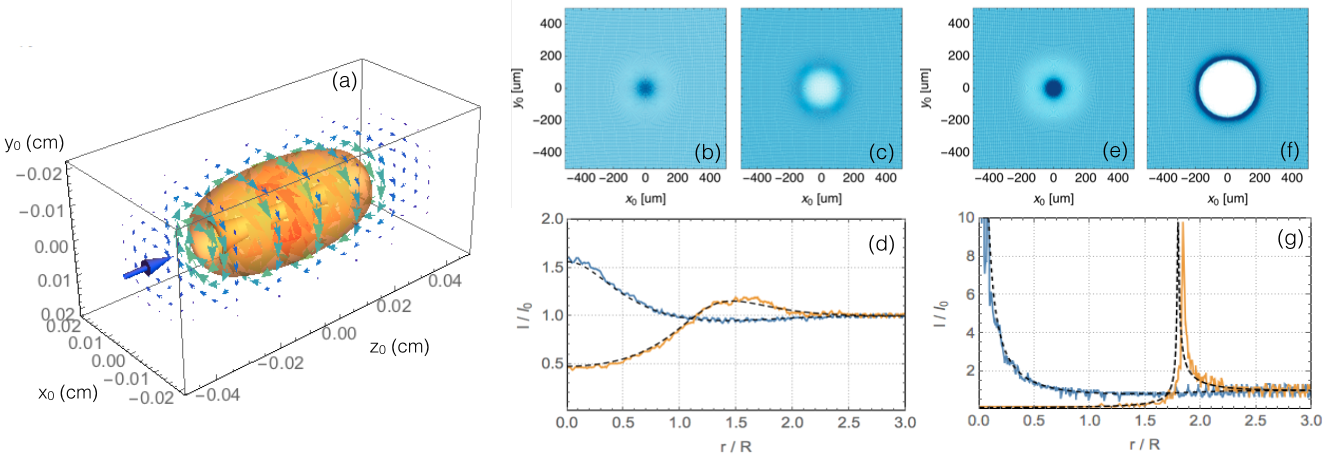}}
  \end{center}
  \caption{Comparison between theory and simulation results in the linear regime.  The situations shown here interact a monoenergetic $\ep = 14.7$ MeV proton source of $|\zs|=1$ cm, $\zi=10$ cm with a single magnetic filament of the form given by equation (\ref{eqn:bPhi}) having $a=100 \mu$m and $b=300 \mu$m.
 \mclR{(a) illustrates the three dimensional proton-field interaction geometry. The transparent orange surface represents an isocontour of the field magnitude $|B|$, and the colored arrows show the vector $B_\varphi$ field, with both arrow size and color corresponding to field strength. The blue three dimensional arrow indicates the axis of proton propagation.} (b-c) show  simulated radiograph results using $B_0 = 0.2 \times B_{0crit}$ for the focusing and defocusing cases, respectively, while (e-f) use  $B_0 = 0.9 \times B_{0crit}$. The color scale is fixed between images with darker (lighter) regions indicating a surplus (deficit) of protons.  (d) and (g) depict normalized lineouts of the proton fluence along $y_0=0$ for the $0.2 \times B_{0crit}$ and  $0.9 \times B_{0crit}$ simulations, respectively. The blue curves correspond to the focusing cases and the yellow curves  to the defocusing cases. The simulations agree with the theory predictions of equations (\ref{eqn:derivR0}-\ref{eqn:derivR}), indicated using dashed black curves, to better than 5\% in all cases.
  }
  \label{fig:bench1} 
\end{figure*}

Consider a Gaussian ellipsoidal `cocoon' filled with magnetic field having only an azimuthal ($\varphi$) component,
\begin{eqnarray}
B_\varphi = B_0 \frac{r}{a} \exp\left( - \frac{r^2}{a^2} -\frac{z^2}{b^2} \right)
\label{eqn:bPhi}
\end{eqnarray}
for radial coordinate $r$, axial coordinate $z$ and semi-major and major axes $b$ and $a$, respectively. For elongated $b>a$ situations this field structure resembles \mcln{a single} Weibel instability-\mcln{driven} magnetic filament\cite{\knat,Kugland2013,Ryutov2013}.
Note that in this representation $B_0$ is not a maximum value of the field; the maximum is reached at $r=a/\sqrt{2}$ and is equal to $B_{\mathrm{peak}}=B_0/\sqrt{2 e} \approx 0.43 B_0$ where $e$ is the natural logarithm base. 
To create this primitive in the radiography tool the user specifies the index $\pindex=4$ in conjunction with equation (\ref{eqn:basicVec}).
We assume that the distance from the source to the center of the object is $|\zs|=1$ cm, the distance from the center to the image plane is $\zi=10$ cm, proton energy is $\ep=1/2 m_p v_p^2 = 14.7$ MeV for proton mass $m_p$ and velocity $v_p$, $a=100 \mu$m and $b=300 \mu$m.

This situation is therefore consistent with the paraxial approximation ($a/|\zs| \sim 10^{-2}$). In the analytic evaluation of the proton deflection we use the smallness of the dimension $b$ compared to the proton gyroradius $\rho \sim 3$ cm for the fields that are needed to form the caustics. This allows us to use a linear approximation: integration of the transverse force over the unperturbed (straight) trajectory within the field structure. The anticipated error of this assumption is less than 10\%. With that, we find that deflection angle $\dang$ is related to the radius $r_0$ of the point where protons intersect the object plane by,
\begin{eqnarray}
\dang = \mu \frac{r_0}{a} \exp\left( -\frac{r_0^2}{a^2} \right)
\label{eqn:thetaDeflect}
\end{eqnarray}
where
\begin{eqnarray}
\mu = \frac{ \sqrt{\pi} |e| B_0 b}{ m_p v_p c}
\label{eqn:muDefn}
\end{eqnarray}
is a dimensionless parameter characterizing the interaction and $e$ is the fundamental charge.  For the 14.7 MeV proton source $v_p/c = 0.177$ and $\mu = 3.2 \times 10^{-6} B_0 \mathrm{[T]} \ b \mathrm{[\mu m]}$.
The position of the point in the image plane is determined by,
\begin{eqnarray}
r=\zi \left( -\frac{r_0}{\zs} \mp \dang(r_0) \right)
\label{eqn:rR0}
\end{eqnarray}
where the sign `minus' corresponds to the focusing case and the sign `plus' to a defocusing case. The derivative  $dr /dr_0$ is,
\begin{eqnarray}
\frac{dr}{dr_0} &=& - \frac{\zi}{\zs} \left[ 1 \mp \frac{\mu |\zs|}{a} f(r_0/a) \right] \\ 
f(r_0/a) &=& \left( 1-2\frac{r_0^2}{a^2} \right) e^{-r_0^2/a^2}
\label{eqn:derivativeR}
\end{eqnarray}
For small $\mu$ (small magnetic field) the second term is negligible and one has just a uniform magnification. When one increases $\mu$, the condition $dr/dr_0 =0$ is finally met at some $\mu_{crit}$ having different values for the focusing and defocusing cases.
For the focusing case the critical value is,
\begin{eqnarray}
\mu_{crit} = -\frac{a}{\zs}
\label{eqn:muCritFocus}
\end{eqnarray}
whereas for the defocusing case,
\begin{eqnarray}
\mu_{crit} = -\frac{a}{\zs} \frac{e^{3/2}}{2} \approx -2.24 \frac{a}{\zs}
\label{eqn:muCritDeFocus}
\end{eqnarray}


Introducing values of the universal constants one arrives at the following expressions for the critical magnetic fields,
\begin{eqnarray}
B_{0crit}\ \mathrm{[T]} = -8.12 \frac{a}{b} \frac{\sqrt{\ep \mathrm{[MeV]}}}{\zs \mathrm{[cm]}}
\label{eqn:bCritFocus}
\end{eqnarray}
and 
\begin{eqnarray}
B_{0crit}\ \mathrm{[T]} = -18.2 \frac{a}{b} \frac{\sqrt{\ep \mathrm{[MeV]}}}{\zs \mathrm{[cm]}}
\label{eqn:bCritDeFocus}
\end{eqnarray}
for the focusing and defocusing cases, respectively. Using the input parameters for these test cases, we find the fields of 10.38 T and 23.26 T, respectively.

Using equations (\ref{eqn:thetaDeflect}-\ref{eqn:derivativeR}) the intensity distribution in the image plane for $\mu$ smaller than critical can be presented in parametric form as,
\begin{eqnarray}
\frac{I}{I_0} &=& \left| e^{-2 t^2} \ \left(\nu  \mp e^{t^2}\right) \
\left(e^{t^2} \mp \nu  \left(1-2 t^2\right)  \right) \right|^{-1} \label{eqn:derivR0} \\  
\frac{r}{R} & = & t \left|1 \mp  \nu e^{-t^2} \right|, \quad \nu \equiv -\frac{\mu \zs}{a} 
\label{eqn:derivR}
\end{eqnarray}
for parameter $t$.
Here $I_0$ is the intensity in the center of the image plane in the absence of an object and $R=-\zi a/\zs$.

\begin{figure}[h]
\begin{center}
\resizebox{8cm}{!}{\includegraphics{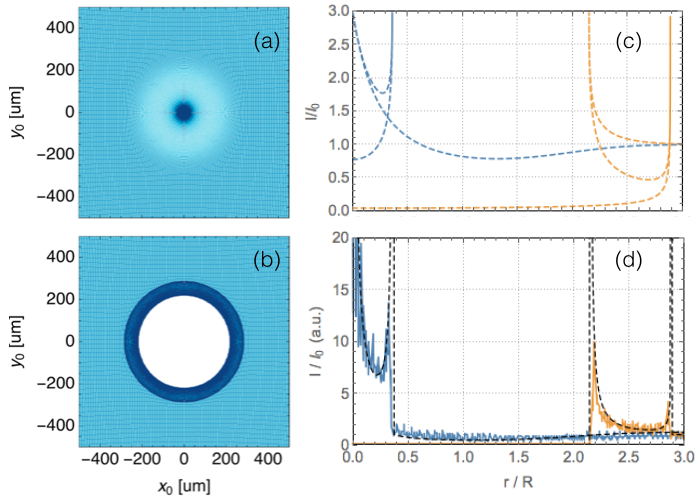}}
  \end{center}
  \caption{Comparison between theory and simulation results in the cau\mcln{s}tic regime.
  In these simulations a monoenergetic $\ep = 14.7$ MeV proton source of $|\zs|=1$ cm, $\zi=10$ cm interacts with a single magnetic filament of the form given by equation (\ref{eqn:bPhi}) having $a=100 \mu$m and $b=300 \mu$m.  \mclR{The interaction geometry is identical to that shown in Fig. \ref{fig:bench1} (a).}
  (a-b) show  simulated radiograph results using $B_0 = 2 \times B_{0crit}$ for the focusing and defocusing cases, respectively. The color scale is fixed between images with darker (lighter) regions indicating a surplus (deficit) of protons. 
  (d) shows normalized lineouts of the simulated proton fluence along $y_0=0$ with the blue curve corresponding to the focusing case and the yellow curve corresponding to the defocusing case.  The complete analytic results formed by summing over all three branches of each curve in (c) are indicated by the dashed b\mcln{l}ack lines in (d), exhibiting close agreement with the simulation results.
  (c) shows the multi-branched caustic structures predicted by the parametric equations (\ref{eqn:derivR}) and (\ref{eqn:derivR0epsilon}) using $\epsilon=0.5$. 
  }
  \label{fig:bench2} 
\end{figure}
	One can also plot intensity distributions for the fields exceeding critical values. In order to do so the amplitude-limiting factor $\epsilon$ as described in ref. \cite{\krsi} must be accounted for in equation (\ref{eqn:derivR0}).  The appropriate parametric relation for the normalized image plane intensity is then given by,
	\begin{eqnarray}
	\frac{I}{I_0} &=& \frac{1+\epsilon}{\epsilon+\left| e^{-2 t^2} \left(\nu  \mp e^{t^2}\right) \
\left(e^{t^2} \mp \nu  \left(1-2 t^2\right)  \right) \right|} \label{eqn:derivR0epsilon} 
\end{eqnarray}
	in concert with equation (\ref{eqn:derivR}) for $r/R$.  The transformation \mcln{$z_s\rightarrow \tilde{z_s}$ where $\tilde{z_s}=z_s z_i/(z_i-z_s)$} enhances the accuracy of equations (\ref{eqn:derivR0}-\ref{eqn:derivR0epsilon}) by relaxing constraints on the relationship between $\zi$ and $\zs$\cite{\krsi}.

	We now validate the synthetic radiographs produced by the numerical tool through comparison to equations (\ref{eqn:derivR0}-\ref{eqn:derivR0epsilon}).
	Fig. \ref{fig:bench1} shows the results of this procedure for four simulations in the linear regime.
	  (b-c) show synthetic proton radiographs generated by the tool using $B_0 = 0.2 \times B_{0crit}$ for the focusing and defocusing cases, respectively. The color scale is fixed between images (and Figs. \ref{fig:bench1}-\ref{fig:filsParam2}), with darker (lighter) regions indicating a surplus (deficit) of protons. The spatial coordinates are  provided in nominal object plane units $x_0$ and $y_0$, i.e., $1/M \times x, y$.  (d) depicts normalized lineouts of the proton fluence along $y_0=0$ with the blue curve corresponding to the focusing simulation and the yellow curve corresponding to the defocusing simulation.  The black dashed curves correspond to analytic theory from equation (\ref{eqn:derivR}).  (e-g) show the same set of plots for simulations and theory corresponding to the field strength  $B_0 = 0.9 \times B_{0crit}$.
	\mcln{Panels} (d) and (g) highlight the excellent agreement between theory and the simulated radiographs \mcln{across conditions}.

	Fig. \ref{fig:bench2} shows results comparing simulations to the predictions of equations (\ref{eqn:derivR}) and (\ref{eqn:derivR0epsilon}) for proton imaging in the nonlinear regime. (a-b) show the synthetic proton radiographs having nonlinear  field strength $B_0 = 2 \times B_{0crit}$ for the focusing and defocusing cases, respectively.  (c) shows the multi-branched caustic structures predicted by the parametric equations (\ref{eqn:derivR}) and (\ref{eqn:derivR0epsilon}).	
(d) shows normalized lineouts of the simulated proton fluence along $y_0=0$ with the blue curve corresponding to the focusing case and the yellow curve corresponding to the defocusing case.  The complete analytic results formed by summing over all three branches of each curve in (c) are indicated by the dashed b\mcln{l}ack lines in (d). 
\mclR{
Plots (c-d) use  $\epsilon=0.5$, a value chosen so that the magnitudes of the $I/I_0$ analytical curves most closely match the simulation data.
This is necessary in this situation since, using a point proton source, for $\epsilon=0$ analytically the caustic intensities tend towards infinity.\cite{\krsi}} \mcln{Recent germane experimental results have  suggested that $I/I_0 \sim 3$ in practice\cite{Huntington2013}, illustrating the importance of  $\epsilon>0$ accounting for finite resolution effects.}  Consistent with this finding (d) shows that the simulation output closely matches the analytics, bolstering confidence in its numerical fidelity.
	
	

\section{Application to the filamentation instability in millimeter-scale HED plasmas\label{sec:example}}

We have developed \pr in connection to laboratory astrophysics experiments performed by the ACSEL collaboration\cite{\knat,Ross2012}. These experiments use powerful lasers to create high velocity plasmas flows by ablating the surface of plastic ($CH_2$) targets.  In a typical experiment two such targets are set up opposing one another and illuminated with laser light to study properties of the colliding plasma plumes.  
For our puposes here the typical plasma parameters\cite{Kugland2012, Ross2012,Ross2013, Ryutov2014} are \mcln{$n_e = 1\times 10^{19} cm^{-3}, T_e = T_i = 1 keV, v_{flow} = 8\times 10^7 cm/s$.} In the interaction between the two flows it is believed that the Weibel filamentation instability\cite{Weibel1959} plays an important role. Indeed, Weibel-like filamentary structures appearing in proton radiographs of the interaction have recently been reported\cite{Huntington2013,Fox2013}.
Yet for the reasons described above the challenge to discern the fields from their radiograph, i.e., to determine the extent to which filamentary magnetic fields produce filamentary radiograph structures, persists. 
Realistic situations introduce further questions: will protons traversing the hundreds of magnetic filaments expected in a realistic situation produce a coherent radiograph, or will they scatter; how important are density and temperature heterogeneities expected in the plasma flows; what is role of field strength as the filaments grow over time; and ultimately if a coherent radiograph can be produced how does its periodicity correspond to that of the underlying fields. 
Resolving these complications will evidently require many simulations, and due to the plasma's \mcln{$\sim 10 mm^3$} scale computational expense implies that multidimensional hydrodynamic and PIC simulations will not be ideally suited to this purpose.
Our purpose here is to show that, using \EM primitives to construct representative filamentary fields, \pr simulations can provide insight into this situation. To this end we address a subset of these questions in this section.

\begin{figure}[h]
\begin{center}
\resizebox{8cm}{!}{\includegraphics{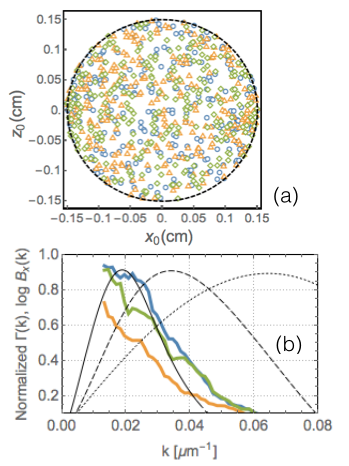}}
  \end{center}
  \caption{
    Comparison of \mcln{simulation fields} to the relevant transverse Weibel instability modes.
  \mclR{
  (a) shows the object plane configuration space of the magnetic filament centroids (at $y_0=0$)  for three simulations: sim. 18 (blue), sim. 45 (orange) and sim. 99 (green), as described in the text.
(b) compares the $k$-space of these filament centroids across $\hat{x}$ at $z_0=0$ to the germane theoretical instability growth rates.
The colored curves correspond to normalized Fourier transformations of the natural logarithm $\log B_x(k)$ from (a).
The black curves correspond to the normalized instability growth rates $\Gamma (k)$ for 
 collisionless Carbon (dotted),
collisional Carbon (dashed) and  
 collisional $C H_2$ flows (solid) from equations (\ref{eqn:kNC}-\ref{eqn:kC}).
}
}
  \label{fig:filamentsK} 
\end{figure}

\begin{figure*}[t]
\begin{center}
\resizebox{17cm}{!}{\includegraphics{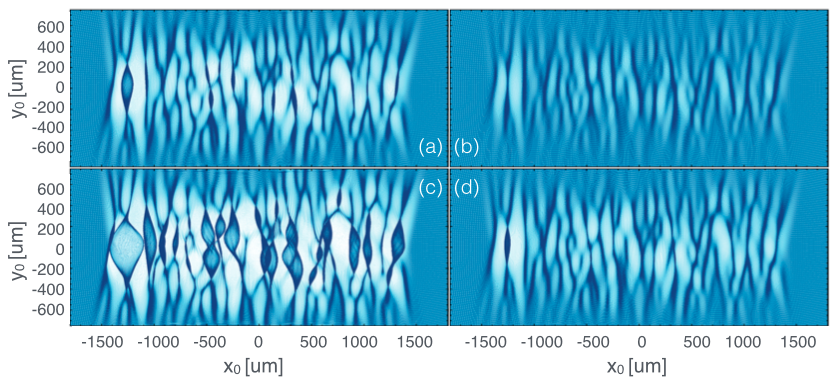}}
  \end{center}
  \caption{Effects of field strength $B_0$ and probing proton energy $\ep$ on the simulated proton radiograph results. All panels here correspond to the sim. 99 configuration indicated by the green curve in Fig. \ref{fig:filamentsK}. (a-b) correspond to probing proton energies of $\ep=$14.7 MeV and (c-d) to $\ep=$3 MeV.  (a) and (c) correspond to $B_0=1 MG$ (meaning a peak simulation field of $0.4 MG$), and (b) and (d) correspond to $B_0=0.3 MG$. }
  \label{fig:filsParam} 
\end{figure*}

We construct a representative field topology, guided by the reported experimental conditions\cite{Huntington2013}, using many dozens of magnetostatic Gaussian `cocoons' of the form given in equation (\ref{eqn:bPhi}).
 The experimental results imply that filaments form within a $\sim 1.5mm$ radius cylinder in the interaction midplane, with axial coordinate directed between the opposing plastic targets.  We model this as a `forest' of 260 filaments  
 each instantiated with a random centroid position in the $x_0-z_0$ plane (at $y_0=0$, oriented along $\hat{y}$) within $x_0^2+z_0^2\leq (1.5mm)^2$.
Experimental conditions also imply that \mcln{$c/\omega_{pi} = 100\mu m$} and the axial length of the cylinder containing the filaments $\sim 0.5mm$, so in the simulation each filament has $a=50 \mu m, b= 500\mu m$, \mcln{meaning that the inverse wavenumber of the filament centroids is nominally $2a = c/\omega_{pi}$.  We further use} randomized tilt parameters \mcln{$\delta \theta,\delta \psi = 15^o$} to account for natural density perturbations occurring in the plasma.  
Since these perturbations affecting the filament growth can be expected to vary between experiments, and since we are interested in determining whether filamentary structures in the radiographs are a robust signature of filamentary magnetic fields, we instantiate this setup in three distinct simulations.  That is, we perform three simulations pursuant to these conditions, meaning that the filament centroid positions in $x_0-z_0$ and the individualized tilts of filaments will vary between simulations, while each filament $a$ and $b$ and the density of filaments across simulations are constant.  \mclR{The three simulation  geometries are shown in Fig. \ref{fig:filamentsK} (a).  This plot shows the positioning of filament centroids, which varies between simulations in a randomized fashion, as well as the high density of filaments, which is held constant at $\sim 75 $ mm$^{-3}$ across simulations.}

To see that these simulation conditions form a reasonable approximation of experimental conditions, it is instructive to consider the relevant Weibel instability growth rates.  For the purely transverse mode the collisionless dispersion relation is given by,
\begin{eqnarray}
k^2 + \frac{\Gamma  \sigma }{\left| k\right|  U_e+\Gamma }+\frac{\Gamma }{\left| k\right|  U_i+\Gamma }=\frac{k^2}{\Gamma ^2+\frac{3 k^2 S}{5}}
\label{eqn:kNC}
\end{eqnarray}
where $\Gamma$ is the growth rate normalized to $v_{flow} \ \omega_{pi}/c$, $k$ is the wave number normalized to $\omega_{pi}/c$, $U_{e,i} = v_{T\ e,i}/( \sqrt{\pi} v_{flow})$ for thermal velocity $v_T$,
$S = 0.014\ T_i\text{ [keV]}$ and $\sigma = A m_p/(Z m_e)$ \mcln{for atomic mass $A$ and charge state $Z$}.\cite{Berger1991,Ryutov2014}  
The dispersion relation accounting for \mclR{inter-flow} collisional effects \cite{Ryutov2014}   can be formulated as,
\begin{eqnarray}
k^2 +\frac{\Gamma }{\Gamma +k^2 V_s}+ \frac{\sigma  \left(\Gamma +k^2 R\right)}{\Gamma +k^2 V_{\text{se}}}= \nonumber \\
\frac{k^2}{\Gamma  \left(k^2 V_b+\Gamma \right)+k^2 S}
\label{eqn:kC}
\end{eqnarray}
where
 $R = 0.00106/(T_e\text{ [keV]})^{3/2}, V_s = 0.0175\ (T_i\text{ [keV]})^{5/2}, V_b = 0.0253\ (T_i\text{ [keV]})^{5/2}, V_{\text{se}}=  64\ (T_e\text{ [keV]})^{5/2}$.

Equations (\ref{eqn:kNC}-\ref{eqn:kC}) provide physical references for filament periodicity in the simulations.
Fig. \ref{fig:filamentsK} (b) shows the simulation fields in relation to the normalized $\Gamma$ curves for 
 collisionless Carbon flows (dotted black),
collisional Carbon flows (dashed black) and  
 collisional $C H_2$ flows (solid black) 
  in which the light ions exhibit a stabilizing effect on the instability growth.
 The $\Gamma$ calculations assume plasma states of full ionization consistent with typical conditions that $T_e = T_i = 1 keV, v_{flow} = 8\times 10^7 cm/s$, and their depictions in Fig. \ref{fig:filamentsK} (b) indicate the transverse Weibel modes which can be expected to grow most rapidly in the plasma.
Since $\hat{z}$ is the axis of proton propagation the protons will deflect most strongly from the filamentary $B_x$ fields.  The colored curves correspond to normalized Fourier transformations $\log B_x(k)$ across $\hat{x}$ at the simulation midplane for the three simulations: sim. 18 (blue), sim. 45 (orange) and sim. 99 (green). 
\mclR{Here $\log$ denotes the natural logarithm and $B_x$ forms the representative field quantity, since the filament orientation along $\hat{y}$ means that $B_y$ reflects only $\delta \theta,\delta \psi$ effects. }  
From Fig. \ref{fig:filamentsK} (b) it is clear that the simulations provide an imperfect but reasonable approximation of the $k$-vectors which can be expected in the experimental situation.

\begin{figure*}[t]
\begin{center}
\resizebox{17cm}{!}{\includegraphics{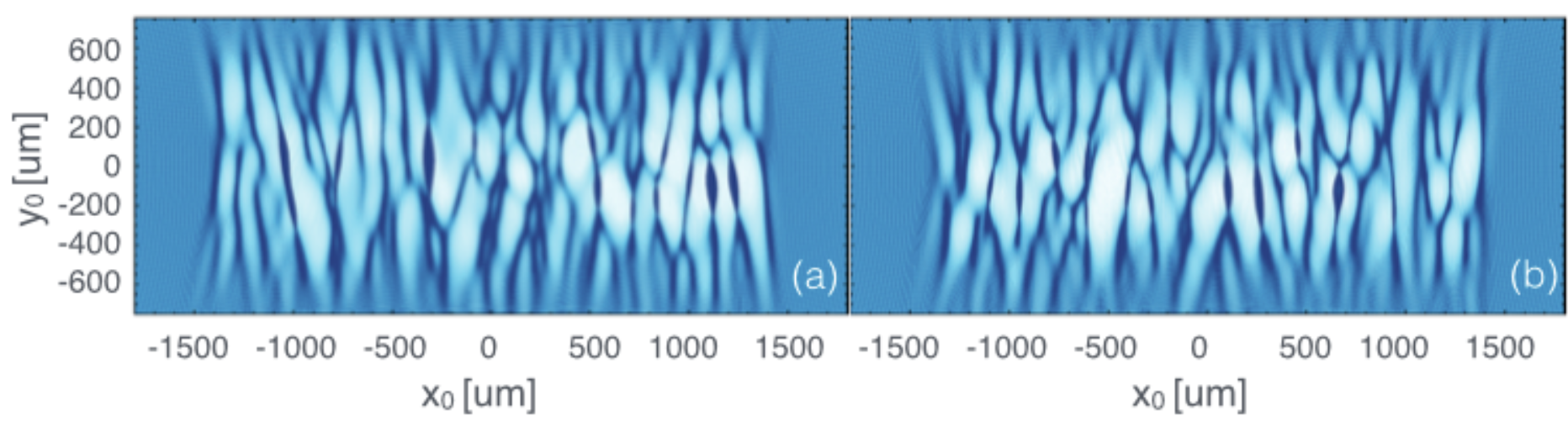}}
  \end{center}
  \caption{Synthetic proton radiographs for (a) sim. 18 and (b) sim. 45.  Across simulations $B_0=1 MG$ and $\ep=$14.7 MeV.  The radiograph corresponding to these conditions for sim. 99 is shown in Fig. \ref{fig:filsParam} (a). }
  \label{fig:filsParam2} 
\end{figure*}

\begin{figure}[t]
\begin{center}
\resizebox{8.5cm}{!}{\includegraphics{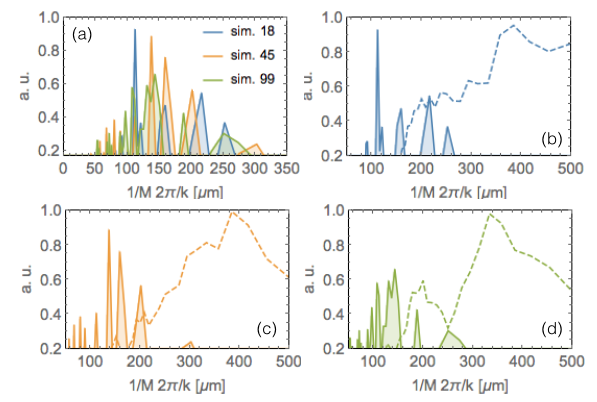}}
  \end{center}
  \caption{Comparison between field periodicity and proton radiograph image periodicity.  Solid lines correspond to Fourier-transformed lineouts at $y_0=0$ of the synthetic radiographs shown in Fig. \ref{fig:filsParam2}.  The dashed curves in (b-d) correspond to $2\pi/k$ for the simulation $k$-vectors shown in Fig. \ref{fig:filamentsK}.}
  \label{fig:filamentsLam} 
\end{figure}

Having described the simulation setup we now analyze the synthetic proton radiographs generated by \pr for these cases.  First we consider the roles of the magnetic field strength and proton beam energy for a single field configuration.  Fig. \ref{fig:filsParam}  shows the simulated proton radiographs for two values of $B_0$ and two values of $\ep$.  (a) corresponding to the $B_0=1MG$ ($B_{peak} \simeq 0.4 MG$) field strength and $\ep=14.7$ MeV proton source closely approximates the calculated field values and the experimental conditions reported on in  ref. \cite{Huntington2013}.  In this simulated radiograph we observe coherent, predominantly vertical filamentary features striated along the plasma flow axis ($\hat{y}$).  This fact is striking since according to \mcln{ref. \cite{\krsi}} protons should deflect in a nonlinear fashion from each of several dozen filamentary field structures on their path to the detector.  Through examination of (b-d) it is clear that these filamentary features persist across a variety of configurations.  
Comparison of (a) and (b) further shows that a reduction in field strength causes an apparent contraction of the  plasma flow interaction region.  The potential conflation in this regard forms an important consideration for experimental diagnosis.  
We also note that the tilting of the field filaments, a feature expected in realistic situations, plays an important role in the simulated radiograph signal.  In additional simulations not presented here we observed that \mcln{$\delta \psi,\delta \theta \to 0$} reduces the fluence amplitude of structures present in the radiograph by a factor of \mcln{three} or more.

To examine the robustness of filamentary radiograph structures we examine the sim. 18 and sim. 45 field configurations.  Fig. \ref{fig:filsParam2}  depicts these images,  which are seen to clearly exhibit similar coherent, predominantly vertical filamentary features.  In order to characterize the relationship between the field periodicity and the radiograph periodicity we have analyzed lineouts of the proton fluence along $y_0=0$ for each of the simulations.  Fig. \ref{fig:filamentsLam}  (a) shows the magnitude of the Fourier-transformed periodicity from each radiograph.  In (b-d) these radiograph periodicities (solid lines) are compared to the underlying magnetic field periodicities (dashed lines).  From these figures it is clear that the radiograph signal is shifted to much shorter wavelengths than those found in the simulation.  Furthermore the radiograph signal is negligible at the low $k$-values which dominate the magnetic field spectra.
These results show that, at minimum for the cases considered here, filamentary structures in proton radiographs are a qualitative signature of Weibel instability-like filamentary magnetic fields.
Future work will focus on parsing the quantitative relationship between the field and radiograph periodicities, a task which 
exceeds the illustrative scope of this section.








\section{Conclusions}
We have presented a new simulation tool for interpreting proton radiography of HED plasmas.
The present tool's numerical approach captures all relevant physics effects, including \mcl{effects related to the formation of caustics}.
Electromagnetic fields can be imported from PIC or hydrodynamic codes in a streamlined fashion. 
A library of \EM field `primitives' is also provided.  \mcl{These primitives can be considered `eigenvectors,' in effect spanning the basis of \EM fields, such that through linear combinations the user may construct realistic field topologies by hand.}
This capability allows users to add a primitive, modify the field strength, rotate a primitive, and so on, while quickly generating a high resolution radiograph at each step.
In this way, \pr enables the user to deconstruct features in a radiograph and interpret them in connection to specific underlying \EM field elements.
We have applied the tool in connection to experimental observations of the Weibel instability in counterstreaming plasmas, using $\sim 10^8$ particles generated from a realistic laser-driven point-like proton source, imaging fields which cover volumes of $\sim10 $mm$^3$.  Insights derived from this application indicate that tilting of magnetic filaments plays a significant role in setting the proton image; field strength tends to affect the apparent axial lengthscale over which the filamentation instability is active; and  coherent imaging is possible in the sense that filamentary structures are observed in radiographs as a signature of the Weibel fields, at least for the cases considered here.
These results show that \pr can support understanding of HED plasmas.

\mclR{M. L. is grateful to Elijah Kemp, Tony Link, Mario Manuel, Chikang Li, Gianluca Gregori and  Anatoly Spitkovsky for useful discussions.
M. L. thanks the LLNL Lawrence Scholarship and Royal Society Newton International Fellowship for support, and the LLNL Institutional Grand Challenge program for computational resources.
F.F. acknowledges the LLNL Lawrence Fellowship for financial support.
}
This work was performed under the auspices of the U.S. Department of Energy by LLNL under Contract DE-AC52-07NA27344.

\begin{appendices}

\mclR{\section{Descriptions, schematics and simulations of electromagnetic field primitives \label{sec:aa}}}
    
    \mclR{In order to develop intuition connecting a proton radiograph image to its underlying electromagnetic field primitive, in this appendix we enumerate the set of available primitives and provide representative schematics and simulations.
    Table A1 represents the four basic field primitives, showing the physical descriptions, unrotated functional forms of the electric potential $\phi$ or magnetic field vector $\vec{B}$, as well as the corresponding $\pindex$ index used to invoke each primitive in the simulation.
     The electric field $\vec{E} = -\nabla \phi$ is obtained in the standard way, and rotation of each primitive along two axes is enabled by specifying the $\theta$ and $\psi$ elements of the field control vectors given by equations (\ref{eqn:basicVec}-\ref{eqn:latticeVec}).
     We maintain the coordinate systems and notations described in the above sections, so that $x_0, y_0, z_0$ denote object plane coordinates, $\phi_0$ ($B_0$) is the nominal peak electric potential (magnetic field) of each primitive, and $a$ and $b$ are the major and semi-major axes of each primitive, respectively.  For the second row of Table A1, the primitive represents a quasi-planar shock propagating along a cylinder in $\hat{y}$ with plasma  density decreasing along the radial coordinate, where
     erf is the Gaussian error function, $\log$ is the natural logarithm and the shock thickness $d = 2 \sqrt{\log 2} F$ for full width at half maximum of the shock potential $F$. \cite{\krsi}     }
  
  \mclR{
  The descriptive information presented in Table A1 is complemented by each primitive's visual representation in the object and image planes. This information is shown in Table A2 for a variety of conditions relevant to HED plasmas.  In this latter table the primitives' index $\pindex$ is shown in the first column,  object plane schematic in the second column and (image plane) simulated proton radiograph in the third column.
 In each schematic the transparent orange surface represents an isocontour of the field magnitude, and the colored arrows show the vector field, with both arrow size and color corresponding to field strength. 
The first row's schematic highlights the geometry of the three dimensional proton-field interaction, with the blue three dimensional arrow at the lower $z$ boundary indicating the direction of proton propagation (along $\hat{z}$). This geometry is maintained for all schematics and simulations shown in the table.
Supporting clarity of interpretation, in each simulation we use an identical point source of monoenergetic 14.7 MeV protons, isotropically emitting 1 billion particles, imitating a realistic \dhe source situated $|z_s|=1$ cm from the object plane.  In \pr this source is instantiated in the simulation using the source control vector $\vec{S}=(2,\ 14.7 \text{MeV},\ 10^9,\ 1  \text{cm})$,
as is covered in section \ref{sec:source}.  
As an example of field instantiation, the electric field shown in the second column of the first row is created in the simulation by specifying the field control vector according to equation (\ref{eqn:basicVec}) of $\vec{G}_1 = (1, 0, 0, 0, 0, 0, 100 \mathrm{\mu m}, 100 \mathrm{\mu m}, -200 \mathrm{kV\ cm^{-1}})$, 
as is covered in section \ref{sec:fieldInput}.
For each case, the simulated proton radiograph is shown using nominal object plane units, using the default detector having a magnification factor of ten.
The fiducial peak field values for each case are labeled in the second column of the table.
The characteristic strengths with which each primitive deflects the protons sampling it is highlighted by the  scales of the radiograph fluence shown in the third column.
Furthermore, the set of characteristic proton radiographs produced by imaging these primitives is expanded  by adjusting the $\theta,\psi$ elements of $\vec{G}_n$.\cite{\krsi}
Through tuning of the elements of the $\vec{G}_n$ field control vectors, the primitives enumerated in this section effectively span the basis of \EM fields, such that by $\sum_n \vec{G}_n$ the user may construct realistic field topologies by hand.}

\end{appendices}


\begin{thebibliography}{72}%
\makeatletter
\providecommand \@ifxundefined [1]{%
 \@ifx{#1\undefined}
}%
\providecommand \@ifnum [1]{%
 \ifnum #1\expandafter \@firstoftwo
 \else \expandafter \@secondoftwo
 \fi
}%
\providecommand \@ifx [1]{%
 \ifx #1\expandafter \@firstoftwo
 \else \expandafter \@secondoftwo
 \fi
}%
\providecommand \natexlab [1]{#1}%
\providecommand \enquote  [1]{``#1''}%
\providecommand \bibnamefont  [1]{#1}%
\providecommand \bibfnamefont [1]{#1}%
\providecommand \citenamefont [1]{#1}%
\providecommand \href@noop [0]{\@secondoftwo}%
\providecommand \href [0]{\begingroup \@sanitize@url \@href}%
\providecommand \@href[1]{\@@startlink{#1}\@@href}%
\providecommand \@@href[1]{\endgroup#1\@@endlink}%
\providecommand \@sanitize@url [0]{\catcode `\\12\catcode `\$12\catcode
  `\&12\catcode `\#12\catcode `\^12\catcode `\_12\catcode `\%12\relax}%
\providecommand \@@startlink[1]{}%
\providecommand \@@endlink[0]{}%
\providecommand \url  [0]{\begingroup\@sanitize@url \@url }%
\providecommand \@url [1]{\endgroup\@href {#1}{\urlprefix }}%
\providecommand \urlprefix  [0]{URL }%
\providecommand \Eprint [0]{\href }%
\providecommand \doibase [0]{http://dx.doi.org/}%
\providecommand \selectlanguage [0]{\@gobble}%
\providecommand \bibinfo  [0]{\@secondoftwo}%
\providecommand \bibfield  [0]{\@secondoftwo}%
\providecommand \translation [1]{[#1]}%
\providecommand \BibitemOpen [0]{}%
\providecommand \bibitemStop [0]{}%
\providecommand \bibitemNoStop [0]{.\EOS\space}%
\providecommand \EOS [0]{\spacefactor3000\relax}%
\providecommand \BibitemShut  [1]{\csname bibitem#1\endcsname}%
\let\auto@bib@innerbib\@empty
\bibitem [{\citenamefont {{Committee on High Energy Density Plasma Physics,
  Plasma Science Committee, and National Research Council}}(2003)}]{HEDP2003}%
  \BibitemOpen
  \bibfield  {author} {\bibinfo {author} {\bibnamefont {{Committee on High
  Energy Density Plasma Physics, Plasma Science Committee, and National
  Research Council}}},\ }\href@noop {} {\emph {\bibinfo {title} {{Frontiers in
  High Energy Density Physics: The X-Games of Contemporary Science}}}}\
  (\bibinfo  {publisher} {The National Academies Press},\ \bibinfo {year}
  {2003})\BibitemShut {NoStop}%
\bibitem [{\citenamefont {Remington}, \citenamefont {Drake},\ and\
  \citenamefont {Ryutov}(2006)}]{Remington2006}%
  \BibitemOpen
  \bibfield  {author} {\bibinfo {author} {\bibfnamefont {B.}~\bibnamefont
  {Remington}}, \bibinfo {author} {\bibfnamefont {R.}~\bibnamefont {Drake}}, \
  and\ \bibinfo {author} {\bibfnamefont {D.}~\bibnamefont {Ryutov}},\
  }\bibfield  {title} {\enquote {\bibinfo {title} {{Experimental astrophysics
  with high power lasers and Z pinches}},}\ }\href {\doibase
  10.1103/RevModPhys.78.755} {\bibfield  {journal} {\bibinfo  {journal}
  {Reviews of Modern Physics}\ }\textbf {\bibinfo {volume} {78}},\ \bibinfo
  {pages} {755--807} (\bibinfo {year} {2006})}\BibitemShut {NoStop}%
\bibitem [{\citenamefont {Drake}(2006)}]{Drake2006}%
  \BibitemOpen
  \bibfield  {author} {\bibinfo {author} {\bibfnamefont {R.~P.}\ \bibnamefont
  {Drake}},\ }\href@noop {} {\emph {\bibinfo {title} {High-Energy-Density
  Physics: Fundamentals, Inertial Fusion, and Experimental Astrophysics}}}\
  (\bibinfo  {publisher} {Springer},\ \bibinfo {year} {2006})\BibitemShut
  {NoStop}%
\bibitem [{\citenamefont {Hogan}\ \emph {et~al.}(1999)\citenamefont {Hogan},
  \citenamefont {Adams}, \citenamefont {Alrick}, \citenamefont {Amann},
  \citenamefont {Boissevain}, \citenamefont {Crow}, \citenamefont {Cushing},
  \citenamefont {Eddelman}, \citenamefont {Espinoza}, \citenamefont {Fife},
  \citenamefont {Gallegos}, \citenamefont {Gomez}, \citenamefont {Gorman},
  \citenamefont {Gray}, \citenamefont {Holmes}, \citenamefont {Jaramillo},
  \citenamefont {King}, \citenamefont {Knudson}, \citenamefont {London},
  \citenamefont {Lopez}, \citenamefont {McClelland}, \citenamefont {Merrill},
  \citenamefont {Morley}, \citenamefont {Morris}, \citenamefont {Mottershead},
  \citenamefont {Mueller}, \citenamefont {Neri}, \citenamefont {Numkena},
  \citenamefont {Pazuchanics}, \citenamefont {Pillai}, \citenamefont {Prael},
  \citenamefont {Riedel}, \citenamefont {Sarracino}, \citenamefont {Saunders},
  \citenamefont {Stacy}, \citenamefont {Takala}, \citenamefont {Thiessen},
  \citenamefont {Tucker}, \citenamefont {Walstrom}, \citenamefont {Yates},
  \citenamefont {Ziock}, \citenamefont {Zumbro}, \citenamefont {Ables},
  \citenamefont {Aufderheide}, \citenamefont {Barnes}, \citenamefont {Bionta},
  \citenamefont {Fujino}, \citenamefont {Hartouni}, \citenamefont {Park},
  \citenamefont {Soltz}, \citenamefont {Wright}, \citenamefont {Balzer},
  \citenamefont {Flores}, \citenamefont {Thompson}, \citenamefont {Pendzick},
  \citenamefont {Prigl}, \citenamefont {Scaduto}, \citenamefont {Schwaner},\
  and\ \citenamefont {O'Donnell}}]{Hogan1999a}%
  \BibitemOpen
  \bibfield  {author} {\bibinfo {author} {\bibfnamefont {G.}~\bibnamefont
  {Hogan}}, \bibinfo {author} {\bibfnamefont {K.}~\bibnamefont {Adams}},
  \bibinfo {author} {\bibfnamefont {K.}~\bibnamefont {Alrick}}, \bibinfo
  {author} {\bibfnamefont {J.}~\bibnamefont {Amann}}, \bibinfo {author}
  {\bibfnamefont {J.}~\bibnamefont {Boissevain}}, \bibinfo {author}
  {\bibfnamefont {M.}~\bibnamefont {Crow}}, \bibinfo {author} {\bibfnamefont
  {S.}~\bibnamefont {Cushing}}, \bibinfo {author} {\bibfnamefont
  {J.}~\bibnamefont {Eddelman}}, \bibinfo {author} {\bibfnamefont
  {C.}~\bibnamefont {Espinoza}}, \bibinfo {author} {\bibfnamefont
  {T.}~\bibnamefont {Fife}}, \bibinfo {author} {\bibfnamefont {R.}~\bibnamefont
  {Gallegos}}, \bibinfo {author} {\bibfnamefont {J.}~\bibnamefont {Gomez}},
  \bibinfo {author} {\bibfnamefont {T.}~\bibnamefont {Gorman}}, \bibinfo
  {author} {\bibfnamefont {N.}~\bibnamefont {Gray}}, \bibinfo {author}
  {\bibfnamefont {V.}~\bibnamefont {Holmes}}, \bibinfo {author} {\bibfnamefont
  {S.}~\bibnamefont {Jaramillo}}, \bibinfo {author} {\bibfnamefont
  {N.}~\bibnamefont {King}}, \bibinfo {author} {\bibfnamefont {J.}~\bibnamefont
  {Knudson}}, \bibinfo {author} {\bibfnamefont {R.}~\bibnamefont {London}},
  \bibinfo {author} {\bibfnamefont {R.}~\bibnamefont {Lopez}}, \bibinfo
  {author} {\bibfnamefont {J.}~\bibnamefont {McClelland}}, \bibinfo {author}
  {\bibfnamefont {F.}~\bibnamefont {Merrill}}, \bibinfo {author} {\bibfnamefont
  {K.}~\bibnamefont {Morley}}, \bibinfo {author} {\bibfnamefont
  {C.}~\bibnamefont {Morris}}, \bibinfo {author} {\bibfnamefont
  {C.}~\bibnamefont {Mottershead}}, \bibinfo {author} {\bibfnamefont
  {K.}~\bibnamefont {Mueller}}, \bibinfo {author} {\bibfnamefont
  {F.}~\bibnamefont {Neri}}, \bibinfo {author} {\bibfnamefont {D.}~\bibnamefont
  {Numkena}}, \bibinfo {author} {\bibfnamefont {P.}~\bibnamefont
  {Pazuchanics}}, \bibinfo {author} {\bibfnamefont {C.}~\bibnamefont {Pillai}},
  \bibinfo {author} {\bibfnamefont {R.}~\bibnamefont {Prael}}, \bibinfo
  {author} {\bibfnamefont {C.}~\bibnamefont {Riedel}}, \bibinfo {author}
  {\bibfnamefont {J.}~\bibnamefont {Sarracino}}, \bibinfo {author}
  {\bibfnamefont {A.}~\bibnamefont {Saunders}}, \bibinfo {author}
  {\bibfnamefont {H.}~\bibnamefont {Stacy}}, \bibinfo {author} {\bibfnamefont
  {B.}~\bibnamefont {Takala}}, \bibinfo {author} {\bibfnamefont
  {H.}~\bibnamefont {Thiessen}}, \bibinfo {author} {\bibfnamefont
  {H.}~\bibnamefont {Tucker}}, \bibinfo {author} {\bibfnamefont
  {P.}~\bibnamefont {Walstrom}}, \bibinfo {author} {\bibfnamefont
  {G.}~\bibnamefont {Yates}}, \bibinfo {author} {\bibfnamefont {H.-J.}\
  \bibnamefont {Ziock}}, \bibinfo {author} {\bibfnamefont {J.}~\bibnamefont
  {Zumbro}}, \bibinfo {author} {\bibfnamefont {E.}~\bibnamefont {Ables}},
  \bibinfo {author} {\bibfnamefont {M.}~\bibnamefont {Aufderheide}}, \bibinfo
  {author} {\bibfnamefont {P.}~\bibnamefont {Barnes}}, \bibinfo {author}
  {\bibfnamefont {R.}~\bibnamefont {Bionta}}, \bibinfo {author} {\bibfnamefont
  {D.}~\bibnamefont {Fujino}}, \bibinfo {author} {\bibfnamefont
  {E.}~\bibnamefont {Hartouni}}, \bibinfo {author} {\bibfnamefont {H.-S.}\
  \bibnamefont {Park}}, \bibinfo {author} {\bibfnamefont {R.}~\bibnamefont
  {Soltz}}, \bibinfo {author} {\bibfnamefont {D.}~\bibnamefont {Wright}},
  \bibinfo {author} {\bibfnamefont {S.}~\bibnamefont {Balzer}}, \bibinfo
  {author} {\bibfnamefont {P.}~\bibnamefont {Flores}}, \bibinfo {author}
  {\bibfnamefont {R.}~\bibnamefont {Thompson}}, \bibinfo {author}
  {\bibfnamefont {A.}~\bibnamefont {Pendzick}}, \bibinfo {author}
  {\bibfnamefont {R.}~\bibnamefont {Prigl}}, \bibinfo {author} {\bibfnamefont
  {J.}~\bibnamefont {Scaduto}}, \bibinfo {author} {\bibfnamefont
  {E.}~\bibnamefont {Schwaner}}, \ and\ \bibinfo {author} {\bibfnamefont
  {J.}~\bibnamefont {O'Donnell}},\ }\bibfield  {title} {\enquote {\bibinfo
  {title} {{Proton radiography}},}\ }in\ \href {\doibase
  10.1109/PAC.1999.795765} {\emph {\bibinfo {booktitle} {Proceedings of the
  1999 Particle Accelerator Conference (Cat. No.99CH36366)}}},\ Vol.~\bibinfo
  {volume} {1}\ (\bibinfo  {publisher} {IEEE},\ \bibinfo {year} {1999})\ pp.\
  \bibinfo {pages} {579--583}\BibitemShut {NoStop}%
\bibitem [{\citenamefont {Borghesi}\ \emph {et~al.}(2001)\citenamefont
  {Borghesi}, \citenamefont {Schiavi}, \citenamefont {Campbell}, \citenamefont
  {Haines}, \citenamefont {Willi}, \citenamefont {MacKinnon}, \citenamefont
  {Gizzi}, \citenamefont {Galimberti}, \citenamefont {Clarke},\ and\
  \citenamefont {Ruhl}}]{Borghesi2001}%
  \BibitemOpen
  \bibfield  {author} {\bibinfo {author} {\bibfnamefont {M.}~\bibnamefont
  {Borghesi}}, \bibinfo {author} {\bibfnamefont {A.}~\bibnamefont {Schiavi}},
  \bibinfo {author} {\bibfnamefont {D.~H.}\ \bibnamefont {Campbell}}, \bibinfo
  {author} {\bibfnamefont {M.~G.}\ \bibnamefont {Haines}}, \bibinfo {author}
  {\bibfnamefont {O.}~\bibnamefont {Willi}}, \bibinfo {author} {\bibfnamefont
  {A.~J.}\ \bibnamefont {MacKinnon}}, \bibinfo {author} {\bibfnamefont {L.~A.}\
  \bibnamefont {Gizzi}}, \bibinfo {author} {\bibfnamefont {M.}~\bibnamefont
  {Galimberti}}, \bibinfo {author} {\bibfnamefont {R.~J.}\ \bibnamefont
  {Clarke}}, \ and\ \bibinfo {author} {\bibfnamefont {H.}~\bibnamefont
  {Ruhl}},\ }\bibfield  {title} {\enquote {\bibinfo {title} {{Proton imaging: a
  diagnostic for inertial confinement fusion/fast ignitor studies}},}\ }\href
  {\doibase 10.1088/0741-3335/43/12A/320} {\bibfield  {journal} {\bibinfo
  {journal} {Plasma Physics and Controlled Fusion}\ }\textbf {\bibinfo {volume}
  {43}},\ \bibinfo {pages} {A267--A276} (\bibinfo {year} {2001})}\BibitemShut
  {NoStop}%
\bibitem [{\citenamefont {BORGHESI}\ \emph {et~al.}(2005)\citenamefont
  {BORGHESI}, \citenamefont {AUDEBERT}, \citenamefont {BULANOV}, \citenamefont
  {COWAN}, \citenamefont {FUCHS}, \citenamefont {GAUTHIER}, \citenamefont
  {MACKINNON}, \citenamefont {PATEL}, \citenamefont {PRETZLER}, \citenamefont
  {ROMAGNANI}, \citenamefont {SCHIAVI}, \citenamefont {TONCIAN},\ and\
  \citenamefont {WILLI}}]{Borghesi2005}%
  \BibitemOpen
  \bibfield  {author} {\bibinfo {author} {\bibfnamefont {M.}~\bibnamefont
  {BORGHESI}}, \bibinfo {author} {\bibfnamefont {P.}~\bibnamefont {AUDEBERT}},
  \bibinfo {author} {\bibfnamefont {S.}~\bibnamefont {BULANOV}}, \bibinfo
  {author} {\bibfnamefont {T.}~\bibnamefont {COWAN}}, \bibinfo {author}
  {\bibfnamefont {J.}~\bibnamefont {FUCHS}}, \bibinfo {author} {\bibfnamefont
  {J.}~\bibnamefont {GAUTHIER}}, \bibinfo {author} {\bibfnamefont
  {A.}~\bibnamefont {MACKINNON}}, \bibinfo {author} {\bibfnamefont
  {P.}~\bibnamefont {PATEL}}, \bibinfo {author} {\bibfnamefont
  {G.}~\bibnamefont {PRETZLER}}, \bibinfo {author} {\bibfnamefont
  {L.}~\bibnamefont {ROMAGNANI}}, \bibinfo {author} {\bibfnamefont
  {A.}~\bibnamefont {SCHIAVI}}, \bibinfo {author} {\bibfnamefont
  {T.}~\bibnamefont {TONCIAN}}, \ and\ \bibinfo {author} {\bibfnamefont
  {O.}~\bibnamefont {WILLI}},\ }\bibfield  {title} {\enquote {\bibinfo {title}
  {{High-intensity laser-plasma interaction studies employing laser-driven
  proton probes}},}\ }\href {\doibase 10.1017/S0263034605050408} {\bibfield
  {journal} {\bibinfo  {journal} {Laser and Particle Beams}\ }\textbf {\bibinfo
  {volume} {23}},\ \bibinfo {pages} {291--295} (\bibinfo {year}
  {2005})}\BibitemShut {NoStop}%
\bibitem [{\citenamefont {Pape}\ \emph {et~al.}(2007)\citenamefont {Pape},
  \citenamefont {Hey}, \citenamefont {Patel}, \citenamefont {Mackinnon},
  \citenamefont {Klein}, \citenamefont {Remington}, \citenamefont {Wilks},
  \citenamefont {Ryutov}, \citenamefont {Moon},\ and\ \citenamefont
  {Foord}}]{Pape2007}%
  \BibitemOpen
  \bibfield  {author} {\bibinfo {author} {\bibfnamefont {S.~L.}\ \bibnamefont
  {Pape}}, \bibinfo {author} {\bibfnamefont {D.}~\bibnamefont {Hey}}, \bibinfo
  {author} {\bibfnamefont {P.}~\bibnamefont {Patel}}, \bibinfo {author}
  {\bibfnamefont {A.}~\bibnamefont {Mackinnon}}, \bibinfo {author}
  {\bibfnamefont {R.}~\bibnamefont {Klein}}, \bibinfo {author} {\bibfnamefont
  {B.}~\bibnamefont {Remington}}, \bibinfo {author} {\bibfnamefont
  {S.}~\bibnamefont {Wilks}}, \bibinfo {author} {\bibfnamefont
  {D.}~\bibnamefont {Ryutov}}, \bibinfo {author} {\bibfnamefont
  {S.}~\bibnamefont {Moon}}, \ and\ \bibinfo {author} {\bibfnamefont
  {M.}~\bibnamefont {Foord}},\ }\bibfield  {title} {\enquote {\bibinfo {title}
  {{Proton Radiography of Megagauss Electromagnetic Fields Generated by the
  Irradiation of a Solid Target by an Ultraintense Laser Pulse}},}\ }\href
  {\doibase 10.1007/s10509-007-9386-y} {\bibfield  {journal} {\bibinfo
  {journal} {Astrophysics and Space Science}\ }\textbf {\bibinfo {volume}
  {307}},\ \bibinfo {pages} {341--345} (\bibinfo {year} {2007})}\BibitemShut
  {NoStop}%
\bibitem [{\citenamefont {Borghesi}\ \emph {et~al.}(2008)\citenamefont
  {Borghesi}, \citenamefont {Cecchetti}, \citenamefont {Toncian}, \citenamefont
  {Fuchs}, \citenamefont {Romagnani}, \citenamefont {Kar}, \citenamefont
  {Wilson}, \citenamefont {Antici}, \citenamefont {Audebert}, \citenamefont
  {Brambrink}, \citenamefont {Pipahl}, \citenamefont {Amin}, \citenamefont
  {Jung}, \citenamefont {Osterholz}, \citenamefont {Willi}, \citenamefont
  {Nazarov}, \citenamefont {Clarke}, \citenamefont {Notley}, \citenamefont
  {Neely}, \citenamefont {Mora}, \citenamefont {Grismayer}, \citenamefont
  {Schurtz}, \citenamefont {Schiavi}, \citenamefont {Sentoku},\ and\
  \citenamefont {D¿Humieres}}]{Borghesi2008a}%
  \BibitemOpen
  \bibfield  {author} {\bibinfo {author} {\bibfnamefont {M.}~\bibnamefont
  {Borghesi}}, \bibinfo {author} {\bibfnamefont {C.~A.}\ \bibnamefont
  {Cecchetti}}, \bibinfo {author} {\bibfnamefont {T.}~\bibnamefont {Toncian}},
  \bibinfo {author} {\bibfnamefont {J.}~\bibnamefont {Fuchs}}, \bibinfo
  {author} {\bibfnamefont {L.}~\bibnamefont {Romagnani}}, \bibinfo {author}
  {\bibfnamefont {S.}~\bibnamefont {Kar}}, \bibinfo {author} {\bibfnamefont
  {P.~A.}\ \bibnamefont {Wilson}}, \bibinfo {author} {\bibfnamefont
  {P.}~\bibnamefont {Antici}}, \bibinfo {author} {\bibfnamefont
  {P.}~\bibnamefont {Audebert}}, \bibinfo {author} {\bibfnamefont
  {E.}~\bibnamefont {Brambrink}}, \bibinfo {author} {\bibfnamefont
  {A.}~\bibnamefont {Pipahl}}, \bibinfo {author} {\bibfnamefont
  {M.}~\bibnamefont {Amin}}, \bibinfo {author} {\bibfnamefont {R.}~\bibnamefont
  {Jung}}, \bibinfo {author} {\bibfnamefont {J.}~\bibnamefont {Osterholz}},
  \bibinfo {author} {\bibfnamefont {O.}~\bibnamefont {Willi}}, \bibinfo
  {author} {\bibfnamefont {W.}~\bibnamefont {Nazarov}}, \bibinfo {author}
  {\bibfnamefont {R.~J.}\ \bibnamefont {Clarke}}, \bibinfo {author}
  {\bibfnamefont {M.}~\bibnamefont {Notley}}, \bibinfo {author} {\bibfnamefont
  {D.}~\bibnamefont {Neely}}, \bibinfo {author} {\bibfnamefont
  {P.}~\bibnamefont {Mora}}, \bibinfo {author} {\bibfnamefont {T.}~\bibnamefont
  {Grismayer}}, \bibinfo {author} {\bibfnamefont {G.}~\bibnamefont {Schurtz}},
  \bibinfo {author} {\bibfnamefont {A.}~\bibnamefont {Schiavi}}, \bibinfo
  {author} {\bibfnamefont {Y.}~\bibnamefont {Sentoku}}, \ and\ \bibinfo
  {author} {\bibfnamefont {E.}~\bibnamefont {D¿Humieres}},\ }\bibfield
  {title} {\enquote {\bibinfo {title} {{Laser-Driven Proton Beams: Acceleration
  Mechanism, Beam Optimization, and Radiographic Applications}},}\ }\href
  {\doibase 10.1109/TPS.2008.927142} {\bibfield  {journal} {\bibinfo  {journal}
  {IEEE Transactions on Plasma Science}\ }\textbf {\bibinfo {volume} {36}},\
  \bibinfo {pages} {1833--1842} (\bibinfo {year} {2008})}\BibitemShut {NoStop}%
\bibitem [{\citenamefont {Mackinnon}\ \emph {et~al.}(2006)\citenamefont
  {Mackinnon}, \citenamefont {Patel}, \citenamefont {Borghesi}, \citenamefont
  {Clarke}, \citenamefont {Freeman}, \citenamefont {Habara}, \citenamefont
  {Hatchett}, \citenamefont {Hey}, \citenamefont {Hicks}, \citenamefont {Kar},
  \citenamefont {Key}, \citenamefont {King}, \citenamefont {Lancaster},
  \citenamefont {Neely}, \citenamefont {Nikkro}, \citenamefont {Norreys},
  \citenamefont {Notley}, \citenamefont {Phillips}, \citenamefont {Romagnani},
  \citenamefont {Snavely}, \citenamefont {Stephens},\ and\ \citenamefont
  {Town}}]{Mackinnon2006}%
  \BibitemOpen
  \bibfield  {author} {\bibinfo {author} {\bibfnamefont {A.}~\bibnamefont
  {Mackinnon}}, \bibinfo {author} {\bibfnamefont {P.}~\bibnamefont {Patel}},
  \bibinfo {author} {\bibfnamefont {M.}~\bibnamefont {Borghesi}}, \bibinfo
  {author} {\bibfnamefont {R.}~\bibnamefont {Clarke}}, \bibinfo {author}
  {\bibfnamefont {R.}~\bibnamefont {Freeman}}, \bibinfo {author} {\bibfnamefont
  {H.}~\bibnamefont {Habara}}, \bibinfo {author} {\bibfnamefont
  {S.}~\bibnamefont {Hatchett}}, \bibinfo {author} {\bibfnamefont
  {D.}~\bibnamefont {Hey}}, \bibinfo {author} {\bibfnamefont {D.}~\bibnamefont
  {Hicks}}, \bibinfo {author} {\bibfnamefont {S.}~\bibnamefont {Kar}}, \bibinfo
  {author} {\bibfnamefont {M.}~\bibnamefont {Key}}, \bibinfo {author}
  {\bibfnamefont {J.}~\bibnamefont {King}}, \bibinfo {author} {\bibfnamefont
  {K.}~\bibnamefont {Lancaster}}, \bibinfo {author} {\bibfnamefont
  {D.}~\bibnamefont {Neely}}, \bibinfo {author} {\bibfnamefont
  {A.}~\bibnamefont {Nikkro}}, \bibinfo {author} {\bibfnamefont
  {P.}~\bibnamefont {Norreys}}, \bibinfo {author} {\bibfnamefont
  {M.}~\bibnamefont {Notley}}, \bibinfo {author} {\bibfnamefont
  {T.}~\bibnamefont {Phillips}}, \bibinfo {author} {\bibfnamefont
  {L.}~\bibnamefont {Romagnani}}, \bibinfo {author} {\bibfnamefont
  {R.}~\bibnamefont {Snavely}}, \bibinfo {author} {\bibfnamefont
  {R.}~\bibnamefont {Stephens}}, \ and\ \bibinfo {author} {\bibfnamefont
  {R.}~\bibnamefont {Town}},\ }\bibfield  {title} {\enquote {\bibinfo {title}
  {{Proton Radiography of a Laser-Driven Implosion}},}\ }\href {\doibase
  10.1103/PhysRevLett.97.045001} {\bibfield  {journal} {\bibinfo  {journal}
  {Physical Review Letters}\ }\textbf {\bibinfo {volume} {97}},\ \bibinfo
  {pages} {045001} (\bibinfo {year} {2006})}\BibitemShut {NoStop}%
\bibitem [{\citenamefont {Li}\ \emph {et~al.}(2006{\natexlab{a}})\citenamefont
  {Li}, \citenamefont {Séguin}, \citenamefont {Frenje}, \citenamefont {Rygg},
  \citenamefont {Petrasso}, \citenamefont {Town}, \citenamefont {Amendt},
  \citenamefont {Hatchett}, \citenamefont {Landen}, \citenamefont {Mackinnon},
  \citenamefont {Patel}, \citenamefont {Smalyuk}, \citenamefont {Knauer},
  \citenamefont {Sangster},\ and\ \citenamefont {Stoeckl}}]{Li2006a}%
  \BibitemOpen
  \bibfield  {author} {\bibinfo {author} {\bibfnamefont {C.~K.}\ \bibnamefont
  {Li}}, \bibinfo {author} {\bibfnamefont {F.~H.}\ \bibnamefont {Séguin}},
  \bibinfo {author} {\bibfnamefont {J.~a.}\ \bibnamefont {Frenje}}, \bibinfo
  {author} {\bibfnamefont {J.~R.}\ \bibnamefont {Rygg}}, \bibinfo {author}
  {\bibfnamefont {R.~D.}\ \bibnamefont {Petrasso}}, \bibinfo {author}
  {\bibfnamefont {R.~P.~J.}\ \bibnamefont {Town}}, \bibinfo {author}
  {\bibfnamefont {P.~a.}\ \bibnamefont {Amendt}}, \bibinfo {author}
  {\bibfnamefont {S.~P.}\ \bibnamefont {Hatchett}}, \bibinfo {author}
  {\bibfnamefont {O.~L.}\ \bibnamefont {Landen}}, \bibinfo {author}
  {\bibfnamefont {a.~J.}\ \bibnamefont {Mackinnon}}, \bibinfo {author}
  {\bibfnamefont {P.~K.}\ \bibnamefont {Patel}}, \bibinfo {author}
  {\bibfnamefont {V.~a.}\ \bibnamefont {Smalyuk}}, \bibinfo {author}
  {\bibfnamefont {J.~P.}\ \bibnamefont {Knauer}}, \bibinfo {author}
  {\bibfnamefont {T.~C.}\ \bibnamefont {Sangster}}, \ and\ \bibinfo {author}
  {\bibfnamefont {C.}~\bibnamefont {Stoeckl}},\ }\bibfield  {title} {\enquote
  {\bibinfo {title} {{Monoenergetic proton backlighter for measuring E and B
  fields and for radiographing implosions and high-energy density plasmas
  (invited)}},}\ }\href {\doibase 10.1063/1.2228252} {\bibfield  {journal}
  {\bibinfo  {journal} {Review of Scientific Instruments}\ }\textbf {\bibinfo
  {volume} {77}},\ \bibinfo {pages} {10E725} (\bibinfo {year}
  {2006}{\natexlab{a}})}\BibitemShut {NoStop}%
\bibitem [{\citenamefont {Li}\ \emph {et~al.}(2006{\natexlab{b}})\citenamefont
  {Li}, \citenamefont {S\'{e}guin}, \citenamefont {Frenje}, \citenamefont
  {Rygg}, \citenamefont {Petrasso}, \citenamefont {Town}, \citenamefont
  {Amendt}, \citenamefont {Hatchett}, \citenamefont {Landen}, \citenamefont
  {Mackinnon}, \citenamefont {Patel}, \citenamefont {Smalyuk}, \citenamefont
  {Sangster},\ and\ \citenamefont {Knauer}}]{Li2006}%
  \BibitemOpen
  \bibfield  {author} {\bibinfo {author} {\bibfnamefont {C.}~\bibnamefont
  {Li}}, \bibinfo {author} {\bibfnamefont {F.}~\bibnamefont {S\'{e}guin}},
  \bibinfo {author} {\bibfnamefont {J.}~\bibnamefont {Frenje}}, \bibinfo
  {author} {\bibfnamefont {J.}~\bibnamefont {Rygg}}, \bibinfo {author}
  {\bibfnamefont {R.}~\bibnamefont {Petrasso}}, \bibinfo {author}
  {\bibfnamefont {R.}~\bibnamefont {Town}}, \bibinfo {author} {\bibfnamefont
  {P.}~\bibnamefont {Amendt}}, \bibinfo {author} {\bibfnamefont
  {S.}~\bibnamefont {Hatchett}}, \bibinfo {author} {\bibfnamefont
  {O.}~\bibnamefont {Landen}}, \bibinfo {author} {\bibfnamefont
  {A.}~\bibnamefont {Mackinnon}}, \bibinfo {author} {\bibfnamefont
  {P.}~\bibnamefont {Patel}}, \bibinfo {author} {\bibfnamefont
  {V.}~\bibnamefont {Smalyuk}}, \bibinfo {author} {\bibfnamefont
  {T.}~\bibnamefont {Sangster}}, \ and\ \bibinfo {author} {\bibfnamefont
  {J.}~\bibnamefont {Knauer}},\ }\bibfield  {title} {\enquote {\bibinfo {title}
  {{Measuring E and B Fields in Laser-Produced Plasmas with Monoenergetic
  Proton Radiography}},}\ }\href {\doibase 10.1103/PhysRevLett.97.135003}
  {\bibfield  {journal} {\bibinfo  {journal} {Physical Review Letters}\
  }\textbf {\bibinfo {volume} {97}},\ \bibinfo {pages} {135003} (\bibinfo
  {year} {2006}{\natexlab{b}})}\BibitemShut {NoStop}%
\bibitem [{\citenamefont {Kar}\ \emph {et~al.}(2008)\citenamefont {Kar},
  \citenamefont {Borghesi}, \citenamefont {Audebert}, \citenamefont
  {Benuzzi-Mounaix}, \citenamefont {Boehly}, \citenamefont {Hicks},
  \citenamefont {Koenig}, \citenamefont {Lancaster}, \citenamefont {Lepape},
  \citenamefont {Mackinnon}, \citenamefont {Norreys}, \citenamefont {Patel},\
  and\ \citenamefont {Romagnani}}]{Kar2008}%
  \BibitemOpen
  \bibfield  {author} {\bibinfo {author} {\bibfnamefont {S.}~\bibnamefont
  {Kar}}, \bibinfo {author} {\bibfnamefont {M.}~\bibnamefont {Borghesi}},
  \bibinfo {author} {\bibfnamefont {P.}~\bibnamefont {Audebert}}, \bibinfo
  {author} {\bibfnamefont {a.}~\bibnamefont {Benuzzi-Mounaix}}, \bibinfo
  {author} {\bibfnamefont {T.}~\bibnamefont {Boehly}}, \bibinfo {author}
  {\bibfnamefont {D.}~\bibnamefont {Hicks}}, \bibinfo {author} {\bibfnamefont
  {M.}~\bibnamefont {Koenig}}, \bibinfo {author} {\bibfnamefont
  {K.}~\bibnamefont {Lancaster}}, \bibinfo {author} {\bibfnamefont
  {S.}~\bibnamefont {Lepape}}, \bibinfo {author} {\bibfnamefont
  {a.}~\bibnamefont {Mackinnon}}, \bibinfo {author} {\bibfnamefont
  {P.}~\bibnamefont {Norreys}}, \bibinfo {author} {\bibfnamefont
  {P.}~\bibnamefont {Patel}}, \ and\ \bibinfo {author} {\bibfnamefont
  {L.}~\bibnamefont {Romagnani}},\ }\bibfield  {title} {\enquote {\bibinfo
  {title} {{Modeling of laser-driven proton radiography of dense matter}},}\
  }\href {\doibase 10.1016/j.hedp.2007.11.002} {\bibfield  {journal} {\bibinfo
  {journal} {High Energy Density Physics}\ }\textbf {\bibinfo {volume} {4}},\
  \bibinfo {pages} {26--40} (\bibinfo {year} {2008})}\BibitemShut {NoStop}%
\bibitem [{\citenamefont {Rygg}\ \emph {et~al.}(2008)\citenamefont {Rygg},
  \citenamefont {S\'{e}guin}, \citenamefont {Li}, \citenamefont {Frenje},
  \citenamefont {Manuel}, \citenamefont {Petrasso}, \citenamefont {Betti},
  \citenamefont {Delettrez}, \citenamefont {Gotchev}, \citenamefont {Knauer},
  \citenamefont {Meyerhofer}, \citenamefont {Marshall}, \citenamefont
  {Stoeckl},\ and\ \citenamefont {Theobald}}]{Rygg2008}%
  \BibitemOpen
  \bibfield  {author} {\bibinfo {author} {\bibfnamefont {J.~R.}\ \bibnamefont
  {Rygg}}, \bibinfo {author} {\bibfnamefont {F.~H.}\ \bibnamefont
  {S\'{e}guin}}, \bibinfo {author} {\bibfnamefont {C.~K.}\ \bibnamefont {Li}},
  \bibinfo {author} {\bibfnamefont {J.~a.}\ \bibnamefont {Frenje}}, \bibinfo
  {author} {\bibfnamefont {M.~J.-E.}\ \bibnamefont {Manuel}}, \bibinfo {author}
  {\bibfnamefont {R.~D.}\ \bibnamefont {Petrasso}}, \bibinfo {author}
  {\bibfnamefont {R.}~\bibnamefont {Betti}}, \bibinfo {author} {\bibfnamefont
  {J.~a.}\ \bibnamefont {Delettrez}}, \bibinfo {author} {\bibfnamefont {O.~V.}\
  \bibnamefont {Gotchev}}, \bibinfo {author} {\bibfnamefont {J.~P.}\
  \bibnamefont {Knauer}}, \bibinfo {author} {\bibfnamefont {D.~D.}\
  \bibnamefont {Meyerhofer}}, \bibinfo {author} {\bibfnamefont {F.~J.}\
  \bibnamefont {Marshall}}, \bibinfo {author} {\bibfnamefont {C.}~\bibnamefont
  {Stoeckl}}, \ and\ \bibinfo {author} {\bibfnamefont {W.}~\bibnamefont
  {Theobald}},\ }\bibfield  {title} {\enquote {\bibinfo {title} {{Proton
  radiography of inertial fusion implosions.}}}\ }\href {\doibase
  10.1126/science.1152640} {\bibfield  {journal} {\bibinfo  {journal} {Science
  (New York, N.Y.)}\ }\textbf {\bibinfo {volume} {319}},\ \bibinfo {pages}
  {1223--5} (\bibinfo {year} {2008})}\BibitemShut {NoStop}%
\bibitem [{\citenamefont {Li}\ \emph {et~al.}(2009{\natexlab{a}})\citenamefont
  {Li}, \citenamefont {S\'{e}guin}, \citenamefont {Frenje}, \citenamefont
  {Manuel}, \citenamefont {Petrasso}, \citenamefont {Smalyuk}, \citenamefont
  {Betti}, \citenamefont {Delettrez}, \citenamefont {Knauer}, \citenamefont
  {Marshall}, \citenamefont {Meyerhofer}, \citenamefont {Shvarts},
  \citenamefont {Stoeckl}, \citenamefont {Theobald}, \citenamefont {Rygg},
  \citenamefont {Landen}, \citenamefont {Town}, \citenamefont {Amendt},
  \citenamefont {Back},\ and\ \citenamefont {Kilkenny}}]{Li2009a}%
  \BibitemOpen
  \bibfield  {author} {\bibinfo {author} {\bibfnamefont {C.~K.}\ \bibnamefont
  {Li}}, \bibinfo {author} {\bibfnamefont {F.~H.}\ \bibnamefont {S\'{e}guin}},
  \bibinfo {author} {\bibfnamefont {J.~A.}\ \bibnamefont {Frenje}}, \bibinfo
  {author} {\bibfnamefont {M.}~\bibnamefont {Manuel}}, \bibinfo {author}
  {\bibfnamefont {R.~D.}\ \bibnamefont {Petrasso}}, \bibinfo {author}
  {\bibfnamefont {V.~a.}\ \bibnamefont {Smalyuk}}, \bibinfo {author}
  {\bibfnamefont {R.}~\bibnamefont {Betti}}, \bibinfo {author} {\bibfnamefont
  {J.}~\bibnamefont {Delettrez}}, \bibinfo {author} {\bibfnamefont {J.~P.}\
  \bibnamefont {Knauer}}, \bibinfo {author} {\bibfnamefont {F.}~\bibnamefont
  {Marshall}}, \bibinfo {author} {\bibfnamefont {D.~D.}\ \bibnamefont
  {Meyerhofer}}, \bibinfo {author} {\bibfnamefont {D.}~\bibnamefont {Shvarts}},
  \bibinfo {author} {\bibfnamefont {C.}~\bibnamefont {Stoeckl}}, \bibinfo
  {author} {\bibfnamefont {W.}~\bibnamefont {Theobald}}, \bibinfo {author}
  {\bibfnamefont {J.~R.}\ \bibnamefont {Rygg}}, \bibinfo {author}
  {\bibfnamefont {O.~L.}\ \bibnamefont {Landen}}, \bibinfo {author}
  {\bibfnamefont {R.~P.~J.}\ \bibnamefont {Town}}, \bibinfo {author}
  {\bibfnamefont {P.~a.}\ \bibnamefont {Amendt}}, \bibinfo {author}
  {\bibfnamefont {C.~a.}\ \bibnamefont {Back}}, \ and\ \bibinfo {author}
  {\bibfnamefont {J.~D.}\ \bibnamefont {Kilkenny}},\ }\bibfield  {title}
  {\enquote {\bibinfo {title} {{Study of direct-drive capsule implosions in
  inertial confinement fusion with proton radiography}},}\ }\href {\doibase
  10.1088/0741-3335/51/1/014003} {\bibfield  {journal} {\bibinfo  {journal}
  {Plasma Physics and Controlled Fusion}\ }\textbf {\bibinfo {volume} {51}},\
  \bibinfo {pages} {014003} (\bibinfo {year} {2009}{\natexlab{a}})}\BibitemShut
  {NoStop}%
\bibitem [{\citenamefont {Li}\ \emph {et~al.}(2009{\natexlab{b}})\citenamefont
  {Li}, \citenamefont {Séguin}, \citenamefont {Frenje}, \citenamefont
  {Manuel}, \citenamefont {Casey}, \citenamefont {Sinenian}, \citenamefont
  {Petrasso}, \citenamefont {Amendt}, \citenamefont {Landen}, \citenamefont
  {Rygg}, \citenamefont {Town}, \citenamefont {Betti}, \citenamefont
  {Delettrez}, \citenamefont {Knauer}, \citenamefont {Marshall}, \citenamefont
  {Meyerhofer}, \citenamefont {Sangster}, \citenamefont {Shvarts},
  \citenamefont {Smalyuk}, \citenamefont {Soures}, \citenamefont {Back},
  \citenamefont {Kilkenny},\ and\ \citenamefont {Nikroo}}]{Li2009}%
  \BibitemOpen
  \bibfield  {author} {\bibinfo {author} {\bibfnamefont {C.~K.}\ \bibnamefont
  {Li}}, \bibinfo {author} {\bibfnamefont {F.~H.}\ \bibnamefont {Séguin}},
  \bibinfo {author} {\bibfnamefont {J.~A.}\ \bibnamefont {Frenje}}, \bibinfo
  {author} {\bibfnamefont {M.}~\bibnamefont {Manuel}}, \bibinfo {author}
  {\bibfnamefont {D.}~\bibnamefont {Casey}}, \bibinfo {author} {\bibfnamefont
  {N.}~\bibnamefont {Sinenian}}, \bibinfo {author} {\bibfnamefont {R.~D.}\
  \bibnamefont {Petrasso}}, \bibinfo {author} {\bibfnamefont {P.~A.}\
  \bibnamefont {Amendt}}, \bibinfo {author} {\bibfnamefont {O.~L.}\
  \bibnamefont {Landen}}, \bibinfo {author} {\bibfnamefont {J.~R.}\
  \bibnamefont {Rygg}}, \bibinfo {author} {\bibfnamefont {R.~P.~J.}\
  \bibnamefont {Town}}, \bibinfo {author} {\bibfnamefont {R.}~\bibnamefont
  {Betti}}, \bibinfo {author} {\bibfnamefont {J.}~\bibnamefont {Delettrez}},
  \bibinfo {author} {\bibfnamefont {J.~P.}\ \bibnamefont {Knauer}}, \bibinfo
  {author} {\bibfnamefont {F.}~\bibnamefont {Marshall}}, \bibinfo {author}
  {\bibfnamefont {D.~D.}\ \bibnamefont {Meyerhofer}}, \bibinfo {author}
  {\bibfnamefont {T.~C.}\ \bibnamefont {Sangster}}, \bibinfo {author}
  {\bibfnamefont {D.}~\bibnamefont {Shvarts}}, \bibinfo {author} {\bibfnamefont
  {V.~A.}\ \bibnamefont {Smalyuk}}, \bibinfo {author} {\bibfnamefont {J.~M.}\
  \bibnamefont {Soures}}, \bibinfo {author} {\bibfnamefont {C.~A.}\
  \bibnamefont {Back}}, \bibinfo {author} {\bibfnamefont {J.~D.}\ \bibnamefont
  {Kilkenny}}, \ and\ \bibinfo {author} {\bibfnamefont {A.}~\bibnamefont
  {Nikroo}},\ }\bibfield  {title} {\enquote {\bibinfo {title} {{Proton
  radiography of dynamic electric and magnetic fields in laser-produced
  high-energy-density plasmas}},}\ }\href {\doibase 10.1063/1.3096781}
  {\bibfield  {journal} {\bibinfo  {journal} {Physics of Plasmas}\ }\textbf
  {\bibinfo {volume} {16}},\ \bibinfo {pages} {056304} (\bibinfo {year}
  {2009}{\natexlab{b}})}\BibitemShut {NoStop}%
\bibitem [{\citenamefont {Gotchev}\ \emph {et~al.}(2009)\citenamefont
  {Gotchev}, \citenamefont {Chang}, \citenamefont {Knauer}, \citenamefont
  {Meyerhofer}, \citenamefont {Polomarov}, \citenamefont {Frenje},
  \citenamefont {Li}, \citenamefont {Manuel}, \citenamefont {Petrasso},
  \citenamefont {Rygg}, \citenamefont {S\'{e}guin},\ and\ \citenamefont
  {Betti}}]{Gotchev2009}%
  \BibitemOpen
  \bibfield  {author} {\bibinfo {author} {\bibfnamefont {O.}~\bibnamefont
  {Gotchev}}, \bibinfo {author} {\bibfnamefont {P.}~\bibnamefont {Chang}},
  \bibinfo {author} {\bibfnamefont {J.}~\bibnamefont {Knauer}}, \bibinfo
  {author} {\bibfnamefont {D.}~\bibnamefont {Meyerhofer}}, \bibinfo {author}
  {\bibfnamefont {O.}~\bibnamefont {Polomarov}}, \bibinfo {author}
  {\bibfnamefont {J.}~\bibnamefont {Frenje}}, \bibinfo {author} {\bibfnamefont
  {C.}~\bibnamefont {Li}}, \bibinfo {author} {\bibfnamefont {M.}~\bibnamefont
  {Manuel}}, \bibinfo {author} {\bibfnamefont {R.}~\bibnamefont {Petrasso}},
  \bibinfo {author} {\bibfnamefont {J.}~\bibnamefont {Rygg}}, \bibinfo {author}
  {\bibfnamefont {F.}~\bibnamefont {S\'{e}guin}}, \ and\ \bibinfo {author}
  {\bibfnamefont {R.}~\bibnamefont {Betti}},\ }\bibfield  {title} {\enquote
  {\bibinfo {title} {{Laser-Driven Magnetic-Flux Compression in
  High-Energy-Density Plasmas}},}\ }\href {\doibase
  10.1103/PhysRevLett.103.215004} {\bibfield  {journal} {\bibinfo  {journal}
  {Physical Review Letters}\ }\textbf {\bibinfo {volume} {103}},\ \bibinfo
  {pages} {215004} (\bibinfo {year} {2009})}\BibitemShut {NoStop}%
\bibitem [{\citenamefont {Borghesi}\ \emph {et~al.}(2010)\citenamefont
  {Borghesi}, \citenamefont {Sarri}, \citenamefont {Cecchetti}, \citenamefont
  {Kourakis}, \citenamefont {Hoarty}, \citenamefont {Stevenson}, \citenamefont
  {James}, \citenamefont {Brown}, \citenamefont {Hobbs}, \citenamefont
  {Lockyear}, \citenamefont {Morton}, \citenamefont {Willi}, \citenamefont
  {Jung},\ and\ \citenamefont {Dieckmann}}]{Borghesi2010}%
  \BibitemOpen
  \bibfield  {author} {\bibinfo {author} {\bibfnamefont {M.}~\bibnamefont
  {Borghesi}}, \bibinfo {author} {\bibfnamefont {G.}~\bibnamefont {Sarri}},
  \bibinfo {author} {\bibfnamefont {C.}~\bibnamefont {Cecchetti}}, \bibinfo
  {author} {\bibfnamefont {I.}~\bibnamefont {Kourakis}}, \bibinfo {author}
  {\bibfnamefont {D.}~\bibnamefont {Hoarty}}, \bibinfo {author} {\bibfnamefont
  {R.}~\bibnamefont {Stevenson}}, \bibinfo {author} {\bibfnamefont
  {S.}~\bibnamefont {James}}, \bibinfo {author} {\bibfnamefont
  {C.}~\bibnamefont {Brown}}, \bibinfo {author} {\bibfnamefont
  {P.}~\bibnamefont {Hobbs}}, \bibinfo {author} {\bibfnamefont
  {J.}~\bibnamefont {Lockyear}}, \bibinfo {author} {\bibfnamefont
  {J.}~\bibnamefont {Morton}}, \bibinfo {author} {\bibfnamefont
  {O.}~\bibnamefont {Willi}}, \bibinfo {author} {\bibfnamefont
  {R.}~\bibnamefont {Jung}}, \ and\ \bibinfo {author} {\bibfnamefont
  {M.}~\bibnamefont {Dieckmann}},\ }\bibfield  {title} {\enquote {\bibinfo
  {title} {{Progress in proton radiography for diagnosis of ICF-relevant
  plasmas}},}\ }\href {\doibase 10.1017/S0263034610000170} {\bibfield
  {journal} {\bibinfo  {journal} {Laser and Particle Beams}\ }\textbf {\bibinfo
  {volume} {28}},\ \bibinfo {pages} {277--284} (\bibinfo {year}
  {2010})}\BibitemShut {NoStop}%
\bibitem [{\citenamefont {Sarri}\ \emph {et~al.}(2010)\citenamefont {Sarri},
  \citenamefont {Cecchetti}, \citenamefont {Romagnani}, \citenamefont {Brown},
  \citenamefont {Hoarty}, \citenamefont {James}, \citenamefont {Morton},
  \citenamefont {Dieckmann}, \citenamefont {Jung}, \citenamefont {Willi},
  \citenamefont {Bulanov}, \citenamefont {Pegoraro},\ and\ \citenamefont
  {Borghesi}}]{Sarri2010}%
  \BibitemOpen
  \bibfield  {author} {\bibinfo {author} {\bibfnamefont {G.}~\bibnamefont
  {Sarri}}, \bibinfo {author} {\bibfnamefont {C.~a.}\ \bibnamefont
  {Cecchetti}}, \bibinfo {author} {\bibfnamefont {L.}~\bibnamefont
  {Romagnani}}, \bibinfo {author} {\bibfnamefont {C.~M.}\ \bibnamefont
  {Brown}}, \bibinfo {author} {\bibfnamefont {D.~J.}\ \bibnamefont {Hoarty}},
  \bibinfo {author} {\bibfnamefont {S.}~\bibnamefont {James}}, \bibinfo
  {author} {\bibfnamefont {J.}~\bibnamefont {Morton}}, \bibinfo {author}
  {\bibfnamefont {M.~E.}\ \bibnamefont {Dieckmann}}, \bibinfo {author}
  {\bibfnamefont {R.}~\bibnamefont {Jung}}, \bibinfo {author} {\bibfnamefont
  {O.}~\bibnamefont {Willi}}, \bibinfo {author} {\bibfnamefont {S.~V.}\
  \bibnamefont {Bulanov}}, \bibinfo {author} {\bibfnamefont {F.}~\bibnamefont
  {Pegoraro}}, \ and\ \bibinfo {author} {\bibfnamefont {M.}~\bibnamefont
  {Borghesi}},\ }\bibfield  {title} {\enquote {\bibinfo {title} {{The
  application of laser-driven proton beams to the radiography of intense
  laser–hohlraum interactions}},}\ }\href {\doibase
  10.1088/1367-2630/12/4/045006} {\bibfield  {journal} {\bibinfo  {journal}
  {New Journal of Physics}\ }\textbf {\bibinfo {volume} {12}},\ \bibinfo
  {pages} {045006} (\bibinfo {year} {2010})}\BibitemShut {NoStop}%
\bibitem [{\citenamefont {Li}\ \emph {et~al.}(2010{\natexlab{a}})\citenamefont
  {Li}, \citenamefont {S\'{e}guin}, \citenamefont {Frenje}, \citenamefont
  {Rosenberg}, \citenamefont {Petrasso}, \citenamefont {Amendt}, \citenamefont
  {Koch}, \citenamefont {Landen}, \citenamefont {Park}, \citenamefont {Robey},
  \citenamefont {Town}, \citenamefont {Casner}, \citenamefont {Philippe},
  \citenamefont {Betti}, \citenamefont {Knauer}, \citenamefont {Meyerhofer},
  \citenamefont {Back}, \citenamefont {Kilkenny},\ and\ \citenamefont
  {Nikroo}}]{Li2010}%
  \BibitemOpen
  \bibfield  {author} {\bibinfo {author} {\bibfnamefont {C.~K.}\ \bibnamefont
  {Li}}, \bibinfo {author} {\bibfnamefont {F.~H.}\ \bibnamefont {S\'{e}guin}},
  \bibinfo {author} {\bibfnamefont {J.~a.}\ \bibnamefont {Frenje}}, \bibinfo
  {author} {\bibfnamefont {M.}~\bibnamefont {Rosenberg}}, \bibinfo {author}
  {\bibfnamefont {R.~D.}\ \bibnamefont {Petrasso}}, \bibinfo {author}
  {\bibfnamefont {P.~a.}\ \bibnamefont {Amendt}}, \bibinfo {author}
  {\bibfnamefont {J.~a.}\ \bibnamefont {Koch}}, \bibinfo {author}
  {\bibfnamefont {O.~L.}\ \bibnamefont {Landen}}, \bibinfo {author}
  {\bibfnamefont {H.~S.}\ \bibnamefont {Park}}, \bibinfo {author}
  {\bibfnamefont {H.~F.}\ \bibnamefont {Robey}}, \bibinfo {author}
  {\bibfnamefont {R.~P.~J.}\ \bibnamefont {Town}}, \bibinfo {author}
  {\bibfnamefont {a.}~\bibnamefont {Casner}}, \bibinfo {author} {\bibfnamefont
  {F.}~\bibnamefont {Philippe}}, \bibinfo {author} {\bibfnamefont
  {R.}~\bibnamefont {Betti}}, \bibinfo {author} {\bibfnamefont {J.~P.}\
  \bibnamefont {Knauer}}, \bibinfo {author} {\bibfnamefont {D.~D.}\
  \bibnamefont {Meyerhofer}}, \bibinfo {author} {\bibfnamefont {C.~a.}\
  \bibnamefont {Back}}, \bibinfo {author} {\bibfnamefont {J.~D.}\ \bibnamefont
  {Kilkenny}}, \ and\ \bibinfo {author} {\bibfnamefont {a.}~\bibnamefont
  {Nikroo}},\ }\bibfield  {title} {\enquote {\bibinfo {title}
  {{Charged-particle probing of x-ray-driven inertial-fusion implosions.}}}\
  }\href {\doibase 10.1126/science.1185747} {\bibfield  {journal} {\bibinfo
  {journal} {Science (New York, N.Y.)}\ }\textbf {\bibinfo {volume} {327}},\
  \bibinfo {pages} {1231--5} (\bibinfo {year}
  {2010}{\natexlab{a}})}\BibitemShut {NoStop}%
\bibitem [{\citenamefont {Manuel}\ \emph
  {et~al.}(2012{\natexlab{a}})\citenamefont {Manuel}, \citenamefont {Li},
  \citenamefont {S\'{e}guin}, \citenamefont {Frenje}, \citenamefont {Casey},
  \citenamefont {Petrasso}, \citenamefont {Hu}, \citenamefont {Betti},
  \citenamefont {Hager}, \citenamefont {Meyerhofer},\ and\ \citenamefont
  {Smalyuk}}]{Manuel2012}%
  \BibitemOpen
  \bibfield  {author} {\bibinfo {author} {\bibfnamefont {M.~J.-E.}\
  \bibnamefont {Manuel}}, \bibinfo {author} {\bibfnamefont {C.~K.}\
  \bibnamefont {Li}}, \bibinfo {author} {\bibfnamefont {F.~H.}\ \bibnamefont
  {S\'{e}guin}}, \bibinfo {author} {\bibfnamefont {J.}~\bibnamefont {Frenje}},
  \bibinfo {author} {\bibfnamefont {D.~T.}\ \bibnamefont {Casey}}, \bibinfo
  {author} {\bibfnamefont {R.~D.}\ \bibnamefont {Petrasso}}, \bibinfo {author}
  {\bibfnamefont {S.~X.}\ \bibnamefont {Hu}}, \bibinfo {author} {\bibfnamefont
  {R.}~\bibnamefont {Betti}}, \bibinfo {author} {\bibfnamefont {J.~D.}\
  \bibnamefont {Hager}}, \bibinfo {author} {\bibfnamefont {D.~D.}\ \bibnamefont
  {Meyerhofer}}, \ and\ \bibinfo {author} {\bibfnamefont {V.~a.}\ \bibnamefont
  {Smalyuk}},\ }\bibfield  {title} {\enquote {\bibinfo {title} {{First
  Measurements of Rayleigh-Taylor-Induced Magnetic Fields in Laser-Produced
  Plasmas}},}\ }\href {\doibase 10.1103/PhysRevLett.108.255006} {\bibfield
  {journal} {\bibinfo  {journal} {Physical Review Letters}\ }\textbf {\bibinfo
  {volume} {108}},\ \bibinfo {pages} {255006} (\bibinfo {year}
  {2012}{\natexlab{a}})}\BibitemShut {NoStop}%
\bibitem [{\citenamefont {Li}\ \emph {et~al.}(2012)\citenamefont {Li},
  \citenamefont {S\'{e}guin}, \citenamefont {Frenje}, \citenamefont
  {Rosenberg}, \citenamefont {Rinderknecht}, \citenamefont {Zylstra},
  \citenamefont {Petrasso}, \citenamefont {Amendt}, \citenamefont {Landen},
  \citenamefont {Mackinnon}, \citenamefont {Town}, \citenamefont {Wilks},
  \citenamefont {Betti}, \citenamefont {Meyerhofer}, \citenamefont {Soures},
  \citenamefont {Hund}, \citenamefont {Kilkenny},\ and\ \citenamefont
  {Nikroo}}]{Li2012}%
  \BibitemOpen
  \bibfield  {author} {\bibinfo {author} {\bibfnamefont {C.~K.}\ \bibnamefont
  {Li}}, \bibinfo {author} {\bibfnamefont {F.~H.}\ \bibnamefont {S\'{e}guin}},
  \bibinfo {author} {\bibfnamefont {J.~a.}\ \bibnamefont {Frenje}}, \bibinfo
  {author} {\bibfnamefont {M.~J.}\ \bibnamefont {Rosenberg}}, \bibinfo {author}
  {\bibfnamefont {H.~G.}\ \bibnamefont {Rinderknecht}}, \bibinfo {author}
  {\bibfnamefont {a.~B.}\ \bibnamefont {Zylstra}}, \bibinfo {author}
  {\bibfnamefont {R.~D.}\ \bibnamefont {Petrasso}}, \bibinfo {author}
  {\bibfnamefont {P.~a.}\ \bibnamefont {Amendt}}, \bibinfo {author}
  {\bibfnamefont {O.~L.}\ \bibnamefont {Landen}}, \bibinfo {author}
  {\bibfnamefont {a.~J.}\ \bibnamefont {Mackinnon}}, \bibinfo {author}
  {\bibfnamefont {R.~P.~J.}\ \bibnamefont {Town}}, \bibinfo {author}
  {\bibfnamefont {S.~C.}\ \bibnamefont {Wilks}}, \bibinfo {author}
  {\bibfnamefont {R.}~\bibnamefont {Betti}}, \bibinfo {author} {\bibfnamefont
  {D.~D.}\ \bibnamefont {Meyerhofer}}, \bibinfo {author} {\bibfnamefont
  {J.~M.}\ \bibnamefont {Soures}}, \bibinfo {author} {\bibfnamefont
  {J.}~\bibnamefont {Hund}}, \bibinfo {author} {\bibfnamefont {J.~D.}\
  \bibnamefont {Kilkenny}}, \ and\ \bibinfo {author} {\bibfnamefont
  {a.}~\bibnamefont {Nikroo}},\ }\bibfield  {title} {\enquote {\bibinfo {title}
  {{Impeding Hohlraum Plasma Stagnation in Inertial-Confinement Fusion}},}\
  }\href {\doibase 10.1103/PhysRevLett.108.025001} {\bibfield  {journal}
  {\bibinfo  {journal} {Physical Review Letters}\ }\textbf {\bibinfo {volume}
  {108}},\ \bibinfo {pages} {025001} (\bibinfo {year} {2012})}\BibitemShut
  {NoStop}%
\bibitem [{\citenamefont {Zylstra}\ \emph {et~al.}(2012)\citenamefont
  {Zylstra}, \citenamefont {Li}, \citenamefont {Rinderknecht}, \citenamefont
  {S\'{e}guin}, \citenamefont {Petrasso}, \citenamefont {Stoeckl},
  \citenamefont {Meyerhofer}, \citenamefont {Nilson}, \citenamefont {Sangster},
  \citenamefont {{Le Pape}}, \citenamefont {Mackinnon},\ and\ \citenamefont
  {Patel}}]{Zylstra2012}%
  \BibitemOpen
  \bibfield  {author} {\bibinfo {author} {\bibfnamefont {A.~B.}\ \bibnamefont
  {Zylstra}}, \bibinfo {author} {\bibfnamefont {C.~K.}\ \bibnamefont {Li}},
  \bibinfo {author} {\bibfnamefont {H.~G.}\ \bibnamefont {Rinderknecht}},
  \bibinfo {author} {\bibfnamefont {F.~H.}\ \bibnamefont {S\'{e}guin}},
  \bibinfo {author} {\bibfnamefont {R.~D.}\ \bibnamefont {Petrasso}}, \bibinfo
  {author} {\bibfnamefont {C.}~\bibnamefont {Stoeckl}}, \bibinfo {author}
  {\bibfnamefont {D.~D.}\ \bibnamefont {Meyerhofer}}, \bibinfo {author}
  {\bibfnamefont {P.}~\bibnamefont {Nilson}}, \bibinfo {author} {\bibfnamefont
  {T.~C.}\ \bibnamefont {Sangster}}, \bibinfo {author} {\bibfnamefont
  {S.}~\bibnamefont {{Le Pape}}}, \bibinfo {author} {\bibfnamefont
  {A.}~\bibnamefont {Mackinnon}}, \ and\ \bibinfo {author} {\bibfnamefont
  {P.}~\bibnamefont {Patel}},\ }\bibfield  {title} {\enquote {\bibinfo {title}
  {{Using high-intensity laser-generated energetic protons to radiograph
  directly driven implosions.}}}\ }\href {\doibase 10.1063/1.3680110}
  {\bibfield  {journal} {\bibinfo  {journal} {The Review of scientific
  instruments}\ }\textbf {\bibinfo {volume} {83}},\ \bibinfo {pages} {013511}
  (\bibinfo {year} {2012})}\BibitemShut {NoStop}%
\bibitem [{\citenamefont {Kugland}\ \emph
  {et~al.}(2012{\natexlab{a}})\citenamefont {Kugland}, \citenamefont {Ryutov},
  \citenamefont {Chang}, \citenamefont {Drake}, \citenamefont {Fiksel},
  \citenamefont {Froula}, \citenamefont {Glenzer}, \citenamefont {Gregori},
  \citenamefont {Grosskopf}, \citenamefont {Koenig}, \citenamefont {Kuramitsu},
  \citenamefont {Kuranz}, \citenamefont {Levy}, \citenamefont {Liang},
  \citenamefont {Meinecke}, \citenamefont {Miniati}, \citenamefont {Morita},
  \citenamefont {Pelka}, \citenamefont {Plechaty}, \citenamefont {Presura},
  \citenamefont {Ravasio}, \citenamefont {Remington}, \citenamefont {Reville},
  \citenamefont {Ross}, \citenamefont {Sakawa}, \citenamefont {Spitkovsky},
  \citenamefont {Takabe},\ and\ \citenamefont {Park}}]{Kugland2012a}%
  \BibitemOpen
  \bibfield  {author} {\bibinfo {author} {\bibfnamefont {N.~L.}\ \bibnamefont
  {Kugland}}, \bibinfo {author} {\bibfnamefont {D.~D.}\ \bibnamefont {Ryutov}},
  \bibinfo {author} {\bibfnamefont {P.-Y.}\ \bibnamefont {Chang}}, \bibinfo
  {author} {\bibfnamefont {R.~P.}\ \bibnamefont {Drake}}, \bibinfo {author}
  {\bibfnamefont {G.}~\bibnamefont {Fiksel}}, \bibinfo {author} {\bibfnamefont
  {D.~H.}\ \bibnamefont {Froula}}, \bibinfo {author} {\bibfnamefont {S.~H.}\
  \bibnamefont {Glenzer}}, \bibinfo {author} {\bibfnamefont {G.}~\bibnamefont
  {Gregori}}, \bibinfo {author} {\bibfnamefont {M.}~\bibnamefont {Grosskopf}},
  \bibinfo {author} {\bibfnamefont {M.}~\bibnamefont {Koenig}}, \bibinfo
  {author} {\bibfnamefont {Y.}~\bibnamefont {Kuramitsu}}, \bibinfo {author}
  {\bibfnamefont {C.}~\bibnamefont {Kuranz}}, \bibinfo {author} {\bibfnamefont
  {M.~C.}\ \bibnamefont {Levy}}, \bibinfo {author} {\bibfnamefont
  {E.}~\bibnamefont {Liang}}, \bibinfo {author} {\bibfnamefont
  {J.}~\bibnamefont {Meinecke}}, \bibinfo {author} {\bibfnamefont
  {F.}~\bibnamefont {Miniati}}, \bibinfo {author} {\bibfnamefont
  {T.}~\bibnamefont {Morita}}, \bibinfo {author} {\bibfnamefont
  {A.}~\bibnamefont {Pelka}}, \bibinfo {author} {\bibfnamefont
  {C.}~\bibnamefont {Plechaty}}, \bibinfo {author} {\bibfnamefont
  {R.}~\bibnamefont {Presura}}, \bibinfo {author} {\bibfnamefont
  {A.}~\bibnamefont {Ravasio}}, \bibinfo {author} {\bibfnamefont {B.~a.}\
  \bibnamefont {Remington}}, \bibinfo {author} {\bibfnamefont {B.}~\bibnamefont
  {Reville}}, \bibinfo {author} {\bibfnamefont {J.~S.}\ \bibnamefont {Ross}},
  \bibinfo {author} {\bibfnamefont {Y.}~\bibnamefont {Sakawa}}, \bibinfo
  {author} {\bibfnamefont {A.}~\bibnamefont {Spitkovsky}}, \bibinfo {author}
  {\bibfnamefont {H.}~\bibnamefont {Takabe}}, \ and\ \bibinfo {author}
  {\bibfnamefont {H.-S.}\ \bibnamefont {Park}},\ }\bibfield  {title} {\enquote
  {\bibinfo {title} {{Self-organized electromagnetic field structures in
  laser-produced counter-streaming plasmas}},}\ }\href {\doibase
  10.1038/nphys2434} {\bibfield  {journal} {\bibinfo  {journal} {Nature
  Physics}\ }\textbf {\bibinfo {volume} {8}},\ \bibinfo {pages} {809--812}
  (\bibinfo {year} {2012}{\natexlab{a}})}\BibitemShut {NoStop}%
\bibitem [{\citenamefont {Nilson}\ \emph {et~al.}(2006)\citenamefont {Nilson},
  \citenamefont {Willingale}, \citenamefont {Kaluza}, \citenamefont
  {Kamperidis}, \citenamefont {Minardi}, \citenamefont {Wei}, \citenamefont
  {Fernandes}, \citenamefont {Notley}, \citenamefont {Bandyopadhyay},
  \citenamefont {Sherlock}, \citenamefont {Kingham}, \citenamefont {Tatarakis},
  \citenamefont {Najmudin}, \citenamefont {Rozmus}, \citenamefont {Evans},
  \citenamefont {Haines}, \citenamefont {Dangor},\ and\ \citenamefont
  {Krushelnick}}]{Nilson2006}%
  \BibitemOpen
  \bibfield  {author} {\bibinfo {author} {\bibfnamefont {P.}~\bibnamefont
  {Nilson}}, \bibinfo {author} {\bibfnamefont {L.}~\bibnamefont {Willingale}},
  \bibinfo {author} {\bibfnamefont {M.}~\bibnamefont {Kaluza}}, \bibinfo
  {author} {\bibfnamefont {C.}~\bibnamefont {Kamperidis}}, \bibinfo {author}
  {\bibfnamefont {S.}~\bibnamefont {Minardi}}, \bibinfo {author} {\bibfnamefont
  {M.}~\bibnamefont {Wei}}, \bibinfo {author} {\bibfnamefont {P.}~\bibnamefont
  {Fernandes}}, \bibinfo {author} {\bibfnamefont {M.}~\bibnamefont {Notley}},
  \bibinfo {author} {\bibfnamefont {S.}~\bibnamefont {Bandyopadhyay}}, \bibinfo
  {author} {\bibfnamefont {M.}~\bibnamefont {Sherlock}}, \bibinfo {author}
  {\bibfnamefont {R.}~\bibnamefont {Kingham}}, \bibinfo {author} {\bibfnamefont
  {M.}~\bibnamefont {Tatarakis}}, \bibinfo {author} {\bibfnamefont
  {Z.}~\bibnamefont {Najmudin}}, \bibinfo {author} {\bibfnamefont
  {W.}~\bibnamefont {Rozmus}}, \bibinfo {author} {\bibfnamefont
  {R.}~\bibnamefont {Evans}}, \bibinfo {author} {\bibfnamefont
  {M.}~\bibnamefont {Haines}}, \bibinfo {author} {\bibfnamefont
  {a.}~\bibnamefont {Dangor}}, \ and\ \bibinfo {author} {\bibfnamefont
  {K.}~\bibnamefont {Krushelnick}},\ }\bibfield  {title} {\enquote {\bibinfo
  {title} {{Magnetic Reconnection and Plasma Dynamics in Two-Beam Laser-Solid
  Interactions}},}\ }\href {\doibase 10.1103/PhysRevLett.97.255001} {\bibfield
  {journal} {\bibinfo  {journal} {Physical Review Letters}\ }\textbf {\bibinfo
  {volume} {97}},\ \bibinfo {pages} {255001} (\bibinfo {year}
  {2006})}\BibitemShut {NoStop}%
\bibitem [{\citenamefont {Li}\ \emph {et~al.}(2007{\natexlab{a}})\citenamefont
  {Li}, \citenamefont {S\'{e}guin}, \citenamefont {Frenje}, \citenamefont
  {Rygg}, \citenamefont {Petrasso}, \citenamefont {Town}, \citenamefont
  {Landen}, \citenamefont {Knauer},\ and\ \citenamefont {Smalyuk}}]{Li2007}%
  \BibitemOpen
  \bibfield  {author} {\bibinfo {author} {\bibfnamefont {C.}~\bibnamefont
  {Li}}, \bibinfo {author} {\bibfnamefont {F.}~\bibnamefont {S\'{e}guin}},
  \bibinfo {author} {\bibfnamefont {J.}~\bibnamefont {Frenje}}, \bibinfo
  {author} {\bibfnamefont {J.}~\bibnamefont {Rygg}}, \bibinfo {author}
  {\bibfnamefont {R.}~\bibnamefont {Petrasso}}, \bibinfo {author}
  {\bibfnamefont {R.}~\bibnamefont {Town}}, \bibinfo {author} {\bibfnamefont
  {O.}~\bibnamefont {Landen}}, \bibinfo {author} {\bibfnamefont
  {J.}~\bibnamefont {Knauer}}, \ and\ \bibinfo {author} {\bibfnamefont
  {V.}~\bibnamefont {Smalyuk}},\ }\bibfield  {title} {\enquote {\bibinfo
  {title} {{Observation of Megagauss-Field Topology Changes due to Magnetic
  Reconnection in Laser-Produced Plasmas}},}\ }\href {\doibase
  10.1103/PhysRevLett.99.055001} {\bibfield  {journal} {\bibinfo  {journal}
  {Physical Review Letters}\ }\textbf {\bibinfo {volume} {99}},\ \bibinfo
  {pages} {055001} (\bibinfo {year} {2007}{\natexlab{a}})}\BibitemShut
  {NoStop}%
\bibitem [{\citenamefont {Willingale}\ \emph {et~al.}(2010)\citenamefont
  {Willingale}, \citenamefont {Nilson}, \citenamefont {Kaluza}, \citenamefont
  {Dangor}, \citenamefont {Evans}, \citenamefont {Fernandes}, \citenamefont
  {Haines}, \citenamefont {Kamperidis}, \citenamefont {Kingham}, \citenamefont
  {Ridgers}, \citenamefont {Sherlock}, \citenamefont {Thomas}, \citenamefont
  {Wei}, \citenamefont {Najmudin}, \citenamefont {Krushelnick}, \citenamefont
  {Bandyopadhyay}, \citenamefont {Notley}, \citenamefont {Minardi},
  \citenamefont {Tatarakis},\ and\ \citenamefont {Rozmus}}]{Willingale2010}%
  \BibitemOpen
  \bibfield  {author} {\bibinfo {author} {\bibfnamefont {L.}~\bibnamefont
  {Willingale}}, \bibinfo {author} {\bibfnamefont {P.~M.}\ \bibnamefont
  {Nilson}}, \bibinfo {author} {\bibfnamefont {M.~C.}\ \bibnamefont {Kaluza}},
  \bibinfo {author} {\bibfnamefont {a.~E.}\ \bibnamefont {Dangor}}, \bibinfo
  {author} {\bibfnamefont {R.~G.}\ \bibnamefont {Evans}}, \bibinfo {author}
  {\bibfnamefont {P.}~\bibnamefont {Fernandes}}, \bibinfo {author}
  {\bibfnamefont {M.~G.}\ \bibnamefont {Haines}}, \bibinfo {author}
  {\bibfnamefont {C.}~\bibnamefont {Kamperidis}}, \bibinfo {author}
  {\bibfnamefont {R.~J.}\ \bibnamefont {Kingham}}, \bibinfo {author}
  {\bibfnamefont {C.~P.}\ \bibnamefont {Ridgers}}, \bibinfo {author}
  {\bibfnamefont {M.}~\bibnamefont {Sherlock}}, \bibinfo {author}
  {\bibfnamefont {a.~G.~R.}\ \bibnamefont {Thomas}}, \bibinfo {author}
  {\bibfnamefont {M.~S.}\ \bibnamefont {Wei}}, \bibinfo {author} {\bibfnamefont
  {Z.}~\bibnamefont {Najmudin}}, \bibinfo {author} {\bibfnamefont
  {K.}~\bibnamefont {Krushelnick}}, \bibinfo {author} {\bibfnamefont
  {S.}~\bibnamefont {Bandyopadhyay}}, \bibinfo {author} {\bibfnamefont
  {M.}~\bibnamefont {Notley}}, \bibinfo {author} {\bibfnamefont
  {S.}~\bibnamefont {Minardi}}, \bibinfo {author} {\bibfnamefont
  {M.}~\bibnamefont {Tatarakis}}, \ and\ \bibinfo {author} {\bibfnamefont
  {W.}~\bibnamefont {Rozmus}},\ }\bibfield  {title} {\enquote {\bibinfo {title}
  {{Proton deflectometry of a magnetic reconnection geometry}},}\ }\href
  {\doibase 10.1063/1.3377787} {\bibfield  {journal} {\bibinfo  {journal}
  {Physics of Plasmas}\ }\textbf {\bibinfo {volume} {17}},\ \bibinfo {pages}
  {043104} (\bibinfo {year} {2010})}\BibitemShut {NoStop}%
\bibitem [{\citenamefont {Li}\ \emph {et~al.}(2007{\natexlab{b}})\citenamefont
  {Li}, \citenamefont {S\'{e}guin}, \citenamefont {Frenje}, \citenamefont
  {Rygg}, \citenamefont {Petrasso}, \citenamefont {Town}, \citenamefont
  {Amendt}, \citenamefont {Hatchett}, \citenamefont {Landen}, \citenamefont
  {Mackinnon}, \citenamefont {Patel}, \citenamefont {Tabak}, \citenamefont
  {Knauer}, \citenamefont {Sangster},\ and\ \citenamefont {Smalyuk}}]{Li2007a}%
  \BibitemOpen
  \bibfield  {author} {\bibinfo {author} {\bibfnamefont {C.}~\bibnamefont
  {Li}}, \bibinfo {author} {\bibfnamefont {F.}~\bibnamefont {S\'{e}guin}},
  \bibinfo {author} {\bibfnamefont {J.}~\bibnamefont {Frenje}}, \bibinfo
  {author} {\bibfnamefont {J.}~\bibnamefont {Rygg}}, \bibinfo {author}
  {\bibfnamefont {R.}~\bibnamefont {Petrasso}}, \bibinfo {author}
  {\bibfnamefont {R.}~\bibnamefont {Town}}, \bibinfo {author} {\bibfnamefont
  {P.}~\bibnamefont {Amendt}}, \bibinfo {author} {\bibfnamefont
  {S.}~\bibnamefont {Hatchett}}, \bibinfo {author} {\bibfnamefont
  {O.}~\bibnamefont {Landen}}, \bibinfo {author} {\bibfnamefont
  {a.}~\bibnamefont {Mackinnon}}, \bibinfo {author} {\bibfnamefont
  {P.}~\bibnamefont {Patel}}, \bibinfo {author} {\bibfnamefont
  {M.}~\bibnamefont {Tabak}}, \bibinfo {author} {\bibfnamefont
  {J.}~\bibnamefont {Knauer}}, \bibinfo {author} {\bibfnamefont
  {T.}~\bibnamefont {Sangster}}, \ and\ \bibinfo {author} {\bibfnamefont
  {V.}~\bibnamefont {Smalyuk}},\ }\bibfield  {title} {\enquote {\bibinfo
  {title} {{Observation of the Decay Dynamics and Instabilities of Megagauss
  Field Structures in Laser-Produced Plasmas}},}\ }\href {\doibase
  10.1103/PhysRevLett.99.015001} {\bibfield  {journal} {\bibinfo  {journal}
  {Physical Review Letters}\ }\textbf {\bibinfo {volume} {99}},\ \bibinfo
  {pages} {015001} (\bibinfo {year} {2007}{\natexlab{b}})}\BibitemShut
  {NoStop}%
\bibitem [{\citenamefont {Huntington}\ \emph {et~al.}(2013)\citenamefont
  {Huntington}, \citenamefont {Fiuza}, \citenamefont {Ross}, \citenamefont
  {Zylstra}, \citenamefont {Drake}, \citenamefont {Froula}, \citenamefont
  {Gregori}, \citenamefont {Kugland}, \citenamefont {Kuranz}, \citenamefont
  {Levy}, \citenamefont {Li}, \citenamefont {Meinecke}, \citenamefont {Morita},
  \citenamefont {Petrasso}, \citenamefont {Plechaty}, \citenamefont
  {Remington}, \citenamefont {Ryutov}, \citenamefont {Sakawa}, \citenamefont
  {Spitkovsky}, \citenamefont {Takabe},\ and\ \citenamefont
  {Park}}]{Huntington2013}%
  \BibitemOpen
  \bibfield  {author} {\bibinfo {author} {\bibfnamefont {C.~M.}\ \bibnamefont
  {Huntington}}, \bibinfo {author} {\bibfnamefont {F.}~\bibnamefont {Fiuza}},
  \bibinfo {author} {\bibfnamefont {J.~S.}\ \bibnamefont {Ross}}, \bibinfo
  {author} {\bibfnamefont {A.~B.}\ \bibnamefont {Zylstra}}, \bibinfo {author}
  {\bibfnamefont {R.~P.}\ \bibnamefont {Drake}}, \bibinfo {author}
  {\bibfnamefont {D.~H.}\ \bibnamefont {Froula}}, \bibinfo {author}
  {\bibfnamefont {G.}~\bibnamefont {Gregori}}, \bibinfo {author} {\bibfnamefont
  {N.~L.}\ \bibnamefont {Kugland}}, \bibinfo {author} {\bibfnamefont {C.~C.}\
  \bibnamefont {Kuranz}}, \bibinfo {author} {\bibfnamefont {M.~C.}\
  \bibnamefont {Levy}}, \bibinfo {author} {\bibfnamefont {C.~K.}\ \bibnamefont
  {Li}}, \bibinfo {author} {\bibfnamefont {J.}~\bibnamefont {Meinecke}},
  \bibinfo {author} {\bibfnamefont {T.}~\bibnamefont {Morita}}, \bibinfo
  {author} {\bibfnamefont {R.}~\bibnamefont {Petrasso}}, \bibinfo {author}
  {\bibfnamefont {C.}~\bibnamefont {Plechaty}}, \bibinfo {author}
  {\bibfnamefont {B.~A.}\ \bibnamefont {Remington}}, \bibinfo {author}
  {\bibfnamefont {D.~D.}\ \bibnamefont {Ryutov}}, \bibinfo {author}
  {\bibfnamefont {Y.}~\bibnamefont {Sakawa}}, \bibinfo {author} {\bibfnamefont
  {A.}~\bibnamefont {Spitkovsky}}, \bibinfo {author} {\bibfnamefont
  {H.}~\bibnamefont {Takabe}}, \ and\ \bibinfo {author} {\bibfnamefont {H.~S.}\
  \bibnamefont {Park}},\ }\bibfield  {title} {\enquote {\bibinfo {title}
  {{Observation of magnetic field generation via the Weibel instability in
  interpenetrating plasma flows}},}\ }\href {http://arxiv.org/abs/1310.3337} {\
  ,\ \bibinfo {pages} {1--13} (\bibinfo {year} {2013})},\ \Eprint
  {http://arxiv.org/abs/1310.3337} {arXiv:1310.3337} \BibitemShut {NoStop}%
\bibitem [{\citenamefont {Fox}\ \emph {et~al.}(2013)\citenamefont {Fox},
  \citenamefont {Fiksel}, \citenamefont {Bhattacharjee}, \citenamefont {Chang},
  \citenamefont {Germaschewski}, \citenamefont {Hu},\ and\ \citenamefont
  {Nilson}}]{Fox2013}%
  \BibitemOpen
  \bibfield  {author} {\bibinfo {author} {\bibfnamefont {W.}~\bibnamefont
  {Fox}}, \bibinfo {author} {\bibfnamefont {G.}~\bibnamefont {Fiksel}},
  \bibinfo {author} {\bibfnamefont {A.}~\bibnamefont {Bhattacharjee}}, \bibinfo
  {author} {\bibfnamefont {P.-Y.}\ \bibnamefont {Chang}}, \bibinfo {author}
  {\bibfnamefont {K.}~\bibnamefont {Germaschewski}}, \bibinfo {author}
  {\bibfnamefont {S.~X.}\ \bibnamefont {Hu}}, \ and\ \bibinfo {author}
  {\bibfnamefont {P.~M.}\ \bibnamefont {Nilson}},\ }\bibfield  {title}
  {\enquote {\bibinfo {title} {{Filamentation Instability of Counterstreaming
  Laser-Driven Plasmas}},}\ }\href {\doibase 10.1103/PhysRevLett.111.225002}
  {\bibfield  {journal} {\bibinfo  {journal} {Physical Review Letters}\
  }\textbf {\bibinfo {volume} {111}},\ \bibinfo {pages} {225002} (\bibinfo
  {year} {2013})}\BibitemShut {NoStop}%
\bibitem [{\citenamefont {Gao}\ \emph {et~al.}(2012)\citenamefont {Gao},
  \citenamefont {Nilson}, \citenamefont {Igumenschev}, \citenamefont {Hu},
  \citenamefont {Davies}, \citenamefont {Stoeckl}, \citenamefont {Haines},
  \citenamefont {Froula}, \citenamefont {Betti},\ and\ \citenamefont
  {Meyerhofer}}]{Gao2012}%
  \BibitemOpen
  \bibfield  {author} {\bibinfo {author} {\bibfnamefont {L.}~\bibnamefont
  {Gao}}, \bibinfo {author} {\bibfnamefont {P.~M.}\ \bibnamefont {Nilson}},
  \bibinfo {author} {\bibfnamefont {I.~V.}\ \bibnamefont {Igumenschev}},
  \bibinfo {author} {\bibfnamefont {S.~X.}\ \bibnamefont {Hu}}, \bibinfo
  {author} {\bibfnamefont {J.~R.}\ \bibnamefont {Davies}}, \bibinfo {author}
  {\bibfnamefont {C.}~\bibnamefont {Stoeckl}}, \bibinfo {author} {\bibfnamefont
  {M.~G.}\ \bibnamefont {Haines}}, \bibinfo {author} {\bibfnamefont {D.~H.}\
  \bibnamefont {Froula}}, \bibinfo {author} {\bibfnamefont {R.}~\bibnamefont
  {Betti}}, \ and\ \bibinfo {author} {\bibfnamefont {D.~D.}\ \bibnamefont
  {Meyerhofer}},\ }\bibfield  {title} {\enquote {\bibinfo {title} {{Magnetic
  Field Generation by the Rayleigh-Taylor Instability in Laser-Driven Planar
  Plastic Targets}},}\ }\href {\doibase 10.1103/PhysRevLett.109.115001}
  {\bibfield  {journal} {\bibinfo  {journal} {Physical Review Letters}\
  }\textbf {\bibinfo {volume} {109}},\ \bibinfo {pages} {115001} (\bibinfo
  {year} {2012})}\BibitemShut {NoStop}%
\bibitem [{\citenamefont {Gao}\ \emph {et~al.}(2013)\citenamefont {Gao},
  \citenamefont {Nilson}, \citenamefont {Igumenschev}, \citenamefont {Fiksel},
  \citenamefont {Yan}, \citenamefont {Davies}, \citenamefont {Martinez},
  \citenamefont {Smalyuk}, \citenamefont {Haines}, \citenamefont {Blackman},
  \citenamefont {Froula}, \citenamefont {Betti},\ and\ \citenamefont
  {Meyerhofer}}]{Gao2013}%
  \BibitemOpen
  \bibfield  {author} {\bibinfo {author} {\bibfnamefont {L.}~\bibnamefont
  {Gao}}, \bibinfo {author} {\bibfnamefont {P.~M.}\ \bibnamefont {Nilson}},
  \bibinfo {author} {\bibfnamefont {I.~V.}\ \bibnamefont {Igumenschev}},
  \bibinfo {author} {\bibfnamefont {G.}~\bibnamefont {Fiksel}}, \bibinfo
  {author} {\bibfnamefont {R.}~\bibnamefont {Yan}}, \bibinfo {author}
  {\bibfnamefont {J.~R.}\ \bibnamefont {Davies}}, \bibinfo {author}
  {\bibfnamefont {D.}~\bibnamefont {Martinez}}, \bibinfo {author}
  {\bibfnamefont {V.}~\bibnamefont {Smalyuk}}, \bibinfo {author} {\bibfnamefont
  {M.~G.}\ \bibnamefont {Haines}}, \bibinfo {author} {\bibfnamefont {E.~G.}\
  \bibnamefont {Blackman}}, \bibinfo {author} {\bibfnamefont {D.~H.}\
  \bibnamefont {Froula}}, \bibinfo {author} {\bibfnamefont {R.}~\bibnamefont
  {Betti}}, \ and\ \bibinfo {author} {\bibfnamefont {D.~D.}\ \bibnamefont
  {Meyerhofer}},\ }\bibfield  {title} {\enquote {\bibinfo {title} {{Observation
  of Self-Similarity in the Magnetic Fields Generated by the Ablative Nonlinear
  Rayleigh-Taylor Instability}},}\ }\href {\doibase
  10.1103/PhysRevLett.110.185003} {\bibfield  {journal} {\bibinfo  {journal}
  {Physical Review Letters}\ }\textbf {\bibinfo {volume} {110}},\ \bibinfo
  {pages} {185003} (\bibinfo {year} {2013})}\BibitemShut {NoStop}%
\bibitem [{\citenamefont {Roth}\ \emph {et~al.}(2002)\citenamefont {Roth},
  \citenamefont {Blazevic}, \citenamefont {Geissel}, \citenamefont {Schlegel},
  \citenamefont {Cowan}, \citenamefont {Allen}, \citenamefont {Gauthier},
  \citenamefont {Audebert}, \citenamefont {Fuchs}, \citenamefont {Meyer-ter
  Vehn}, \citenamefont {Hegelich}, \citenamefont {Karsch},\ and\ \citenamefont
  {Pukhov}}]{Roth2002}%
  \BibitemOpen
  \bibfield  {author} {\bibinfo {author} {\bibfnamefont {M.}~\bibnamefont
  {Roth}}, \bibinfo {author} {\bibfnamefont {A.}~\bibnamefont {Blazevic}},
  \bibinfo {author} {\bibfnamefont {M.}~\bibnamefont {Geissel}}, \bibinfo
  {author} {\bibfnamefont {T.}~\bibnamefont {Schlegel}}, \bibinfo {author}
  {\bibfnamefont {T.}~\bibnamefont {Cowan}}, \bibinfo {author} {\bibfnamefont
  {M.}~\bibnamefont {Allen}}, \bibinfo {author} {\bibfnamefont {J.-C.}\
  \bibnamefont {Gauthier}}, \bibinfo {author} {\bibfnamefont {P.}~\bibnamefont
  {Audebert}}, \bibinfo {author} {\bibfnamefont {J.}~\bibnamefont {Fuchs}},
  \bibinfo {author} {\bibfnamefont {J.}~\bibnamefont {Meyer-ter Vehn}},
  \bibinfo {author} {\bibfnamefont {M.}~\bibnamefont {Hegelich}}, \bibinfo
  {author} {\bibfnamefont {S.}~\bibnamefont {Karsch}}, \ and\ \bibinfo {author}
  {\bibfnamefont {A.}~\bibnamefont {Pukhov}},\ }\bibfield  {title} {\enquote
  {\bibinfo {title} {{Energetic ions generated by laser pulses: A detailed
  study on target properties}},}\ }\href {\doibase
  10.1103/PhysRevSTAB.5.061301} {\bibfield  {journal} {\bibinfo  {journal}
  {Physical Review Special Topics - Accelerators and Beams}\ }\textbf {\bibinfo
  {volume} {5}},\ \bibinfo {pages} {061301} (\bibinfo {year}
  {2002})}\BibitemShut {NoStop}%
\bibitem [{\citenamefont {Borghesi}\ \emph {et~al.}(2003)\citenamefont
  {Borghesi}, \citenamefont {Romagnani}, \citenamefont {Schiavi}, \citenamefont
  {Campbell}, \citenamefont {Haines}, \citenamefont {Willi}, \citenamefont
  {Mackinnon}, \citenamefont {Galimberti}, \citenamefont {Gizzi}, \citenamefont
  {Clarke},\ and\ \citenamefont {Hawkes}}]{Borghesi2003}%
  \BibitemOpen
  \bibfield  {author} {\bibinfo {author} {\bibfnamefont {M.}~\bibnamefont
  {Borghesi}}, \bibinfo {author} {\bibfnamefont {L.}~\bibnamefont {Romagnani}},
  \bibinfo {author} {\bibfnamefont {A.}~\bibnamefont {Schiavi}}, \bibinfo
  {author} {\bibfnamefont {D.~H.}\ \bibnamefont {Campbell}}, \bibinfo {author}
  {\bibfnamefont {M.~G.}\ \bibnamefont {Haines}}, \bibinfo {author}
  {\bibfnamefont {O.}~\bibnamefont {Willi}}, \bibinfo {author} {\bibfnamefont
  {a.~J.}\ \bibnamefont {Mackinnon}}, \bibinfo {author} {\bibfnamefont
  {M.}~\bibnamefont {Galimberti}}, \bibinfo {author} {\bibfnamefont
  {L.}~\bibnamefont {Gizzi}}, \bibinfo {author} {\bibfnamefont {R.~J.}\
  \bibnamefont {Clarke}}, \ and\ \bibinfo {author} {\bibfnamefont
  {S.}~\bibnamefont {Hawkes}},\ }\bibfield  {title} {\enquote {\bibinfo {title}
  {{Measurement of highly transient electrical charging following
  high-intensity laser–solid interaction}},}\ }\href {\doibase
  10.1063/1.1560554} {\bibfield  {journal} {\bibinfo  {journal} {Applied
  Physics Letters}\ }\textbf {\bibinfo {volume} {82}},\ \bibinfo {pages} {1529}
  (\bibinfo {year} {2003})}\BibitemShut {NoStop}%
\bibitem [{\citenamefont {Mackinnon}\ \emph {et~al.}(2004)\citenamefont
  {Mackinnon}, \citenamefont {Patel}, \citenamefont {Town}, \citenamefont
  {Edwards}, \citenamefont {Phillips}, \citenamefont {Lerner}, \citenamefont
  {Price}, \citenamefont {Hicks}, \citenamefont {Key}, \citenamefont
  {Hatchett}, \citenamefont {Wilks}, \citenamefont {Borghesi}, \citenamefont
  {Romagnani}, \citenamefont {Kar}, \citenamefont {Toncian}, \citenamefont
  {Pretzler}, \citenamefont {Willi}, \citenamefont {Koenig}, \citenamefont
  {Martinolli}, \citenamefont {Lepape}, \citenamefont {Benuzzi-Mounaix},
  \citenamefont {Audebert}, \citenamefont {Gauthier}, \citenamefont {King},
  \citenamefont {Snavely}, \citenamefont {Freeman},\ and\ \citenamefont
  {Boehlly}}]{Mackinnon2004}%
  \BibitemOpen
  \bibfield  {author} {\bibinfo {author} {\bibfnamefont {A.~J.}\ \bibnamefont
  {Mackinnon}}, \bibinfo {author} {\bibfnamefont {P.~K.}\ \bibnamefont
  {Patel}}, \bibinfo {author} {\bibfnamefont {R.~P.}\ \bibnamefont {Town}},
  \bibinfo {author} {\bibfnamefont {M.~J.}\ \bibnamefont {Edwards}}, \bibinfo
  {author} {\bibfnamefont {T.}~\bibnamefont {Phillips}}, \bibinfo {author}
  {\bibfnamefont {S.~C.}\ \bibnamefont {Lerner}}, \bibinfo {author}
  {\bibfnamefont {D.~W.}\ \bibnamefont {Price}}, \bibinfo {author}
  {\bibfnamefont {D.}~\bibnamefont {Hicks}}, \bibinfo {author} {\bibfnamefont
  {M.~H.}\ \bibnamefont {Key}}, \bibinfo {author} {\bibfnamefont
  {S.}~\bibnamefont {Hatchett}}, \bibinfo {author} {\bibfnamefont {S.~C.}\
  \bibnamefont {Wilks}}, \bibinfo {author} {\bibfnamefont {M.}~\bibnamefont
  {Borghesi}}, \bibinfo {author} {\bibfnamefont {L.}~\bibnamefont {Romagnani}},
  \bibinfo {author} {\bibfnamefont {S.}~\bibnamefont {Kar}}, \bibinfo {author}
  {\bibfnamefont {T.}~\bibnamefont {Toncian}}, \bibinfo {author} {\bibfnamefont
  {G.}~\bibnamefont {Pretzler}}, \bibinfo {author} {\bibfnamefont
  {O.}~\bibnamefont {Willi}}, \bibinfo {author} {\bibfnamefont
  {M.}~\bibnamefont {Koenig}}, \bibinfo {author} {\bibfnamefont
  {E.}~\bibnamefont {Martinolli}}, \bibinfo {author} {\bibfnamefont
  {S.}~\bibnamefont {Lepape}}, \bibinfo {author} {\bibfnamefont
  {A.}~\bibnamefont {Benuzzi-Mounaix}}, \bibinfo {author} {\bibfnamefont
  {P.}~\bibnamefont {Audebert}}, \bibinfo {author} {\bibfnamefont {J.~C.}\
  \bibnamefont {Gauthier}}, \bibinfo {author} {\bibfnamefont {J.}~\bibnamefont
  {King}}, \bibinfo {author} {\bibfnamefont {R.}~\bibnamefont {Snavely}},
  \bibinfo {author} {\bibfnamefont {R.~R.}\ \bibnamefont {Freeman}}, \ and\
  \bibinfo {author} {\bibfnamefont {T.}~\bibnamefont {Boehlly}},\ }\bibfield
  {title} {\enquote {\bibinfo {title} {{Proton radiography as an
  electromagnetic field and density perturbation diagnostic (invited)}},}\
  }\href {\doibase 10.1063/1.1788893} {\bibfield  {journal} {\bibinfo
  {journal} {Review of Scientific Instruments}\ }\textbf {\bibinfo {volume}
  {75}},\ \bibinfo {pages} {3531} (\bibinfo {year} {2004})}\BibitemShut
  {NoStop}%
\bibitem [{\citenamefont {Borghesi}\ \emph {et~al.}(2007)\citenamefont
  {Borghesi}, \citenamefont {Kar}, \citenamefont {Romagnani}, \citenamefont
  {Toncian}, \citenamefont {Antici}, \citenamefont {Audebert}, \citenamefont
  {Brambrink}, \citenamefont {Ceccherini}, \citenamefont {Cecchetti},
  \citenamefont {Fuchs}, \citenamefont {Galimberti}, \citenamefont {Gizzi},
  \citenamefont {Grismayer}, \citenamefont {Lyseikina}, \citenamefont {Jung},
  \citenamefont {Macchi}, \citenamefont {Mora}, \citenamefont {Osterholtz},
  \citenamefont {Schiavi},\ and\ \citenamefont {Willi}}]{Borghesi2007}%
  \BibitemOpen
  \bibfield  {author} {\bibinfo {author} {\bibfnamefont {M.}~\bibnamefont
  {Borghesi}}, \bibinfo {author} {\bibfnamefont {S.}~\bibnamefont {Kar}},
  \bibinfo {author} {\bibfnamefont {L.}~\bibnamefont {Romagnani}}, \bibinfo
  {author} {\bibfnamefont {T.}~\bibnamefont {Toncian}}, \bibinfo {author}
  {\bibfnamefont {P.}~\bibnamefont {Antici}}, \bibinfo {author} {\bibfnamefont
  {P.}~\bibnamefont {Audebert}}, \bibinfo {author} {\bibfnamefont
  {E.}~\bibnamefont {Brambrink}}, \bibinfo {author} {\bibfnamefont
  {F.}~\bibnamefont {Ceccherini}}, \bibinfo {author} {\bibfnamefont
  {C.}~\bibnamefont {Cecchetti}}, \bibinfo {author} {\bibfnamefont
  {J.}~\bibnamefont {Fuchs}}, \bibinfo {author} {\bibfnamefont
  {M.}~\bibnamefont {Galimberti}}, \bibinfo {author} {\bibfnamefont
  {L.}~\bibnamefont {Gizzi}}, \bibinfo {author} {\bibfnamefont
  {T.}~\bibnamefont {Grismayer}}, \bibinfo {author} {\bibfnamefont
  {T.}~\bibnamefont {Lyseikina}}, \bibinfo {author} {\bibfnamefont
  {R.}~\bibnamefont {Jung}}, \bibinfo {author} {\bibfnamefont {a.}~\bibnamefont
  {Macchi}}, \bibinfo {author} {\bibfnamefont {P.}~\bibnamefont {Mora}},
  \bibinfo {author} {\bibfnamefont {J.}~\bibnamefont {Osterholtz}}, \bibinfo
  {author} {\bibfnamefont {a.}~\bibnamefont {Schiavi}}, \ and\ \bibinfo
  {author} {\bibfnamefont {O.}~\bibnamefont {Willi}},\ }\bibfield  {title}
  {\enquote {\bibinfo {title} {{Impulsive electric fields driven by
  high-intensity laser matter interactions}},}\ }\href {\doibase
  10.1017/S0263034607070218} {\bibfield  {journal} {\bibinfo  {journal} {Laser
  and Particle Beams}\ }\textbf {\bibinfo {volume} {25}},\ \bibinfo {pages}
  {161--167} (\bibinfo {year} {2007})}\BibitemShut {NoStop}%
\bibitem [{\citenamefont {Romagnani}\ \emph {et~al.}(2008)\citenamefont
  {Romagnani}, \citenamefont {Borghesi}, \citenamefont {Cecchetti},
  \citenamefont {Kar}, \citenamefont {Antici}, \citenamefont {Audebert},
  \citenamefont {Bandhoupadjay}, \citenamefont {Ceccherini}, \citenamefont
  {Cowan}, \citenamefont {Fuchs}, \citenamefont {Galimberti}, \citenamefont
  {Gizzi}, \citenamefont {Grismayer}, \citenamefont {Heathcote}, \citenamefont
  {Jung}, \citenamefont {Liseykina}, \citenamefont {Macchi}, \citenamefont
  {Mora}, \citenamefont {Neely}, \citenamefont {Notley}, \citenamefont
  {Osterholtz}, \citenamefont {Pipahl}, \citenamefont {Pretzler}, \citenamefont
  {Schiavi}, \citenamefont {Schurtz}, \citenamefont {Toncian}, \citenamefont
  {Wilson},\ and\ \citenamefont {Willi}}]{Romagnani2008}%
  \BibitemOpen
  \bibfield  {author} {\bibinfo {author} {\bibfnamefont {L.}~\bibnamefont
  {Romagnani}}, \bibinfo {author} {\bibfnamefont {M.}~\bibnamefont {Borghesi}},
  \bibinfo {author} {\bibfnamefont {C.}~\bibnamefont {Cecchetti}}, \bibinfo
  {author} {\bibfnamefont {S.}~\bibnamefont {Kar}}, \bibinfo {author}
  {\bibfnamefont {P.}~\bibnamefont {Antici}}, \bibinfo {author} {\bibfnamefont
  {P.}~\bibnamefont {Audebert}}, \bibinfo {author} {\bibfnamefont
  {S.}~\bibnamefont {Bandhoupadjay}}, \bibinfo {author} {\bibfnamefont
  {F.}~\bibnamefont {Ceccherini}}, \bibinfo {author} {\bibfnamefont
  {T.}~\bibnamefont {Cowan}}, \bibinfo {author} {\bibfnamefont
  {J.}~\bibnamefont {Fuchs}}, \bibinfo {author} {\bibfnamefont
  {M.}~\bibnamefont {Galimberti}}, \bibinfo {author} {\bibfnamefont
  {L.}~\bibnamefont {Gizzi}}, \bibinfo {author} {\bibfnamefont
  {T.}~\bibnamefont {Grismayer}}, \bibinfo {author} {\bibfnamefont
  {R.}~\bibnamefont {Heathcote}}, \bibinfo {author} {\bibfnamefont
  {R.}~\bibnamefont {Jung}}, \bibinfo {author} {\bibfnamefont {T.}~\bibnamefont
  {Liseykina}}, \bibinfo {author} {\bibfnamefont {a.}~\bibnamefont {Macchi}},
  \bibinfo {author} {\bibfnamefont {P.}~\bibnamefont {Mora}}, \bibinfo {author}
  {\bibfnamefont {D.}~\bibnamefont {Neely}}, \bibinfo {author} {\bibfnamefont
  {M.}~\bibnamefont {Notley}}, \bibinfo {author} {\bibfnamefont
  {J.}~\bibnamefont {Osterholtz}}, \bibinfo {author} {\bibfnamefont
  {C.}~\bibnamefont {Pipahl}}, \bibinfo {author} {\bibfnamefont
  {G.}~\bibnamefont {Pretzler}}, \bibinfo {author} {\bibfnamefont
  {a.}~\bibnamefont {Schiavi}}, \bibinfo {author} {\bibfnamefont
  {G.}~\bibnamefont {Schurtz}}, \bibinfo {author} {\bibfnamefont
  {T.}~\bibnamefont {Toncian}}, \bibinfo {author} {\bibfnamefont
  {P.}~\bibnamefont {Wilson}}, \ and\ \bibinfo {author} {\bibfnamefont
  {O.}~\bibnamefont {Willi}},\ }\bibfield  {title} {\enquote {\bibinfo {title}
  {{Proton probing measurement of electric and magnetic fields generated by ns
  and ps laser-matter interactions}},}\ }\href {\doibase
  10.1017/S0263034608000281} {\bibfield  {journal} {\bibinfo  {journal} {Laser
  and Particle Beams}\ }\textbf {\bibinfo {volume} {26}},\ \bibinfo {pages}
  {241--248} (\bibinfo {year} {2008})}\BibitemShut {NoStop}%
\bibitem [{\citenamefont {Cecchetti}\ \emph {et~al.}(2009)\citenamefont
  {Cecchetti}, \citenamefont {Borghesi}, \citenamefont {Fuchs}, \citenamefont
  {Schurtz}, \citenamefont {Kar}, \citenamefont {Macchi}, \citenamefont
  {Romagnani}, \citenamefont {Wilson}, \citenamefont {Antici}, \citenamefont
  {Jung}, \citenamefont {Osterholtz}, \citenamefont {Pipahl}, \citenamefont
  {Willi}, \citenamefont {Schiavi}, \citenamefont {Notley},\ and\ \citenamefont
  {Neely}}]{Cecchetti2009}%
  \BibitemOpen
  \bibfield  {author} {\bibinfo {author} {\bibfnamefont {C.~a.}\ \bibnamefont
  {Cecchetti}}, \bibinfo {author} {\bibfnamefont {M.}~\bibnamefont {Borghesi}},
  \bibinfo {author} {\bibfnamefont {J.}~\bibnamefont {Fuchs}}, \bibinfo
  {author} {\bibfnamefont {G.}~\bibnamefont {Schurtz}}, \bibinfo {author}
  {\bibfnamefont {S.}~\bibnamefont {Kar}}, \bibinfo {author} {\bibfnamefont
  {a.}~\bibnamefont {Macchi}}, \bibinfo {author} {\bibfnamefont
  {L.}~\bibnamefont {Romagnani}}, \bibinfo {author} {\bibfnamefont {P.~a.}\
  \bibnamefont {Wilson}}, \bibinfo {author} {\bibfnamefont {P.}~\bibnamefont
  {Antici}}, \bibinfo {author} {\bibfnamefont {R.}~\bibnamefont {Jung}},
  \bibinfo {author} {\bibfnamefont {J.}~\bibnamefont {Osterholtz}}, \bibinfo
  {author} {\bibfnamefont {C.~a.}\ \bibnamefont {Pipahl}}, \bibinfo {author}
  {\bibfnamefont {O.}~\bibnamefont {Willi}}, \bibinfo {author} {\bibfnamefont
  {a.}~\bibnamefont {Schiavi}}, \bibinfo {author} {\bibfnamefont
  {M.}~\bibnamefont {Notley}}, \ and\ \bibinfo {author} {\bibfnamefont
  {D.}~\bibnamefont {Neely}},\ }\bibfield  {title} {\enquote {\bibinfo {title}
  {{Magnetic field measurements in laser-produced plasmas via proton
  deflectometry}},}\ }\href {\doibase 10.1063/1.3097899} {\bibfield  {journal}
  {\bibinfo  {journal} {Physics of Plasmas}\ }\textbf {\bibinfo {volume}
  {16}},\ \bibinfo {pages} {043102} (\bibinfo {year} {2009})}\BibitemShut
  {NoStop}%
\bibitem [{\citenamefont {Sokollik}\ \emph {et~al.}(2009)\citenamefont
  {Sokollik}, \citenamefont {Schnürer}, \citenamefont {Ter-Avetisyan},
  \citenamefont {Steinke}, \citenamefont {Nickles}, \citenamefont {Sandner},
  \citenamefont {Amin}, \citenamefont {Toncian}, \citenamefont {Willi},
  \citenamefont {Andreev}, \citenamefont {Bolton}, \citenamefont {Daido},\ and\
  \citenamefont {Bulanov}}]{Sokollik2009}%
  \BibitemOpen
  \bibfield  {author} {\bibinfo {author} {\bibfnamefont {T.}~\bibnamefont
  {Sokollik}}, \bibinfo {author} {\bibfnamefont {M.}~\bibnamefont
  {Schnürer}}, \bibinfo {author} {\bibfnamefont {S.}~\bibnamefont
  {Ter-Avetisyan}}, \bibinfo {author} {\bibfnamefont {S.}~\bibnamefont
  {Steinke}}, \bibinfo {author} {\bibfnamefont {P.~V.}\ \bibnamefont
  {Nickles}}, \bibinfo {author} {\bibfnamefont {W.}~\bibnamefont {Sandner}},
  \bibinfo {author} {\bibfnamefont {M.}~\bibnamefont {Amin}}, \bibinfo {author}
  {\bibfnamefont {T.}~\bibnamefont {Toncian}}, \bibinfo {author} {\bibfnamefont
  {O.}~\bibnamefont {Willi}}, \bibinfo {author} {\bibfnamefont {a.~a.}\
  \bibnamefont {Andreev}}, \bibinfo {author} {\bibfnamefont {P.~R.}\
  \bibnamefont {Bolton}}, \bibinfo {author} {\bibfnamefont {H.}~\bibnamefont
  {Daido}}, \ and\ \bibinfo {author} {\bibfnamefont {S.~V.}\ \bibnamefont
  {Bulanov}},\ }\bibfield  {title} {\enquote {\bibinfo {title} {{Proton Imaging
  Of Laser Irradiated Foils And Mass-Limited Targets}},}\ }\href {\doibase
  10.1063/1.3204546} {\bibfield  {journal} {\bibinfo  {journal} {AIP Conference
  Proceedings}\ }\textbf {\bibinfo {volume} {364}},\ \bibinfo {pages}
  {364--373} (\bibinfo {year} {2009})}\BibitemShut {NoStop}%
\bibitem [{\citenamefont {Volpe}\ \emph {et~al.}(2011)\citenamefont {Volpe},
  \citenamefont {Jafer}, \citenamefont {Vauzour}, \citenamefont {Nicolai},
  \citenamefont {Santos}, \citenamefont {Dorchies}, \citenamefont {Fourment},
  \citenamefont {Hulin}, \citenamefont {Regan}, \citenamefont {Perez},
  \citenamefont {Baton}, \citenamefont {Lancaster}, \citenamefont {Galimberti},
  \citenamefont {Heathcote}, \citenamefont {Tolley}, \citenamefont {Spindloe},
  \citenamefont {Nazarov}, \citenamefont {Koester}, \citenamefont {Labate},
  \citenamefont {Gizzi}, \citenamefont {Benedetti}, \citenamefont {Sgattoni},
  \citenamefont {Richetta}, \citenamefont {Pasley}, \citenamefont {Beg},
  \citenamefont {Chawla}, \citenamefont {Higginson}, \citenamefont {MacPhee},\
  and\ \citenamefont {Batani}}]{Volpe2011}%
  \BibitemOpen
  \bibfield  {author} {\bibinfo {author} {\bibfnamefont {L.}~\bibnamefont
  {Volpe}}, \bibinfo {author} {\bibfnamefont {R.}~\bibnamefont {Jafer}},
  \bibinfo {author} {\bibfnamefont {B.}~\bibnamefont {Vauzour}}, \bibinfo
  {author} {\bibfnamefont {P.}~\bibnamefont {Nicolai}}, \bibinfo {author}
  {\bibfnamefont {J.~J.}\ \bibnamefont {Santos}}, \bibinfo {author}
  {\bibfnamefont {F.}~\bibnamefont {Dorchies}}, \bibinfo {author}
  {\bibfnamefont {C.}~\bibnamefont {Fourment}}, \bibinfo {author}
  {\bibfnamefont {S.}~\bibnamefont {Hulin}}, \bibinfo {author} {\bibfnamefont
  {C.}~\bibnamefont {Regan}}, \bibinfo {author} {\bibfnamefont
  {F.}~\bibnamefont {Perez}}, \bibinfo {author} {\bibfnamefont
  {S.}~\bibnamefont {Baton}}, \bibinfo {author} {\bibfnamefont
  {K.}~\bibnamefont {Lancaster}}, \bibinfo {author} {\bibfnamefont
  {M.}~\bibnamefont {Galimberti}}, \bibinfo {author} {\bibfnamefont
  {R.}~\bibnamefont {Heathcote}}, \bibinfo {author} {\bibfnamefont
  {M.}~\bibnamefont {Tolley}}, \bibinfo {author} {\bibfnamefont
  {C.}~\bibnamefont {Spindloe}}, \bibinfo {author} {\bibfnamefont
  {W.}~\bibnamefont {Nazarov}}, \bibinfo {author} {\bibfnamefont
  {P.}~\bibnamefont {Koester}}, \bibinfo {author} {\bibfnamefont
  {L.}~\bibnamefont {Labate}}, \bibinfo {author} {\bibfnamefont {L.~a.}\
  \bibnamefont {Gizzi}}, \bibinfo {author} {\bibfnamefont {C.}~\bibnamefont
  {Benedetti}}, \bibinfo {author} {\bibfnamefont {A.}~\bibnamefont {Sgattoni}},
  \bibinfo {author} {\bibfnamefont {M.}~\bibnamefont {Richetta}}, \bibinfo
  {author} {\bibfnamefont {J.}~\bibnamefont {Pasley}}, \bibinfo {author}
  {\bibfnamefont {F.~N.}\ \bibnamefont {Beg}}, \bibinfo {author} {\bibfnamefont
  {S.}~\bibnamefont {Chawla}}, \bibinfo {author} {\bibfnamefont {D.~P.}\
  \bibnamefont {Higginson}}, \bibinfo {author} {\bibfnamefont {A.~G.}\
  \bibnamefont {MacPhee}}, \ and\ \bibinfo {author} {\bibfnamefont
  {D.}~\bibnamefont {Batani}},\ }\bibfield  {title} {\enquote {\bibinfo {title}
  {{Proton radiography of cylindrical laser-driven implosions}},}\ }\href
  {\doibase 10.1088/0741-3335/53/3/032003} {\bibfield  {journal} {\bibinfo
  {journal} {Plasma Physics and Controlled Fusion}\ }\textbf {\bibinfo {volume}
  {53}},\ \bibinfo {pages} {032003} (\bibinfo {year} {2011})}\BibitemShut
  {NoStop}%
\bibitem [{\citenamefont {Quinn}\ \emph {et~al.}(2012)\citenamefont {Quinn},
  \citenamefont {Romagnani}, \citenamefont {Ramakrishna}, \citenamefont
  {Sarri}, \citenamefont {Dieckmann}, \citenamefont {Wilson}, \citenamefont
  {Fuchs}, \citenamefont {Lancia}, \citenamefont {Pipahl}, \citenamefont
  {Toncian}, \citenamefont {Willi}, \citenamefont {Clarke}, \citenamefont
  {Notley}, \citenamefont {Macchi},\ and\ \citenamefont
  {Borghesi}}]{Quinn2012}%
  \BibitemOpen
  \bibfield  {author} {\bibinfo {author} {\bibfnamefont {K.}~\bibnamefont
  {Quinn}}, \bibinfo {author} {\bibfnamefont {L.}~\bibnamefont {Romagnani}},
  \bibinfo {author} {\bibfnamefont {B.}~\bibnamefont {Ramakrishna}}, \bibinfo
  {author} {\bibfnamefont {G.}~\bibnamefont {Sarri}}, \bibinfo {author}
  {\bibfnamefont {M.~E.}\ \bibnamefont {Dieckmann}}, \bibinfo {author}
  {\bibfnamefont {P.~a.}\ \bibnamefont {Wilson}}, \bibinfo {author}
  {\bibfnamefont {J.}~\bibnamefont {Fuchs}}, \bibinfo {author} {\bibfnamefont
  {L.}~\bibnamefont {Lancia}}, \bibinfo {author} {\bibfnamefont
  {a.}~\bibnamefont {Pipahl}}, \bibinfo {author} {\bibfnamefont
  {T.}~\bibnamefont {Toncian}}, \bibinfo {author} {\bibfnamefont
  {O.}~\bibnamefont {Willi}}, \bibinfo {author} {\bibfnamefont {R.~J.}\
  \bibnamefont {Clarke}}, \bibinfo {author} {\bibfnamefont {M.}~\bibnamefont
  {Notley}}, \bibinfo {author} {\bibfnamefont {a.}~\bibnamefont {Macchi}}, \
  and\ \bibinfo {author} {\bibfnamefont {M.}~\bibnamefont {Borghesi}},\
  }\bibfield  {title} {\enquote {\bibinfo {title} {{Weibel-Induced
  Filamentation during an Ultrafast Laser-Driven Plasma Expansion}},}\ }\href
  {\doibase 10.1103/PhysRevLett.108.135001} {\bibfield  {journal} {\bibinfo
  {journal} {Physical Review Letters}\ }\textbf {\bibinfo {volume} {108}},\
  \bibinfo {pages} {135001} (\bibinfo {year} {2012})}\BibitemShut {NoStop}%
\bibitem [{\citenamefont {Wilks}\ \emph {et~al.}(2001)\citenamefont {Wilks},
  \citenamefont {Langdon}, \citenamefont {Cowan}, \citenamefont {Roth},
  \citenamefont {Singh}, \citenamefont {Hatchett}, \citenamefont {Key},
  \citenamefont {Pennington}, \citenamefont {MacKinnon},\ and\ \citenamefont
  {Snavely}}]{Wilks2001}%
  \BibitemOpen
  \bibfield  {author} {\bibinfo {author} {\bibfnamefont {S.~C.}\ \bibnamefont
  {Wilks}}, \bibinfo {author} {\bibfnamefont {A.~B.}\ \bibnamefont {Langdon}},
  \bibinfo {author} {\bibfnamefont {T.~E.}\ \bibnamefont {Cowan}}, \bibinfo
  {author} {\bibfnamefont {M.}~\bibnamefont {Roth}}, \bibinfo {author}
  {\bibfnamefont {M.}~\bibnamefont {Singh}}, \bibinfo {author} {\bibfnamefont
  {S.}~\bibnamefont {Hatchett}}, \bibinfo {author} {\bibfnamefont {M.~H.}\
  \bibnamefont {Key}}, \bibinfo {author} {\bibfnamefont {D.}~\bibnamefont
  {Pennington}}, \bibinfo {author} {\bibfnamefont {A.}~\bibnamefont
  {MacKinnon}}, \ and\ \bibinfo {author} {\bibfnamefont {R.~a.}\ \bibnamefont
  {Snavely}},\ }\bibfield  {title} {\enquote {\bibinfo {title} {{Energetic
  proton generation in ultra-intense laser–solid interactions}},}\ }\href
  {\doibase 10.1063/1.1333697} {\bibfield  {journal} {\bibinfo  {journal}
  {Physics of Plasmas}\ }\textbf {\bibinfo {volume} {8}},\ \bibinfo {pages}
  {542} (\bibinfo {year} {2001})}\BibitemShut {NoStop}%
\bibitem [{\citenamefont {Berger}\ \emph {et~al.}(2014)\citenamefont {Berger},
  \citenamefont {Coursey}, \citenamefont {Zucker},\ and\ \citenamefont
  {Chang}}]{PSTAR}%
  \BibitemOpen
  \bibfield  {author} {\bibinfo {author} {\bibfnamefont {M.}~\bibnamefont
  {Berger}}, \bibinfo {author} {\bibfnamefont {J.}~\bibnamefont {Coursey}},
  \bibinfo {author} {\bibfnamefont {M.}~\bibnamefont {Zucker}}, \ and\ \bibinfo
  {author} {\bibfnamefont {J.}~\bibnamefont {Chang}},\ }\href
  {http://physics.nist.gov/Star} {\enquote {\bibinfo {title} {{ESTAR, PSTAR,
  and ASTAR: Computer Programs for Calculating Stopping-Power and Range Tables
  for Electrons, Protons, and Helium Ions (version 1.2.3)}},}\ } (\bibinfo
  {year} {2014})\BibitemShut {NoStop}%
\bibitem [{\citenamefont {Maksimchuk}\ \emph {et~al.}(2000)\citenamefont
  {Maksimchuk}, \citenamefont {Gu}, \citenamefont {Flippo},\ and\ \citenamefont
  {Umstadter}}]{Maksimchuk2000}%
  \BibitemOpen
  \bibfield  {author} {\bibinfo {author} {\bibfnamefont {a.}~\bibnamefont
  {Maksimchuk}}, \bibinfo {author} {\bibfnamefont {S.}~\bibnamefont {Gu}},
  \bibinfo {author} {\bibfnamefont {K.}~\bibnamefont {Flippo}}, \ and\ \bibinfo
  {author} {\bibfnamefont {D.}~\bibnamefont {Umstadter}},\ }\bibfield  {title}
  {\enquote {\bibinfo {title} {{Forward Ion Acceleration in Thin Films Driven
  by a High-Intensity Laser}},}\ }\href {\doibase 10.1103/PhysRevLett.84.4108}
  {\bibfield  {journal} {\bibinfo  {journal} {Physical Review Letters}\
  }\textbf {\bibinfo {volume} {84}},\ \bibinfo {pages} {4108--4111} (\bibinfo
  {year} {2000})}\BibitemShut {NoStop}%
\bibitem [{\citenamefont {Borghesi}\ \emph
  {et~al.}(2002{\natexlab{a}})\citenamefont {Borghesi}, \citenamefont
  {Campbell}, \citenamefont {Schiavi}, \citenamefont {Haines}, \citenamefont
  {Willi}, \citenamefont {MacKinnon}, \citenamefont {Patel}, \citenamefont
  {Gizzi}, \citenamefont {Galimberti}, \citenamefont {Clarke}, \citenamefont
  {Pegoraro}, \citenamefont {Ruhl},\ and\ \citenamefont
  {Bulanov}}]{Borghesi2002}%
  \BibitemOpen
  \bibfield  {author} {\bibinfo {author} {\bibfnamefont {M.}~\bibnamefont
  {Borghesi}}, \bibinfo {author} {\bibfnamefont {D.~H.}\ \bibnamefont
  {Campbell}}, \bibinfo {author} {\bibfnamefont {a.}~\bibnamefont {Schiavi}},
  \bibinfo {author} {\bibfnamefont {M.~G.}\ \bibnamefont {Haines}}, \bibinfo
  {author} {\bibfnamefont {O.}~\bibnamefont {Willi}}, \bibinfo {author}
  {\bibfnamefont {a.~J.}\ \bibnamefont {MacKinnon}}, \bibinfo {author}
  {\bibfnamefont {P.}~\bibnamefont {Patel}}, \bibinfo {author} {\bibfnamefont
  {L.~a.}\ \bibnamefont {Gizzi}}, \bibinfo {author} {\bibfnamefont
  {M.}~\bibnamefont {Galimberti}}, \bibinfo {author} {\bibfnamefont {R.~J.}\
  \bibnamefont {Clarke}}, \bibinfo {author} {\bibfnamefont {F.}~\bibnamefont
  {Pegoraro}}, \bibinfo {author} {\bibfnamefont {H.}~\bibnamefont {Ruhl}}, \
  and\ \bibinfo {author} {\bibfnamefont {S.}~\bibnamefont {Bulanov}},\
  }\bibfield  {title} {\enquote {\bibinfo {title} {{Electric field detection in
  laser-plasma interaction experiments via the proton imaging technique}},}\
  }\href {\doibase 10.1063/1.1459457} {\bibfield  {journal} {\bibinfo
  {journal} {Physics of Plasmas}\ }\textbf {\bibinfo {volume} {9}},\ \bibinfo
  {pages} {2214} (\bibinfo {year} {2002}{\natexlab{a}})}\BibitemShut {NoStop}%
\bibitem [{\citenamefont {Borghesi}\ \emph
  {et~al.}(2002{\natexlab{b}})\citenamefont {Borghesi}, \citenamefont
  {Campbell}, \citenamefont {Schiavi}, \citenamefont {Willi}, \citenamefont
  {a.J. Mackinnon}, \citenamefont {Hicks}, \citenamefont {Patel}, \citenamefont
  {Gizzi}, \citenamefont {Galimberti},\ and\ \citenamefont
  {Clarke}}]{Borghesi2002a}%
  \BibitemOpen
  \bibfield  {author} {\bibinfo {author} {\bibfnamefont {M.}~\bibnamefont
  {Borghesi}}, \bibinfo {author} {\bibfnamefont {D.}~\bibnamefont {Campbell}},
  \bibinfo {author} {\bibfnamefont {a.}~\bibnamefont {Schiavi}}, \bibinfo
  {author} {\bibfnamefont {O.}~\bibnamefont {Willi}}, \bibinfo {author}
  {\bibnamefont {a.J. Mackinnon}}, \bibinfo {author} {\bibfnamefont
  {D.}~\bibnamefont {Hicks}}, \bibinfo {author} {\bibfnamefont
  {P.}~\bibnamefont {Patel}}, \bibinfo {author} {\bibfnamefont
  {L.}~\bibnamefont {Gizzi}}, \bibinfo {author} {\bibfnamefont
  {M.}~\bibnamefont {Galimberti}}, \ and\ \bibinfo {author} {\bibfnamefont
  {R.}~\bibnamefont {Clarke}},\ }\bibfield  {title} {\enquote {\bibinfo {title}
  {{Laser-produced protons and their application as a particle probe}},}\
  }\href {\doibase 10.1017/S0263034602202177} {\bibfield  {journal} {\bibinfo
  {journal} {Laser and Particle Beams}\ }\textbf {\bibinfo {volume} {20}},\
  \bibinfo {pages} {269--275} (\bibinfo {year}
  {2002}{\natexlab{b}})}\BibitemShut {NoStop}%
\bibitem [{\citenamefont {Borghesi}\ \emph {et~al.}(2006)\citenamefont
  {Borghesi}, \citenamefont {Fuchs}, \citenamefont {Bulanov}, \citenamefont
  {Mackinnon}, \citenamefont {Patel},\ and\ \citenamefont
  {Roth}}]{borghesi2005fast}%
  \BibitemOpen
  \bibfield  {author} {\bibinfo {author} {\bibfnamefont {M.}~\bibnamefont
  {Borghesi}}, \bibinfo {author} {\bibfnamefont {J.}~\bibnamefont {Fuchs}},
  \bibinfo {author} {\bibfnamefont {S.}~\bibnamefont {Bulanov}}, \bibinfo
  {author} {\bibfnamefont {A.}~\bibnamefont {Mackinnon}}, \bibinfo {author}
  {\bibfnamefont {P.}~\bibnamefont {Patel}}, \ and\ \bibinfo {author}
  {\bibfnamefont {M.}~\bibnamefont {Roth}},\ }\bibfield  {title} {\enquote
  {\bibinfo {title} {Fast ion generation by high-intensity laser irradiation of
  solid targets and applications},}\ }\href@noop {} {\bibfield  {journal}
  {\bibinfo  {journal} {Fusion Science and Technology}\ }\textbf {\bibinfo
  {volume} {49}},\ \bibinfo {pages} {412--439} (\bibinfo {year}
  {2006})}\BibitemShut {NoStop}%
\bibitem [{\citenamefont {Loupias}\ \emph {et~al.}(2009)\citenamefont
  {Loupias}, \citenamefont {Gregory}, \citenamefont {Falize}, \citenamefont
  {Waugh}, \citenamefont {Seiichi}, \citenamefont {Pikuz}, \citenamefont
  {Kuramitsu}, \citenamefont {Ravasio}, \citenamefont {Bouquet}, \citenamefont
  {Michaut}, \citenamefont {Barroso}, \citenamefont {{Rabec le Gloahec}},
  \citenamefont {Nazarov}, \citenamefont {Takabe}, \citenamefont {Sakawa},
  \citenamefont {Woolsey},\ and\ \citenamefont {Koenig}}]{Loupias2009}%
  \BibitemOpen
  \bibfield  {author} {\bibinfo {author} {\bibfnamefont {B.}~\bibnamefont
  {Loupias}}, \bibinfo {author} {\bibfnamefont {C.~D.}\ \bibnamefont
  {Gregory}}, \bibinfo {author} {\bibfnamefont {E.}~\bibnamefont {Falize}},
  \bibinfo {author} {\bibfnamefont {J.}~\bibnamefont {Waugh}}, \bibinfo
  {author} {\bibfnamefont {D.}~\bibnamefont {Seiichi}}, \bibinfo {author}
  {\bibfnamefont {S.}~\bibnamefont {Pikuz}}, \bibinfo {author} {\bibfnamefont
  {Y.}~\bibnamefont {Kuramitsu}}, \bibinfo {author} {\bibfnamefont
  {a.}~\bibnamefont {Ravasio}}, \bibinfo {author} {\bibfnamefont
  {S.}~\bibnamefont {Bouquet}}, \bibinfo {author} {\bibfnamefont
  {C.}~\bibnamefont {Michaut}}, \bibinfo {author} {\bibfnamefont
  {P.}~\bibnamefont {Barroso}}, \bibinfo {author} {\bibfnamefont
  {M.}~\bibnamefont {{Rabec le Gloahec}}}, \bibinfo {author} {\bibfnamefont
  {W.}~\bibnamefont {Nazarov}}, \bibinfo {author} {\bibfnamefont
  {H.}~\bibnamefont {Takabe}}, \bibinfo {author} {\bibfnamefont
  {Y.}~\bibnamefont {Sakawa}}, \bibinfo {author} {\bibfnamefont
  {N.}~\bibnamefont {Woolsey}}, \ and\ \bibinfo {author} {\bibfnamefont
  {M.}~\bibnamefont {Koenig}},\ }\bibfield  {title} {\enquote {\bibinfo {title}
  {{Experimental results to study astrophysical plasma jets using Intense
  Lasers}},}\ }\href {\doibase 10.1007/s10509-009-0025-7} {\bibfield  {journal}
  {\bibinfo  {journal} {Astrophysics and Space Science}\ }\textbf {\bibinfo
  {volume} {322}},\ \bibinfo {pages} {25--29} (\bibinfo {year}
  {2009})}\BibitemShut {NoStop}%
\bibitem [{\citenamefont {Gregory}\ \emph {et~al.}(2010)\citenamefont
  {Gregory}, \citenamefont {Loupias}, \citenamefont {Waugh}, \citenamefont
  {Dono}, \citenamefont {Bouquet}, \citenamefont {Falize}, \citenamefont
  {Kuramitsu}, \citenamefont {Michaut}, \citenamefont {Nazarov}, \citenamefont
  {Pikuz}, \citenamefont {Sakawa}, \citenamefont {Woolsey},\ and\ \citenamefont
  {Koenig}}]{Gregory2010}%
  \BibitemOpen
  \bibfield  {author} {\bibinfo {author} {\bibfnamefont {C.~D.}\ \bibnamefont
  {Gregory}}, \bibinfo {author} {\bibfnamefont {B.}~\bibnamefont {Loupias}},
  \bibinfo {author} {\bibfnamefont {J.}~\bibnamefont {Waugh}}, \bibinfo
  {author} {\bibfnamefont {S.}~\bibnamefont {Dono}}, \bibinfo {author}
  {\bibfnamefont {S.}~\bibnamefont {Bouquet}}, \bibinfo {author} {\bibfnamefont
  {E.}~\bibnamefont {Falize}}, \bibinfo {author} {\bibfnamefont
  {Y.}~\bibnamefont {Kuramitsu}}, \bibinfo {author} {\bibfnamefont
  {C.}~\bibnamefont {Michaut}}, \bibinfo {author} {\bibfnamefont
  {W.}~\bibnamefont {Nazarov}}, \bibinfo {author} {\bibfnamefont {S.~a.}\
  \bibnamefont {Pikuz}}, \bibinfo {author} {\bibfnamefont {Y.}~\bibnamefont
  {Sakawa}}, \bibinfo {author} {\bibfnamefont {N.~C.}\ \bibnamefont {Woolsey}},
  \ and\ \bibinfo {author} {\bibfnamefont {M.}~\bibnamefont {Koenig}},\
  }\bibfield  {title} {\enquote {\bibinfo {title} {{Laser-driven plasma jets
  propagating in an ambient gas studied with optical and proton
  diagnostics}},}\ }\href {\doibase 10.1063/1.3431094} {\bibfield  {journal}
  {\bibinfo  {journal} {Physics of Plasmas}\ }\textbf {\bibinfo {volume}
  {17}},\ \bibinfo {pages} {052708} (\bibinfo {year} {2010})}\BibitemShut
  {NoStop}%
\bibitem [{\citenamefont {Chen}\ \emph {et~al.}(2012)\citenamefont {Chen},
  \citenamefont {D’Humi\`{e}res}, \citenamefont {Lefebvre}, \citenamefont
  {Romagnani}, \citenamefont {Toncian}, \citenamefont {Antici}, \citenamefont
  {Audebert}, \citenamefont {Brambrink}, \citenamefont {Cecchetti},
  \citenamefont {Kudyakov}, \citenamefont {Pipahl}, \citenamefont {Sentoku},
  \citenamefont {Borghesi}, \citenamefont {Willi},\ and\ \citenamefont
  {Fuchs}}]{Chen2012}%
  \BibitemOpen
  \bibfield  {author} {\bibinfo {author} {\bibfnamefont {S.~N.}\ \bibnamefont
  {Chen}}, \bibinfo {author} {\bibfnamefont {E.}~\bibnamefont
  {D’Humi\`{e}res}}, \bibinfo {author} {\bibfnamefont {E.}~\bibnamefont
  {Lefebvre}}, \bibinfo {author} {\bibfnamefont {L.}~\bibnamefont {Romagnani}},
  \bibinfo {author} {\bibfnamefont {T.}~\bibnamefont {Toncian}}, \bibinfo
  {author} {\bibfnamefont {P.}~\bibnamefont {Antici}}, \bibinfo {author}
  {\bibfnamefont {P.}~\bibnamefont {Audebert}}, \bibinfo {author}
  {\bibfnamefont {E.}~\bibnamefont {Brambrink}}, \bibinfo {author}
  {\bibfnamefont {C.~a.}\ \bibnamefont {Cecchetti}}, \bibinfo {author}
  {\bibfnamefont {T.}~\bibnamefont {Kudyakov}}, \bibinfo {author}
  {\bibfnamefont {a.}~\bibnamefont {Pipahl}}, \bibinfo {author} {\bibfnamefont
  {Y.}~\bibnamefont {Sentoku}}, \bibinfo {author} {\bibfnamefont
  {M.}~\bibnamefont {Borghesi}}, \bibinfo {author} {\bibfnamefont
  {O.}~\bibnamefont {Willi}}, \ and\ \bibinfo {author} {\bibfnamefont
  {J.}~\bibnamefont {Fuchs}},\ }\bibfield  {title} {\enquote {\bibinfo {title}
  {{Focusing Dynamics of High-Energy Density, Laser-Driven Ion Beams}},}\
  }\href {\doibase 10.1103/PhysRevLett.108.055001} {\bibfield  {journal}
  {\bibinfo  {journal} {Physical Review Letters}\ }\textbf {\bibinfo {volume}
  {108}},\ \bibinfo {pages} {055001} (\bibinfo {year} {2012})}\BibitemShut
  {NoStop}%
\bibitem [{\citenamefont {Jackson}\ and\ \citenamefont
  {Fox}(1999)}]{jackson1999classical}%
  \BibitemOpen
  \bibfield  {author} {\bibinfo {author} {\bibfnamefont {J.~D.}\ \bibnamefont
  {Jackson}}\ and\ \bibinfo {author} {\bibfnamefont {R.~F.}\ \bibnamefont
  {Fox}},\ }\bibfield  {title} {\enquote {\bibinfo {title} {Classical
  electrodynamics},}\ }\href@noop {} {\bibfield  {journal} {\bibinfo  {journal}
  {American Journal of Physics}\ }\textbf {\bibinfo {volume} {67}},\ \bibinfo
  {pages} {841--842} (\bibinfo {year} {1999})}\BibitemShut {NoStop}%
\bibitem [{\citenamefont {Li}, \citenamefont {Yan},\ and\ \citenamefont
  {Ren}(2008)}]{Li2008}%
  \BibitemOpen
  \bibfield  {author} {\bibinfo {author} {\bibfnamefont {G.}~\bibnamefont
  {Li}}, \bibinfo {author} {\bibfnamefont {R.}~\bibnamefont {Yan}}, \ and\
  \bibinfo {author} {\bibfnamefont {C.}~\bibnamefont {Ren}},\ }\bibfield
  {title} {\enquote {\bibinfo {title} {{Laser Channeling in Millimeter-Scale
  Underdense Plasmas of Fast-Ignition Targets}},}\ }\href {\doibase
  10.1103/PhysRevLett.100.125002} {\bibfield  {journal} {\bibinfo  {journal}
  {Physical Review Letters}\ }\textbf {\bibinfo {volume} {125002}},\ \bibinfo
  {pages} {1--4} (\bibinfo {year} {2008})}\BibitemShut {NoStop}%
\bibitem [{\citenamefont {Petrasso}\ \emph {et~al.}(2009)\citenamefont
  {Petrasso}, \citenamefont {Li}, \citenamefont {Seguin}, \citenamefont {Rygg},
  \citenamefont {Frenje}, \citenamefont {Betti}, \citenamefont {Knauer},
  \citenamefont {Meyerhofer}, \citenamefont {Amendt}, \citenamefont {Froula},
  \citenamefont {Landen}, \citenamefont {Patel}, \citenamefont {Ross},\ and\
  \citenamefont {Town}}]{Petrasso2009}%
  \BibitemOpen
  \bibfield  {author} {\bibinfo {author} {\bibfnamefont {R.}~\bibnamefont
  {Petrasso}}, \bibinfo {author} {\bibfnamefont {C.}~\bibnamefont {Li}},
  \bibinfo {author} {\bibfnamefont {F.}~\bibnamefont {Seguin}}, \bibinfo
  {author} {\bibfnamefont {J.}~\bibnamefont {Rygg}}, \bibinfo {author}
  {\bibfnamefont {J.}~\bibnamefont {Frenje}}, \bibinfo {author} {\bibfnamefont
  {R.}~\bibnamefont {Betti}}, \bibinfo {author} {\bibfnamefont
  {J.}~\bibnamefont {Knauer}}, \bibinfo {author} {\bibfnamefont
  {D.}~\bibnamefont {Meyerhofer}}, \bibinfo {author} {\bibfnamefont
  {P.}~\bibnamefont {Amendt}}, \bibinfo {author} {\bibfnamefont
  {D.}~\bibnamefont {Froula}}, \bibinfo {author} {\bibfnamefont
  {O.}~\bibnamefont {Landen}}, \bibinfo {author} {\bibfnamefont
  {P.}~\bibnamefont {Patel}}, \bibinfo {author} {\bibfnamefont
  {J.}~\bibnamefont {Ross}}, \ and\ \bibinfo {author} {\bibfnamefont
  {R.}~\bibnamefont {Town}},\ }\bibfield  {title} {\enquote {\bibinfo {title}
  {{Lorentz Mapping of Magnetic Fields in Hot Dense Plasmas}},}\ }\href
  {\doibase 10.1103/PhysRevLett.103.085001} {\bibfield  {journal} {\bibinfo
  {journal} {Physical Review Letters}\ }\textbf {\bibinfo {volume} {103}},\
  \bibinfo {pages} {085001} (\bibinfo {year} {2009})}\BibitemShut {NoStop}%
\bibitem [{\citenamefont {Sarri}\ \emph {et~al.}(2009)\citenamefont {Sarri},
  \citenamefont {Borghesi}, \citenamefont {Cecchetti}, \citenamefont
  {Romagnani}, \citenamefont {Jung}, \citenamefont {Willi}, \citenamefont
  {Hoarty}, \citenamefont {Stevenson}, \citenamefont {Brown}, \citenamefont
  {James}, \citenamefont {Hobbs}, \citenamefont {Lockyear}, \citenamefont
  {Bulanov},\ and\ \citenamefont {Pegoraro}}]{Sarri2009}%
  \BibitemOpen
  \bibfield  {author} {\bibinfo {author} {\bibfnamefont {G.}~\bibnamefont
  {Sarri}}, \bibinfo {author} {\bibfnamefont {M.}~\bibnamefont {Borghesi}},
  \bibinfo {author} {\bibfnamefont {C.~A.}\ \bibnamefont {Cecchetti}}, \bibinfo
  {author} {\bibfnamefont {L.}~\bibnamefont {Romagnani}}, \bibinfo {author}
  {\bibfnamefont {R.}~\bibnamefont {Jung}}, \bibinfo {author} {\bibfnamefont
  {O.}~\bibnamefont {Willi}}, \bibinfo {author} {\bibfnamefont {D.~J.}\
  \bibnamefont {Hoarty}}, \bibinfo {author} {\bibfnamefont {R.~M.}\
  \bibnamefont {Stevenson}}, \bibinfo {author} {\bibfnamefont {C.~R.}\
  \bibnamefont {Brown}}, \bibinfo {author} {\bibfnamefont {S.~F.}\ \bibnamefont
  {James}}, \bibinfo {author} {\bibfnamefont {P.}~\bibnamefont {Hobbs}},
  \bibinfo {author} {\bibfnamefont {J.}~\bibnamefont {Lockyear}}, \bibinfo
  {author} {\bibfnamefont {S.~V.}\ \bibnamefont {Bulanov}}, \ and\ \bibinfo
  {author} {\bibfnamefont {F.}~\bibnamefont {Pegoraro}},\ }\bibfield  {title}
  {\enquote {\bibinfo {title} {{Application of proton radiography in
  experiments of relevance to inertial confinement fusion}},}\ }\href {\doibase
  10.1140/epjd/e2009-00115-8} {\bibfield  {journal} {\bibinfo  {journal} {The
  European Physical Journal D}\ }\textbf {\bibinfo {volume} {55}},\ \bibinfo
  {pages} {299--303} (\bibinfo {year} {2009})}\BibitemShut {NoStop}%
\bibitem [{\citenamefont {Li}\ \emph {et~al.}(2010{\natexlab{b}})\citenamefont
  {Li}, \citenamefont {S\'{e}guin}, \citenamefont {Frenje}, \citenamefont
  {Rosenberg}, \citenamefont {Petrasso}, \citenamefont {Amendt}, \citenamefont
  {Koch}, \citenamefont {Landen}, \citenamefont {Park}, \citenamefont {Robey},
  \citenamefont {Town}, \citenamefont {Casner}, \citenamefont {Philippe},
  \citenamefont {Betti}, \citenamefont {Knauer}, \citenamefont {Meyerhofer},
  \citenamefont {Back}, \citenamefont {Kilkenny},\ and\ \citenamefont
  {Nikroo}}]{Li2010a}%
  \BibitemOpen
  \bibfield  {author} {\bibinfo {author} {\bibfnamefont {C.~K.}\ \bibnamefont
  {Li}}, \bibinfo {author} {\bibfnamefont {F.~H.}\ \bibnamefont {S\'{e}guin}},
  \bibinfo {author} {\bibfnamefont {J.~a.}\ \bibnamefont {Frenje}}, \bibinfo
  {author} {\bibfnamefont {M.}~\bibnamefont {Rosenberg}}, \bibinfo {author}
  {\bibfnamefont {R.~D.}\ \bibnamefont {Petrasso}}, \bibinfo {author}
  {\bibfnamefont {P.~a.}\ \bibnamefont {Amendt}}, \bibinfo {author}
  {\bibfnamefont {J.~a.}\ \bibnamefont {Koch}}, \bibinfo {author}
  {\bibfnamefont {O.~L.}\ \bibnamefont {Landen}}, \bibinfo {author}
  {\bibfnamefont {H.~S.}\ \bibnamefont {Park}}, \bibinfo {author}
  {\bibfnamefont {H.~F.}\ \bibnamefont {Robey}}, \bibinfo {author}
  {\bibfnamefont {R.~P.~J.}\ \bibnamefont {Town}}, \bibinfo {author}
  {\bibfnamefont {a.}~\bibnamefont {Casner}}, \bibinfo {author} {\bibfnamefont
  {F.}~\bibnamefont {Philippe}}, \bibinfo {author} {\bibfnamefont
  {R.}~\bibnamefont {Betti}}, \bibinfo {author} {\bibfnamefont {J.~P.}\
  \bibnamefont {Knauer}}, \bibinfo {author} {\bibfnamefont {D.~D.}\
  \bibnamefont {Meyerhofer}}, \bibinfo {author} {\bibfnamefont {C.~a.}\
  \bibnamefont {Back}}, \bibinfo {author} {\bibfnamefont {J.~D.}\ \bibnamefont
  {Kilkenny}}, \ and\ \bibinfo {author} {\bibfnamefont {a.}~\bibnamefont
  {Nikroo}},\ }\bibfield  {title} {\enquote {\bibinfo {title}
  {{Charged-particle probing of x-ray-driven inertial-fusion implosions.}}}\
  }\href {\doibase 10.1126/science.1185747} {\bibfield  {journal} {\bibinfo
  {journal} {Science (New York, N.Y.)}\ }\textbf {\bibinfo {volume} {327}},\
  \bibinfo {pages} {1231--5} (\bibinfo {year}
  {2010}{\natexlab{b}})}\BibitemShut {NoStop}%
\bibitem [{\citenamefont {Willingale}\ \emph {et~al.}(2011)\citenamefont
  {Willingale}, \citenamefont {Nilson}, \citenamefont {Thomas}, \citenamefont
  {Cobble}, \citenamefont {Craxton}, \citenamefont {Maksimchuk}, \citenamefont
  {Norreys}, \citenamefont {Sangster}, \citenamefont {Scott}, \citenamefont
  {Stoeckl}, \citenamefont {Zulick},\ and\ \citenamefont
  {Krushelnick}}]{Willingale2011}%
  \BibitemOpen
  \bibfield  {author} {\bibinfo {author} {\bibfnamefont {L.}~\bibnamefont
  {Willingale}}, \bibinfo {author} {\bibfnamefont {P.~M.}\ \bibnamefont
  {Nilson}}, \bibinfo {author} {\bibfnamefont {a.~G.~R.}\ \bibnamefont
  {Thomas}}, \bibinfo {author} {\bibfnamefont {J.}~\bibnamefont {Cobble}},
  \bibinfo {author} {\bibfnamefont {R.~S.}\ \bibnamefont {Craxton}}, \bibinfo
  {author} {\bibfnamefont {a.}~\bibnamefont {Maksimchuk}}, \bibinfo {author}
  {\bibfnamefont {P.~a.}\ \bibnamefont {Norreys}}, \bibinfo {author}
  {\bibfnamefont {T.~C.}\ \bibnamefont {Sangster}}, \bibinfo {author}
  {\bibfnamefont {R.~H.~H.}\ \bibnamefont {Scott}}, \bibinfo {author}
  {\bibfnamefont {C.}~\bibnamefont {Stoeckl}}, \bibinfo {author} {\bibfnamefont
  {C.}~\bibnamefont {Zulick}}, \ and\ \bibinfo {author} {\bibfnamefont
  {K.}~\bibnamefont {Krushelnick}},\ }\bibfield  {title} {\enquote {\bibinfo
  {title} {{High-Power, Kilojoule Class Laser Channeling in Millimeter-Scale
  Underdense Plasma}},}\ }\href {\doibase 10.1103/PhysRevLett.106.105002}
  {\bibfield  {journal} {\bibinfo  {journal} {Physical Review Letters}\
  }\textbf {\bibinfo {volume} {106}},\ \bibinfo {pages} {105002} (\bibinfo
  {year} {2011})}\BibitemShut {NoStop}%
\bibitem [{\citenamefont {Séguin}\ \emph {et~al.}(2012)\citenamefont
  {Séguin}, \citenamefont {Li}, \citenamefont {Manuel}, \citenamefont
  {Rinderknecht}, \citenamefont {Sinenian}, \citenamefont {Frenje},
  \citenamefont {Rygg}, \citenamefont {Hicks}, \citenamefont {Petrasso},
  \citenamefont {Delettrez}, \citenamefont {Betti}, \citenamefont {Marshall},\
  and\ \citenamefont {Smalyuk}}]{Seguin2012}%
  \BibitemOpen
  \bibfield  {author} {\bibinfo {author} {\bibfnamefont {F.~H.}\ \bibnamefont
  {Séguin}}, \bibinfo {author} {\bibfnamefont {C.~K.}\ \bibnamefont {Li}},
  \bibinfo {author} {\bibfnamefont {M.~J.-E.}\ \bibnamefont {Manuel}}, \bibinfo
  {author} {\bibfnamefont {H.~G.}\ \bibnamefont {Rinderknecht}}, \bibinfo
  {author} {\bibfnamefont {N.}~\bibnamefont {Sinenian}}, \bibinfo {author}
  {\bibfnamefont {J.~a.}\ \bibnamefont {Frenje}}, \bibinfo {author}
  {\bibfnamefont {J.~R.}\ \bibnamefont {Rygg}}, \bibinfo {author}
  {\bibfnamefont {D.~G.}\ \bibnamefont {Hicks}}, \bibinfo {author}
  {\bibfnamefont {R.~D.}\ \bibnamefont {Petrasso}}, \bibinfo {author}
  {\bibfnamefont {J.}~\bibnamefont {Delettrez}}, \bibinfo {author}
  {\bibfnamefont {R.}~\bibnamefont {Betti}}, \bibinfo {author} {\bibfnamefont
  {F.~J.}\ \bibnamefont {Marshall}}, \ and\ \bibinfo {author} {\bibfnamefont
  {V.~a.}\ \bibnamefont {Smalyuk}},\ }\bibfield  {title} {\enquote {\bibinfo
  {title} {{Time evolution of filamentation and self-generated fields in the
  coronae of directly driven inertial-confinement fusion capsules}},}\ }\href
  {\doibase 10.1063/1.3671908} {\bibfield  {journal} {\bibinfo  {journal}
  {Physics of Plasmas}\ }\textbf {\bibinfo {volume} {19}},\ \bibinfo {pages}
  {012701} (\bibinfo {year} {2012})}\BibitemShut {NoStop}%
\bibitem [{\citenamefont {Kugland}\ \emph
  {et~al.}(2012{\natexlab{b}})\citenamefont {Kugland}, \citenamefont {Ryutov},
  \citenamefont {Plechaty}, \citenamefont {Ross},\ and\ \citenamefont
  {Park}}]{Kugland2012}%
  \BibitemOpen
  \bibfield  {author} {\bibinfo {author} {\bibfnamefont {N.~L.}\ \bibnamefont
  {Kugland}}, \bibinfo {author} {\bibfnamefont {D.~D.}\ \bibnamefont {Ryutov}},
  \bibinfo {author} {\bibfnamefont {C.}~\bibnamefont {Plechaty}}, \bibinfo
  {author} {\bibfnamefont {J.~S.}\ \bibnamefont {Ross}}, \ and\ \bibinfo
  {author} {\bibfnamefont {H.-S.}\ \bibnamefont {Park}},\ }\bibfield  {title}
  {\enquote {\bibinfo {title} {{Invited Article: Relation between electric and
  magnetic field structures and their proton-beam images}},}\ }\href {\doibase
  10.1063/1.4750234} {\bibfield  {journal} {\bibinfo  {journal} {Review of
  Scientific Instruments}\ }\textbf {\bibinfo {volume} {83}},\ \bibinfo {pages}
  {101301} (\bibinfo {year} {2012}{\natexlab{b}})}\BibitemShut {NoStop}%
\bibitem [{\citenamefont {Aufderheide}, \citenamefont {Slone},\ and\
  \citenamefont {{Schach von Wittenau}}(2000)}]{Aufderheide2000}%
  \BibitemOpen
  \bibfield  {author} {\bibinfo {author} {\bibfnamefont {M.~B.}\ \bibnamefont
  {Aufderheide}}, \bibinfo {author} {\bibfnamefont {D.~M.}\ \bibnamefont
  {Slone}}, \ and\ \bibinfo {author} {\bibfnamefont {A.~E.}\ \bibnamefont
  {{Schach von Wittenau}}},\ }\bibfield  {title} {\enquote {\bibinfo {title}
  {{Hades , A Radiographic Simulation Code}},}\ }\href@noop {} {\bibfield
  {journal} {\bibinfo  {journal} {Annual Review of Progress in Quantitative
  Nondestructive Evaluation}\ } (\bibinfo {year} {2000})}\BibitemShut {NoStop}%
\bibitem [{\citenamefont {{Los Alamos National Laboratory}}(2005)}]{MCNP5}%
  \BibitemOpen
  \bibfield  {author} {\bibinfo {author} {\bibnamefont {{Los Alamos National
  Laboratory}}},\ }\href {http://mcnp.lanl.gov} {\enquote {\bibinfo {title} {{A
  general Monte Carlo N-particle (MCNP) transport code. Release MCNP5-1.40}},}\
  } (\bibinfo {year} {2005})\BibitemShut {NoStop}%
\bibitem [{\citenamefont {Manuel}\ \emph
  {et~al.}(2012{\natexlab{b}})\citenamefont {Manuel}, \citenamefont {Zylstra},
  \citenamefont {Rinderknecht}, \citenamefont {Casey}, \citenamefont
  {Rosenberg}, \citenamefont {Sinenian}, \citenamefont {Li}, \citenamefont
  {Frenje}, \citenamefont {S\'{e}guin},\ and\ \citenamefont
  {Petrasso}}]{Manuel2012a}%
  \BibitemOpen
  \bibfield  {author} {\bibinfo {author} {\bibfnamefont {M.~J.-E.}\
  \bibnamefont {Manuel}}, \bibinfo {author} {\bibfnamefont {a.~B.}\
  \bibnamefont {Zylstra}}, \bibinfo {author} {\bibfnamefont {H.~G.}\
  \bibnamefont {Rinderknecht}}, \bibinfo {author} {\bibfnamefont {D.~T.}\
  \bibnamefont {Casey}}, \bibinfo {author} {\bibfnamefont {M.~J.}\ \bibnamefont
  {Rosenberg}}, \bibinfo {author} {\bibfnamefont {N.}~\bibnamefont {Sinenian}},
  \bibinfo {author} {\bibfnamefont {C.~K.}\ \bibnamefont {Li}}, \bibinfo
  {author} {\bibfnamefont {J.~a.}\ \bibnamefont {Frenje}}, \bibinfo {author}
  {\bibfnamefont {F.~H.}\ \bibnamefont {S\'{e}guin}}, \ and\ \bibinfo {author}
  {\bibfnamefont {R.~D.}\ \bibnamefont {Petrasso}},\ }\bibfield  {title}
  {\enquote {\bibinfo {title} {{Source characterization and modeling
  development for monoenergetic-proton radiography experiments on OMEGA.}}}\
  }\href {\doibase 10.1063/1.4730336} {\bibfield  {journal} {\bibinfo
  {journal} {The Review of scientific instruments}\ }\textbf {\bibinfo {volume}
  {83}},\ \bibinfo {pages} {063506} (\bibinfo {year}
  {2012}{\natexlab{b}})}\BibitemShut {NoStop}%
\bibitem [{\citenamefont {Manuel}\ \emph
  {et~al.}(2012{\natexlab{c}})\citenamefont {Manuel}, \citenamefont {Sinenian},
  \citenamefont {Séguin}, \citenamefont {Li}, \citenamefont {Frenje},
  \citenamefont {Rinderknecht}, \citenamefont {Casey}, \citenamefont {Zylstra},
  \citenamefont {Petrasso},\ and\ \citenamefont {Beg}}]{Manuel2012b}%
  \BibitemOpen
  \bibfield  {author} {\bibinfo {author} {\bibfnamefont {M.~J.-E.}\
  \bibnamefont {Manuel}}, \bibinfo {author} {\bibfnamefont {N.}~\bibnamefont
  {Sinenian}}, \bibinfo {author} {\bibfnamefont {F.~H.}\ \bibnamefont
  {Séguin}}, \bibinfo {author} {\bibfnamefont {C.~K.}\ \bibnamefont {Li}},
  \bibinfo {author} {\bibfnamefont {J.~a.}\ \bibnamefont {Frenje}}, \bibinfo
  {author} {\bibfnamefont {H.~G.}\ \bibnamefont {Rinderknecht}}, \bibinfo
  {author} {\bibfnamefont {D.~T.}\ \bibnamefont {Casey}}, \bibinfo {author}
  {\bibfnamefont {a.~B.}\ \bibnamefont {Zylstra}}, \bibinfo {author}
  {\bibfnamefont {R.~D.}\ \bibnamefont {Petrasso}}, \ and\ \bibinfo {author}
  {\bibfnamefont {F.~N.}\ \bibnamefont {Beg}},\ }\bibfield  {title} {\enquote
  {\bibinfo {title} {{Mapping return currents in laser-generated Z-pinch
  plasmas using proton deflectometry}},}\ }\href {\doibase 10.1063/1.4718425}
  {\bibfield  {journal} {\bibinfo  {journal} {Applied Physics Letters}\
  }\textbf {\bibinfo {volume} {100}},\ \bibinfo {pages} {203505} (\bibinfo
  {year} {2012}{\natexlab{c}})}\BibitemShut {NoStop}%
\bibitem [{\citenamefont {Fonseca}\ \emph {et~al.}(2008)\citenamefont
  {Fonseca}, \citenamefont {Martins}, \citenamefont {Silva}, \citenamefont
  {Tonge}, \citenamefont {Tsung},\ and\ \citenamefont {Mori}}]{Fonseca2008}%
  \BibitemOpen
  \bibfield  {author} {\bibinfo {author} {\bibfnamefont {R.~A.}\ \bibnamefont
  {Fonseca}}, \bibinfo {author} {\bibfnamefont {S.~F.}\ \bibnamefont
  {Martins}}, \bibinfo {author} {\bibfnamefont {L.~O.}\ \bibnamefont {Silva}},
  \bibinfo {author} {\bibfnamefont {J.~W.}\ \bibnamefont {Tonge}}, \bibinfo
  {author} {\bibfnamefont {F.~S.}\ \bibnamefont {Tsung}}, \ and\ \bibinfo
  {author} {\bibfnamefont {W.~B.}\ \bibnamefont {Mori}},\ }\bibfield  {title}
  {\enquote {\bibinfo {title} {{One-to-one direct modeling of experiments and
  astrophysical scenarios: pushing the envelope on kinetic plasma
  simulations}},}\ }\href {\doibase 10.1088/0741-3335/50/12/124034} {\bibfield
  {journal} {\bibinfo  {journal} {Plasma Physics and Controlled Fusion}\
  }\textbf {\bibinfo {volume} {50}},\ \bibinfo {pages} {124034} (\bibinfo
  {year} {2008})}\BibitemShut {NoStop}%
\bibitem [{\citenamefont {Ryutov}\ \emph {et~al.}(2011)\citenamefont {Ryutov},
  \citenamefont {Kugland}, \citenamefont {Park}, \citenamefont {Pollaine},
  \citenamefont {Remington},\ and\ \citenamefont {Ross}}]{Ryutov2011}%
  \BibitemOpen
  \bibfield  {author} {\bibinfo {author} {\bibfnamefont {D.~D.}\ \bibnamefont
  {Ryutov}}, \bibinfo {author} {\bibfnamefont {N.~L.}\ \bibnamefont {Kugland}},
  \bibinfo {author} {\bibfnamefont {H.-S.}\ \bibnamefont {Park}}, \bibinfo
  {author} {\bibfnamefont {S.~M.}\ \bibnamefont {Pollaine}}, \bibinfo {author}
  {\bibfnamefont {B.~a.}\ \bibnamefont {Remington}}, \ and\ \bibinfo {author}
  {\bibfnamefont {J.~S.}\ \bibnamefont {Ross}},\ }\bibfield  {title} {\enquote
  {\bibinfo {title} {{Collisional current drive in two interpenetrating plasma
  jets}},}\ }\href {\doibase 10.1063/1.3646325} {\bibfield  {journal} {\bibinfo
   {journal} {Physics of Plasmas}\ }\textbf {\bibinfo {volume} {18}},\ \bibinfo
  {pages} {104504} (\bibinfo {year} {2011})}\BibitemShut {NoStop}%
\bibitem [{\citenamefont {Ryutov}\ \emph {et~al.}(2013)\citenamefont {Ryutov},
  \citenamefont {Kugland}, \citenamefont {Levy}, \citenamefont {Plechaty},
  \citenamefont {Ross},\ and\ \citenamefont {Park}}]{Ryutov2013}%
  \BibitemOpen
  \bibfield  {author} {\bibinfo {author} {\bibfnamefont {D.~D.}\ \bibnamefont
  {Ryutov}}, \bibinfo {author} {\bibfnamefont {N.~L.}\ \bibnamefont {Kugland}},
  \bibinfo {author} {\bibfnamefont {M.~C.}\ \bibnamefont {Levy}}, \bibinfo
  {author} {\bibfnamefont {C.}~\bibnamefont {Plechaty}}, \bibinfo {author}
  {\bibfnamefont {J.~S.}\ \bibnamefont {Ross}}, \ and\ \bibinfo {author}
  {\bibfnamefont {H.~S.}\ \bibnamefont {Park}},\ }\bibfield  {title} {\enquote
  {\bibinfo {title} {{Magnetic field advection in two interpenetrating plasma
  streams}},}\ }\href {\doibase 10.1063/1.4794200} {\bibfield  {journal}
  {\bibinfo  {journal} {Physics of Plasmas}\ }\textbf {\bibinfo {volume}
  {20}},\ \bibinfo {pages} {032703} (\bibinfo {year} {2013})}\BibitemShut
  {NoStop}%
\bibitem [{\citenamefont {Kugland}\ \emph {et~al.}(2013)\citenamefont
  {Kugland}, \citenamefont {Ross}, \citenamefont {Chang}, \citenamefont
  {Drake}, \citenamefont {Fiksel}, \citenamefont {Froula}, \citenamefont
  {Glenzer}, \citenamefont {Gregori}, \citenamefont {Grosskopf}, \citenamefont
  {Huntington}, \citenamefont {Koenig}, \citenamefont {Kuramitsu},
  \citenamefont {Kuranz}, \citenamefont {Levy}, \citenamefont {Liang},
  \citenamefont {Martinez}, \citenamefont {Meinecke}, \citenamefont {Miniati},
  \citenamefont {Morita}, \citenamefont {Pelka}, \citenamefont {Plechaty},
  \citenamefont {Presura}, \citenamefont {Ravasio}, \citenamefont {Remington},
  \citenamefont {Reville}, \citenamefont {Ryutov}, \citenamefont {Sakawa},
  \citenamefont {Spitkovsky}, \citenamefont {Takabe},\ and\ \citenamefont
  {Park}}]{Kugland2013}%
  \BibitemOpen
  \bibfield  {author} {\bibinfo {author} {\bibfnamefont {N.~L.}\ \bibnamefont
  {Kugland}}, \bibinfo {author} {\bibfnamefont {J.~S.}\ \bibnamefont {Ross}},
  \bibinfo {author} {\bibfnamefont {P.-Y.}\ \bibnamefont {Chang}}, \bibinfo
  {author} {\bibfnamefont {R.~P.}\ \bibnamefont {Drake}}, \bibinfo {author}
  {\bibfnamefont {G.}~\bibnamefont {Fiksel}}, \bibinfo {author} {\bibfnamefont
  {D.~H.}\ \bibnamefont {Froula}}, \bibinfo {author} {\bibfnamefont {S.~H.}\
  \bibnamefont {Glenzer}}, \bibinfo {author} {\bibfnamefont {G.}~\bibnamefont
  {Gregori}}, \bibinfo {author} {\bibfnamefont {M.}~\bibnamefont {Grosskopf}},
  \bibinfo {author} {\bibfnamefont {C.}~\bibnamefont {Huntington}}, \bibinfo
  {author} {\bibfnamefont {M.}~\bibnamefont {Koenig}}, \bibinfo {author}
  {\bibfnamefont {Y.}~\bibnamefont {Kuramitsu}}, \bibinfo {author}
  {\bibfnamefont {C.}~\bibnamefont {Kuranz}}, \bibinfo {author} {\bibfnamefont
  {M.~C.}\ \bibnamefont {Levy}}, \bibinfo {author} {\bibfnamefont
  {E.}~\bibnamefont {Liang}}, \bibinfo {author} {\bibfnamefont
  {D.}~\bibnamefont {Martinez}}, \bibinfo {author} {\bibfnamefont
  {J.}~\bibnamefont {Meinecke}}, \bibinfo {author} {\bibfnamefont
  {F.}~\bibnamefont {Miniati}}, \bibinfo {author} {\bibfnamefont
  {T.}~\bibnamefont {Morita}}, \bibinfo {author} {\bibfnamefont
  {A.}~\bibnamefont {Pelka}}, \bibinfo {author} {\bibfnamefont
  {C.}~\bibnamefont {Plechaty}}, \bibinfo {author} {\bibfnamefont
  {R.}~\bibnamefont {Presura}}, \bibinfo {author} {\bibfnamefont
  {A.}~\bibnamefont {Ravasio}}, \bibinfo {author} {\bibfnamefont {B.~A.}\
  \bibnamefont {Remington}}, \bibinfo {author} {\bibfnamefont {B.}~\bibnamefont
  {Reville}}, \bibinfo {author} {\bibfnamefont {D.~D.}\ \bibnamefont {Ryutov}},
  \bibinfo {author} {\bibfnamefont {Y.}~\bibnamefont {Sakawa}}, \bibinfo
  {author} {\bibfnamefont {A.}~\bibnamefont {Spitkovsky}}, \bibinfo {author}
  {\bibfnamefont {H.}~\bibnamefont {Takabe}}, \ and\ \bibinfo {author}
  {\bibfnamefont {H.-S.}\ \bibnamefont {Park}},\ }\bibfield  {title} {\enquote
  {\bibinfo {title} {{Visualizing electromagnetic fields in laser-produced
  counter-streaming plasma experiments for collisionless shock laboratory
  astrophysics}},}\ }\href {\doibase 10.1063/1.4804548} {\bibfield  {journal}
  {\bibinfo  {journal} {Physics of Plasmas}\ }\textbf {\bibinfo {volume}
  {20}},\ \bibinfo {pages} {056313} (\bibinfo {year} {2013})}\BibitemShut
  {NoStop}%
\bibitem [{\citenamefont {Welch}\ \emph {et~al.}(2004)\citenamefont {Welch},
  \citenamefont {Rose}, \citenamefont {Clark}, \citenamefont {Genoni},\ and\
  \citenamefont {Hughes}}]{Welch2004}%
  \BibitemOpen
  \bibfield  {author} {\bibinfo {author} {\bibfnamefont {D.}~\bibnamefont
  {Welch}}, \bibinfo {author} {\bibfnamefont {D.}~\bibnamefont {Rose}},
  \bibinfo {author} {\bibfnamefont {R.}~\bibnamefont {Clark}}, \bibinfo
  {author} {\bibfnamefont {T.}~\bibnamefont {Genoni}}, \ and\ \bibinfo {author}
  {\bibfnamefont {T.}~\bibnamefont {Hughes}},\ }\bibfield  {title} {\enquote
  {\bibinfo {title} {{Implementation of an non-iterative implicit
  electromagnetic field solver for dense plasma simulation}},}\ }\href
  {\doibase 10.1016/j.cpc.2004.06.028} {\bibfield  {journal} {\bibinfo
  {journal} {Computer Physics Communications}\ }\textbf {\bibinfo {volume}
  {164}},\ \bibinfo {pages} {183--188} (\bibinfo {year} {2004})}\BibitemShut
  {NoStop}%
\bibitem [{\citenamefont {Birdsall}\ and\ \citenamefont
  {Langdon}(2004)}]{birdsall2004plasma}%
  \BibitemOpen
  \bibfield  {author} {\bibinfo {author} {\bibfnamefont {C.~K.}\ \bibnamefont
  {Birdsall}}\ and\ \bibinfo {author} {\bibfnamefont {A.~B.}\ \bibnamefont
  {Langdon}},\ }\href@noop {} {\emph {\bibinfo {title} {Plasma physics via
  computer simulation}}}\ (\bibinfo  {publisher} {CRC Press},\ \bibinfo {year}
  {2004})\BibitemShut {NoStop}%
\bibitem [{\citenamefont {Ross}\ \emph {et~al.}(2012)\citenamefont {Ross},
  \citenamefont {Glenzer}, \citenamefont {Amendt}, \citenamefont {Berger},\
  and\ \citenamefont {Divol}}]{Ross2012}%
  \BibitemOpen
  \bibfield  {author} {\bibinfo {author} {\bibfnamefont {J.~S.}\ \bibnamefont
  {Ross}}, \bibinfo {author} {\bibfnamefont {S.~H.}\ \bibnamefont {Glenzer}},
  \bibinfo {author} {\bibfnamefont {P.}~\bibnamefont {Amendt}}, \bibinfo
  {author} {\bibfnamefont {R.}~\bibnamefont {Berger}}, \ and\ \bibinfo {author}
  {\bibfnamefont {L.}~\bibnamefont {Divol}},\ }\bibfield  {title} {\enquote
  {\bibinfo {title} {{Characterizing counter-streaming interpenetrating plasmas
  relevant to astrophysical collisionless shocks to astrophysical collisionless
  shocks a )}},}\ }\href {\doibase 10.1063/1.3694124} {\bibfield  {journal}
  {\bibinfo  {journal} {Physics of Plasmas}\ }\textbf {\bibinfo {volume}
  {056501}} (\bibinfo {year} {2012}),\ 10.1063/1.3694124}\BibitemShut {NoStop}%
\bibitem [{\citenamefont {Ross}\ \emph {et~al.}(2013)\citenamefont {Ross},
  \citenamefont {Park}, \citenamefont {Berger}, \citenamefont {Divol},
  \citenamefont {Kugland}, \citenamefont {Rozmus}, \citenamefont {Ryutov},\
  and\ \citenamefont {Glenzer}}]{Ross2013}%
  \BibitemOpen
  \bibfield  {author} {\bibinfo {author} {\bibfnamefont {J.~S.}\ \bibnamefont
  {Ross}}, \bibinfo {author} {\bibfnamefont {H.-S.}\ \bibnamefont {Park}},
  \bibinfo {author} {\bibfnamefont {R.}~\bibnamefont {Berger}}, \bibinfo
  {author} {\bibfnamefont {L.}~\bibnamefont {Divol}}, \bibinfo {author}
  {\bibfnamefont {N.~L.}\ \bibnamefont {Kugland}}, \bibinfo {author}
  {\bibfnamefont {W.}~\bibnamefont {Rozmus}}, \bibinfo {author} {\bibfnamefont
  {D.}~\bibnamefont {Ryutov}}, \ and\ \bibinfo {author} {\bibfnamefont {S.~H.}\
  \bibnamefont {Glenzer}},\ }\bibfield  {title} {\enquote {\bibinfo {title}
  {{Collisionless Coupling of Ion and Electron Temperatures in Counterstreaming
  Plasma Flows}},}\ }\href {\doibase 10.1103/PhysRevLett.110.145005} {\bibfield
   {journal} {\bibinfo  {journal} {Physical Review Letters}\ }\textbf {\bibinfo
  {volume} {110}},\ \bibinfo {pages} {145005} (\bibinfo {year}
  {2013})}\BibitemShut {NoStop}%
\bibitem [{\citenamefont {Ryutov}\ \emph {et~al.}(2014)\citenamefont {Ryutov},
  \citenamefont {Fiuza}, \citenamefont {Huntington}, \citenamefont {Ross},\
  and\ \citenamefont {Park}}]{Ryutov2014}%
  \BibitemOpen
  \bibfield  {author} {\bibinfo {author} {\bibfnamefont {D.~D.}\ \bibnamefont
  {Ryutov}}, \bibinfo {author} {\bibfnamefont {F.}~\bibnamefont {Fiuza}},
  \bibinfo {author} {\bibfnamefont {C.~M.}\ \bibnamefont {Huntington}},
  \bibinfo {author} {\bibfnamefont {J.~S.}\ \bibnamefont {Ross}}, \ and\
  \bibinfo {author} {\bibfnamefont {H.-S.}\ \bibnamefont {Park}},\ }\bibfield
  {title} {\enquote {\bibinfo {title} {{Collisional effects in the ion Weibel
  instability for two counter-propagating plasma streams}},}\ }\href {\doibase
  10.1063/1.4867062} {\bibfield  {journal} {\bibinfo  {journal} {Physics of
  Plasmas}\ }\textbf {\bibinfo {volume} {21}},\ \bibinfo {pages} {032701}
  (\bibinfo {year} {2014})}\BibitemShut {NoStop}%
\bibitem [{\citenamefont {Weibel}(1959)}]{Weibel1959}%
  \BibitemOpen
  \bibfield  {author} {\bibinfo {author} {\bibfnamefont {E.}~\bibnamefont
  {Weibel}},\ }\bibfield  {title} {\enquote {\bibinfo {title} {{Weibel
  Instability}},}\ }\href@noop {} {\bibfield  {journal} {\bibinfo  {journal}
  {Physical Review Letters}\ }\textbf {\bibinfo {volume} {2}} (\bibinfo {year}
  {1959})}\BibitemShut {NoStop}%
\bibitem [{\citenamefont {Berger}\ \emph {et~al.}(1991)\citenamefont {Berger},
  \citenamefont {Albritton}, \citenamefont {Randall}, \citenamefont {Williams},
  \citenamefont {Kruer}, \citenamefont {Langdon},\ and\ \citenamefont
  {Hanna}}]{Berger1991}%
  \BibitemOpen
  \bibfield  {author} {\bibinfo {author} {\bibfnamefont {R.~L.}\ \bibnamefont
  {Berger}}, \bibinfo {author} {\bibfnamefont {J.~R.}\ \bibnamefont
  {Albritton}}, \bibinfo {author} {\bibfnamefont {C.~J.}\ \bibnamefont
  {Randall}}, \bibinfo {author} {\bibfnamefont {E.~a.}\ \bibnamefont
  {Williams}}, \bibinfo {author} {\bibfnamefont {W.~L.}\ \bibnamefont {Kruer}},
  \bibinfo {author} {\bibfnamefont {a.~B.}\ \bibnamefont {Langdon}}, \ and\
  \bibinfo {author} {\bibfnamefont {C.~J.}\ \bibnamefont {Hanna}},\ }\bibfield
  {title} {\enquote {\bibinfo {title} {{Stopping and thermalization of
  interpenetrating plasma streams}},}\ }\href {\doibase 10.1063/1.859954}
  {\bibfield  {journal} {\bibinfo  {journal} {Physics of Fluids B: Plasma
  Physics}\ }\textbf {\bibinfo {volume} {3}},\ \bibinfo {pages} {3--12}
  (\bibinfo {year} {1991})}\BibitemShut {NoStop}%
\end{thebibliography}
\end{document}